\documentclass[twocolumn,showpacs,preprintnumbers,amsmath,amssymb,pra]{revtex4}
\usepackage{graphicx}
\usepackage{dcolumn}
\usepackage[tight]{subfigure}
\usepackage{amsmath}
\usepackage{verbatim}
\usepackage[usenames,dvipsnames]{color}
\usepackage{bm} 
\usepackage{bbm}

\newcommand{\Figure}[2]{
  \begin{figure}[ht]
    \includegraphics[width=0.9\linewidth]{#1}
    \caption{#2}
  \end{figure}
}

\newcommand{\Bigfigure}[2]{
  \begin{figure*}[ht]
    \includegraphics[width=0.9\linewidth]{#1}
    \caption{#2}
  \end{figure*}
}

\newcommand{\wideeq}[1]{\begin{widetext}
    #1
  \end{widetext}}

\newcommand{\Fig}[1]{Fig.~\ref{#1}}

\newcommand{\Eq}[1]{Eq.~(\ref{#1})}
\newcommand{\eq}[1]{(\ref{#1})}
\newcommand{\Sec}[1]{Sec.~\ref{#1}}
\renewcommand{\sec}[1]{\ref{#1}}
\newcommand{\App}[1]{App.~\ref{#1}}

\newcommand{\Cite}[1]{Ref.~\onlinecite{#1}}

\newcommand{\new}[1]{\textcolor{blue}{#1}}
\newcommand{\newer}[1]{\textcolor{blue}{#1}}
\newcommand{\cut}[1]{\textcolor{red}{cut: #1}}
\newcommand{\todo}[1]{ \textbf{\textcolor{Bittersweet}{TODO: #1}} }
\newcommand{\hide}[1]{ \textbf{\textcolor{Gray}{#1}} }

\newcommand{\nocomma}{}
\newcommand{\tmem}[1]{{\em #1\/}}
\newcommand{\tmop}[1]{\ensuremath{\operatorname{#1}}}

\newcommand{\nobracket}{}

\renewcommand{\new}[1]{{#1}}
\renewcommand{\newer}[1]{{#1}}
\renewcommand{\cut}[1]{{}}
\renewcommand{\todo}[1]{}
\renewcommand{\hide}[1]{}
\newcommand{\op}[1]{\hat{#1}}

\newcommand{\vecg}[1]{{\bm #1}}

\newcommand{\ket}[1]{|#1 \rangle}

\begin{document}
\title{
  Time scales for Majorana manipulation\\
 using Coulomb blockade in gate-controlled superconducting nanowires
}

\author{Michael Hell$^{(1,2)}$}
\author{Jeroen Danon$^{(2,3)}$}
\author{Karsten Flensberg$^{(2)}$}
\author{Martin Leijnse$^{(1,2)}$}
\affiliation{\\
(1) Division of Solid State Physics and NanoLund, Lund University, Box.~118, S-22100, Lund, Sweden \\
(2) Center for Quantum Devices and Station Q Copenhagen, Niels Bohr Institute, University of Copenhagen, Copenhagen, Denmark \\
(3) Niels Bohr International Academy, Niels Bohr Institute, University of Copenhagen, Copenhagen, Denmark
}

\date{\today}

\begin{abstract}
  We numerically compute the low-energy spectrum of a gate-controlled
  superconducting topological nanowire segmented into two islands, each
  Josephson-coupled to a bulk superconductor. This device may host two pairs
  of Majorana bound states and could provide a platform for testing Majorana
  fusion rules. We analyze the crossover between (i) a
  charge-dominated regime utilizable for initialization and readout of
  Majorana bound states, (ii) a single-island regime for dominating
  inter-island Majorana coupling, (iii) a Josephson-plasmon regime for large
  coupling to the bulk superconductors, and (iv) a regime of four Majorana
  bound states allowing for topologically protected Majorana manipulations.
  From the energy spectrum, we derive conservative estimates for the time
  scales of a fusion-rule testing protocol proposed recently
  [arXiv:1511.05153]. We also analyze the steps needed for 
  basic Majorana braiding operations in branched nanowire structures.
\end{abstract}

 \maketitle

  \section{Introduction}\label{sec:intro}
  
  Systems with topologically nontrivial phases have become a focal point of condensed-matter
  research over the past decade \cite{BernevigBook} and especially systems
  hosting Majorana bound states (MBS) {\cite{KitaevWireMajorana:01}}
  have been heavily pursued
  {\cite{AliceaReview,FlensbergReview,BeenakkerReview,TewariReview,FranzReview,DasSarmaReview15}}.
Two MBS can form a fermionic mode, which can be occupied at the cost of 
zero energy, that is, the ground state has a
  fermion-parity degeneracy. This degeneracy is topologically protected against perturbations, which implies that MBS obey
  non-Abelian exchange statistics {\cite{Rowell15}}. Hence, exchanging MBS (braiding) 
  changes the ground-state in a nontrivial way, a key
  ingredient for topological quantum computation
  {\cite{DasSarmaReview08,kitaev,HasslerSpringSchool,Ivanov01,TQCreview}}. 

Identifying a suitable platform for realizing and manipulating MBS, however, remains challenging.
MBS exist only in superconductors with triplet pairing \cite{Read00}, which appears intrinsically in Sr$_2$RuO$_4$ \cite{DasSarma06} or can be induced extrinsically as a proximity effect \cite{FuKane}. 
The first candidate systems for MBS were vortices of 2D triplet superconductors, in which the MBS might be manipulated through gate-voltage controlled point contacts \cite{Liang12} or supercurrents \cite{InterdigitatedDevice}.
As an arguably more feasible alternative, 1D systems have been considered {\cite{MajoranaQSHedge,Sau,NoSOC1} and among those magnetic atom chains \cite{Yazdani} and semiconductor nanowires {\cite{1DwiresLutchyn,1DwiresOreg}} have been suggested and seem experimentally promising. Here, the combined effect of strong
  spin-orbit coupling, (proximity-induced) superconductivity, and exchange interactions or Zeeman splitting \cite{Alicea} induces MBS located at the opposite ends of a topological phase region. Experiments
  have so far focused on probing transport signatures of MBS
  {\cite{Rokhinson,Nadj-Perge,finck12}}, such as a zero-bias conductance peak
  {\cite{mourik12,das12,deng12,Churchill}}, but they could not conclusively
  rule out other topologically trivial origins.
  \begin{center}
    \Figure{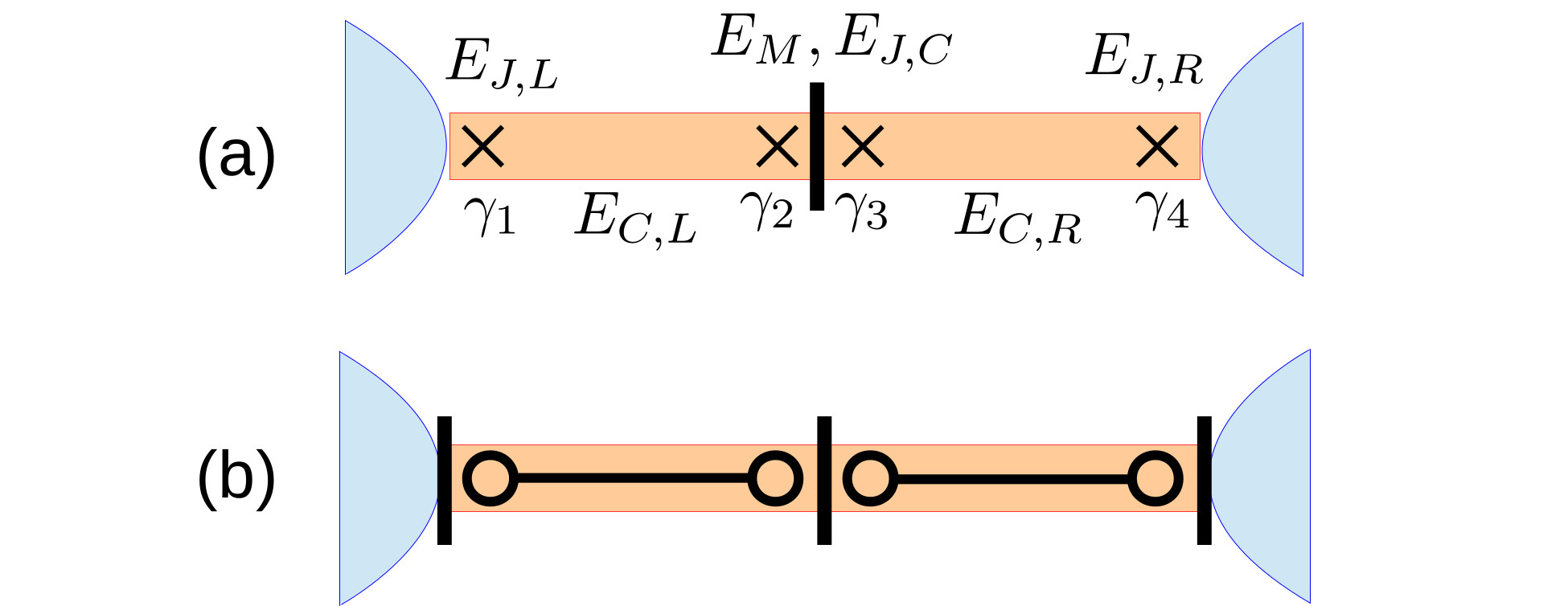}
     {Segmented nanowire setup for gate-controlled fusion of MBS. The device consists of
    two tunnel-coupled superconducting islands (orange), formed by a nanowire in
    proximity to a superconductor (not shown). In addition to that, both islands are connected to nontopological
    bulk superconductors (blue) with a tunable coupling. In (a), when the outer valves are maximally
    open (largest coupling) and the central valve is closed, the device hosts four MBS at zero
    energy (crosses). In (b), when all junctions are closed (minimal coupling), the pairs of MBS on each
    island are {\tmem{fused}} (connected circles). In this situation, the
    device may still possess a subgap state provided $E_{C, \alpha} < \Delta$
    ($\alpha = L, R$), rendering the sketched setup distinct from a
    conventional nontopological Cooper pair box.\label{fig:model}}
  \end{center}
Clear evidence in favor of
  the topological nature of MBS would instead be provided by verifying their
  distinctive exchange characteristics. Various approaches to braiding have been suggested, which fall into two categories: One way is to move topological phase boundaries, for example, in nanowire-based proposals through keyboard gates \cite{AliceaBraiding,BondersonBraiding,ClarkeBraiding,HalperinBraiding} or supercurrents \cite{Romito}. Another way is to slowly change the couplings between the MBS to adiabatically manipulate the ground state \cite{Burello13}. This  
 can be achieved, for example, by magnetic-field control for magnetic atom rings \cite{Li14} or by tuning electric gates \cite{SauBraiding} as well as magnetic fluxes \cite{BeenakkerBraiding} in nanowire devices. With the ongoing progress
 in nanowire fabrication
  {\cite{Gul,Marcus15_NatureMat_14_400,Marcus15_NatureNano_10_232}}, especially the fabrication of 
  branched structures {\cite{Dick04,WireNetworks}}  essential for braiding,
  experiments have begun to move forward in this direction.

The successful implementation of braiding necessitates, however, also initialization
  and readout of MBS and this requires lifting the ground-state degeneracy --- a
  process called {\tmem{fusion}} of MBS {\cite{TQCreview}}. MBS obey nontrivial fusion rules that, in fact, imply braiding \cite{Rowell15} and are therefore an interesting subject in itself.
Avoiding errors by tuning
  between the fused and degenerate regime will practically limit the operation
  speed of topological devices in experiments. The limitations arising from these steps may even be more restrictive than those for manipulating MBS during braiding as we specifically show for nanowire setups.
Estimating such time scales,
  inferred from studying the behavior of the energy spectrum, is thus of the
  utmost importance to devise future experiments.

  A viable strategy to fuse MBS controllably in nanowire setups
  is to form {\tmem{mesoscopic}} superconducting islands (see Fig. \ref{fig:model}), which form the basis of various nontopological qubits \cite{Makhlin01,Devoret04}, including Cooper pair boxes and transmon qubits {\cite{CottetPhD,Koch}}. In the topological case, the charging energy $E_C$ of the islands introduces an energy splitting $\varepsilon_P \sim E_C$}
  between states of different fermion parity. This fuses the MBS as indicated in Fig.~\ref{fig:model}(b) by connected circles. By coupling the island through a junction to
  a bulk superconductor, the parity splitting $\varepsilon_P \sim ( E_J^3
  E_C)^{1 / 4} e^{ - \sqrt{8 E_J / E_C}}$ can be made
  exponentially small by tuning the Josephson energy $E_J$ of this junction~
  {\cite{Hassler}}. This introduces MBS at zero energy, at
  least up to exponential accuracy as indicated in Fig. \ref{fig:model}(a) by crosses.
  The first proposals of this kind envisaged the Josephson couplings to be
  controlled by {\tmem{magnetic fluxes}}
  {\cite{BraidingWithoutTransport,BeenakkerBraiding}}, and parity readout to
  be accomplished by a cavity, analogous to transmon qubits
  {\cite{TopTransmon}}. 
  
A complementary, all-{\tmem{electrical}}
  proposal was put forward in {\color{black} Ref. {\cite{Aasen15}}. In contrast to superconducting qubits typically implemented in metallic systems \cite{Makhlin01}, 
 this approach is based on gateable semiconductor nanowire
 Josephson junctions. Such junctions have been demonstrated
  experimentally recently for nontopological devices
  {\cite{LangeNanowireJosephson,Gatemon}} with prospects also for
  nontopological quantum computation {\cite{Shim}}. This approach allows the
  application of experimental tools from quantum-dot experiments, including
  parity readout by charge sensing and charge pumping. Based on this, a sequence
  of stepping stones interpolating between MBS detection and quantum computing
  was suggested, among these the detection of MBS {\tmem{fusion rules}}
  and {\tmem{braiding}}~\cite{Aasen15}. While braiding requires branched nanowire structures,
  a fusion-rule test could already be realized with a {\tmem{single}}
  nanowire: A prototypical device would consist of a superconducting nanowire
  hosting two islands in series, which are each coupled to a bulk superconductor as
  sketched in {\color{black} Fig. \ref{fig:model}}.

  In this paper, we numerically compute the low-energy spectrum for this
  setup, extending prior studies on coupled Cooper-pair boxes {\cite{Shim}}
  by accounting for both the parity degree of freedom and the charge state of
  the islands. This goes beyond several other studies of MBS devices for braiding that either
  exclude charging energy
  {\cite{AliceaBraiding,ClarkeBraiding,SauBraiding,HalperinBraiding,BondersonBraiding}}, or
  treat charging only effectively
  {\cite{BeenakkerBraiding,BraidingWithoutTransport}}.

Accounting for
  charging effects in the entire range between $E_C \ll E_J$ and $E_C > E_J$
  is necessary to study the crossover between the degenerate and the fused
  regime both used in the proposal of {\color{black} Ref. {\cite{Aasen15}}}. In
  the latter case, the charge stored in the superconducting Cooper pair
  condensate cannot be disregarded any more.
  
  In our study, we will identify four different operating regimes for the coupled topological
  superconducting islands: (i) a charging-dominated regime [see
  {\color{black} Fig. \ref{fig:model}}(b)], (ii) a single-island regime, (iii) a double-island
  regime, and (iv) a regime of four MBS [see {\color{black} Fig.
  \ref{fig:model}}(a)]. These regimes can be divided into
  several subregimes depending on the ``fine structure'' of the energy
  spectrum. Mapping out these regimes is useful for the experimental
  characterization and needed to study the time scales for operating this
  device.
  
  We further derive {\tmem{conservative}} time-scale
  conditions for the fusion-rule testing protocol suggested in Ref. {\cite{Aasen15}}}. In this protocol, one changes the Majorana couplings ($E_M$) and the Josephson couplings ($E_{J, L / R}$), which are indicated in Fig.~\ref{fig:model}(a), in order to tune the system through the above-mentioned
  regimes (i) $\rightarrow$ (iv). The manipulations have to be made on a time scale
  $\Delta t$ that is slow in the sense that
  \begin{eqnarray}
    \Delta t & \gg & \frac{\ln [ \max ( E_{J, L / R}^{\max}, E_M^{\max}) /
    E_C]}{E_C} \nocomma,  \label{eq:timescalelower}
  \end{eqnarray}
  but at the same time fast in the sense that
  \begin{eqnarray}
    \Delta t & \ll & \frac{1}{\max ( \varepsilon_P^{\min}, E_M^{\min})} .  \label{eq:timescaleupper}
  \end{eqnarray}
  (We set $\hbar = e = k_B = 1$.) The above conditions depend on the minimal and maximal values of the couplings, which have to satisfy the condition
  \begin{eqnarray}
    E_C \text{ \ } \ll & E_J^{\max} & \ll \text{ \ } ( E_M^{\max})^2 / E_C
    \text{ \ } \ll \text{ \ } \Delta^2 / E_C.  \label{eq:condscales}
  \end{eqnarray}
  Moreover, Eqs. (\ref{eq:timescaleupper}) and (\ref{eq:condscales}) incorporate the superconducting gap $\Delta$ in the nanowires and the bulk superconductors.
The first criterion
  (\ref{eq:timescalelower}) guarantees that the evolution proceeds
  {\tmem{adiabatically}}, that is, transitions from the ground-state manifold
  into any of the excited state are suppressed. 
However, there is also a {\tmem{diabaticity}} condition because the ground-state degeneracy is changed during the protocol. This degeneracy is not perfect in practice (as for any realistic topological device) but a small energy splitting $\varepsilon_P^{\min}$ remains. Proceeding adiabatically with respect to this (unwanted) remaining splitting could take the system to the lowest of all energy eigenstates.
 This has to be avoided by proceeding fast enough so that the system state has no time to evolve within the ground-state manifold, which leads to the criterion $\Delta t \ll 1/\varepsilon_P^{\min}$ in \Eq{eq:timescaleupper}.
The second condition in Eq. (\ref{eq:timescaleupper}) is needed to avoid relaxation processes (tunneling of single electrons between the islands) when they are unwanted.
 By contrast, resetting the system by going back from (iv) to its initial state at (i) involves a charge-relaxation processes, which contributes to the period of a full cycle [not accounted for by \Eq{eq:timescaleupper}].

With these insights from the single-wire geometry, we additionally analyze
  basic operations on MBS in nanowire networks essential for braiding.
  We show that similar time-scale conditions as Eqs. (\ref{eq:timescalelower})
  and (\ref{eq:timescaleupper}) have to be satisfied. \new{In the context of braiding, diabatic corrections due to a finite operation time have been discussed in the literature \cite{Cheng11}, mostly when the MBS are braided through moving domain walls \cite{Scheurer13,Karzig13,Karzig15}, but also through changing their couplings \cite{Karzig15b,Knapp16}. The latter works also suggest error correction procedures based on introducing counterdiabatic correction terms in the Hamiltonian \cite{Karzig15b}, or smoother parameter changes as well as intermediate measurements \cite{Knapp16}. The focus of our work is on the time scales that are necessary to suppress transition rates into excited states from the start.}

  Our study does not include the effect of quasiparticles which can lead to
  parity flips detrimental to the operation of the devices. Thus, $\Delta t$
  should also be much smaller than quasiparticle poising times and at least
  for closed islands this seems to satisfied in view of recent experimental
  results {\cite{PoisoningTime}}. Our results thus apply when the low-energy
  spectrum lies below the superconducting gap $\Delta$, which is ensured by
  the last inequality in Eq. (\ref{eq:condscales}). In particular, the charging energies $E_{C,\alpha}$ has to be smaller than the superconducting gap $\Delta$. Moreover, $E_{C,\alpha}$ has to be much larger than temperature since thermal fluctuations of the charge states have to be suppressed for initialization and readout.
  
  The paper is structured as follows: In {\color{black} Sec.
  \ref{sec:model}}, we introduce our model for the double-island setup shown in
  {\color{black} Fig. \ref{fig:model}}. After briefly discussing our numerical
  approach in {\color{black} Sec. \ref{sec:numerics}}, we analyze the
  numerically computed energy spectrum in {\color{black} Sec.
  \ref{sec:spectrum}}. We find different operating regimes, which can be
  understood in simple limits from analytic approximations, closely resembling
  the behavior of coupled oscillators. Based on the energy spectrum, we
  discuss the time-scale conditions for the fusion-rule testing protocol in
  {\color{black} Sec. \ref{sec:fusion}} and for basic manipulations in
  branched nanowire structures in {\color{black} Sec. \ref{sec:braiding}}. 
  Finally, Sec. \ref{sec:summary} concludes and summarizes our findings. 
Our analysis is complemented by several appendices, in which we discuss among other things our parameter choices (\App{app:parameter-estimation}), extend our analysis of the parameter regimes mentioned above, e.g., for asymmetric setups (\App{app:low-energy}), and analyze the time scales for the readout (\App{app:readout}).

\section{Model: Gate-tunable coupled superconducting Majorana
islands}\label{sec:model}

The setup we investigate in this work is sketched in {\color{black} Fig.
\ref{fig:model}}: It consists of a segmented superconducting nanowire, which
is at both ends coupled to a bulk superconductor. The superconductivity in
the nanowire is proximity induced, for example, by metal deposition
{\cite{mourik12}} or an epitaxially grown shell coating the nanowire
{\cite{Marcus15_NatureMat_14_400,Marcus15_NatureNano_10_232}}. The combined
proximal superconductor and the nanowire form what we refer to as
{\tmem{superconducting islands}} (red in {\color{black} Fig.
\ref{fig:model}}). The coupling of these islands to the bulk superconductors
and the coupling between the two islands can be controlled electrically by
nearby gates. The junctions can thus be operated as valves, which are open for maximal coupling and closed for minimal coupling. When the nanowire is driven into a topologically nontrivial
regime, as already mentioned in Sec. \ref{sec:intro}, the islands possess an additional ground-state degeneracy
associated with the MBS (see {\color{black} Sec.
\ref{sec:model-mbs}}).

We note that the analysis in this paper is not necessarily tied to
the specific physical realization mentioned above. One might, for example,
also control the Josephson energies by magnetic fluxes {\cite{BeenakkerBraiding}}.
However, it is important that all the couplings can be tuned over a wide range
from $E_J \lesssim E_C$ to $E_J \gg E_C$. To achieve large ratios $E_J/E_C$ with gate control but without perturbing the MBS, it may be experimentally useful to deviate from the geometry shown in Fig.~\ref{fig:model} and instead form a contact from the middle of the nanowire to the ground via a second nanowire allowing gate control of the Josephson energy.

Our goal is to investigate the energy spectrum of this device at energies
below $\Delta = \min ( \Delta_{\tmop{island}}, \Delta_{\tmop{bulk}})$, the
minimum of the superconducting gaps on the island and in the bulk. We model
the device by the following Hamiltonian,
\begin{eqnarray}
  H & = & \sum_{\alpha = L, R} H_{\alpha} + H_T,  \label{eq:hfull}
\end{eqnarray}
which consists of three parts: The first two parts, $H_L$ and $H_R$, describe
the two individual superconducting islands and their coupling to the bulk superconductors, 
and the third part, $H_T$, accounts
for the tunnel coupling between the islands when opening the central junction. 
We neglect here
capacitive coupling between the islands since there are indications that they are much smaller than the
local charging energies {\cite{Aasen15}}. We have verified that small cross-capacitive couplings (much smaller than the local charging energies) do not impair our results for the operating regimes and the time scales discussed below. The reason is that capacitive couplings are relevant only when the charging energies are the dominating energy scale; and otherwise they introduce only minor corrections for the energy spectrum. 

\subsection{Hamiltonian for superconducting islands}\label{sec:singleham}

The superconducting islands $( \alpha = L, R)$ are modeled in the standard way
for a Cooper pair box {\cite{Koch}}:
\begin{eqnarray}
  H_{\alpha} & = & H_{C, \alpha} + H_{J, \alpha} .  \label{eq:hi}
\end{eqnarray}
The first term incorporates the classical Coulomb interaction between the
electrons on the islands,
\begin{eqnarray}
  H_{C, \alpha} & = & E_{C, \alpha} \left( \op{n}_{\alpha} - n_{g, \alpha}
  \right)^2,  \label{eq:hc}
\end{eqnarray}
while the second term accounts for the Josephson couplings to the bulk
superconductors:
\begin{eqnarray}
  H_{J, \alpha} & = & E_{J, \alpha} \left( 1 - \cos \left(
  \op{\varphi}_{\alpha} \right) \right) .  \label{eq:hj}
\end{eqnarray}
Here, the operator $\op{n}_{\alpha}$ counts the number of excess
{\tmem{electrons}} on island $\alpha = L, R$ relative to an arbitrary offset.
The electron number with the minimal energy can be tuned by nearby gates
changing $n_{g, \alpha}$. The operator $\op{n}_{\alpha}$ is conjugate to the
operator $\op{\varphi}_{\alpha}$ of the phase difference between island and
bulk:
\begin{eqnarray}
  \left[ \op{\varphi}_{\alpha}, \op{n}_{\beta} \right] & = & 2 i
  \delta_{\alpha \beta}.  \label{eq:comm}
\end{eqnarray}
This means that the phase operator
generates changes of the charge {\cite{MajoranaTransportWithInteractions1}},
\begin{eqnarray}
  e^{\pm i \op{\varphi}_{\alpha}} | n_{\alpha} \rangle & = & | n_{\alpha} \pm
  2 \rangle,  \label{eq:displ}
\end{eqnarray}
where $\{ | n_{\alpha} \rangle : n_{\alpha} \in \mathbbm{Z} \}$ denotes the
orthonormal number basis consisting of states with $n_{\alpha}$ electrons on
island $\alpha = L, R$.

One may question at this point whether Eq. (\ref{eq:hj}) is an appropriate
model for the Josephson energy for semiconductor nanowire junctions. Recent
experiments on gatemons {\cite{Gatemon,LangeNanowireJosephson}} indicate that
such junctions connect the island with the bulk through a few, say $N$,
channels with large transmission probability $T_i \sim 1$ when the valve is
opened. Then a different expression for the Josephson Hamiltonian involving
higher harmonics in $\op{\varphi}_{\alpha}$ should be used as discussed in
Refs. {\cite{Aasen15,LangeNanowireJosephson}}. 
Moreover, the charging energy might be renormalized when the transmission amplitudes are not small, similar to Refs. \cite{Molenkamp95ChargeFluctuations,Schoeller94QuantumFluctuations}.
 Our model is thus strictly
valid only if the transmission probabilities of all channels are small.
 This
implies that we can study the regime $E_{J, \alpha} \gg E_{C, \alpha}$ only
under the assumption $E_{C, \alpha} \ll \Delta$ because the Josephson
couplings scale as $E_{J, \alpha} \propto \sum_{i = 1}^N T_i \Delta$
{\cite{NazarovBook}}.
However, we emphasize that all the above-mentioned effects will serve to enlarge the window for the time scales compared with our derivation, which therefore remains a useful conservative estimate.

\subsection{Majorana bound states and basis}\label{sec:model-mbs}

At first sight, Eq. (\ref{eq:hi}) does not seem to differ from a standard
Cooper pair box for a topologically trivial superconducting island. What is
different here, however, is that the fermion number can be both even {\tmem{and}}
odd, i.e., we have to account for both {\tmem{fermion parity}} sectors.
Introducing the fermion-parity operator,
\begin{eqnarray}
  P & = & \tfrac{1}{2}(1-( - 1)^{\op{n}_{\alpha}}),
\end{eqnarray}
one can project Eq. (\ref{eq:hi}) onto the two subspaces according to its eigenvalues $p=0$ (even fermion parity) and $p=1$ (odd fermion
parity), respectively, which yields
\begin{eqnarray}
  H_{\alpha} & = & H_{\alpha}^{p = 0} + H_{\alpha}^{p = 1} .
\end{eqnarray}
There are no off-diagonal blocks between even and odd
parity because the Hamiltonians $H_{\alpha}$ ($\alpha = L, R$) conserve the
fermion parity for each island {\footnote{This holds irrespective of the
specific form of the Josephson coupling. The bulk superconductors can only
transfer electrons in (Cooper) {\tmem{pairs}} to the island, which does not
change the fermion parity of the islands.}}:
\begin{eqnarray}
  \left[ H_{\alpha}, P \right] & = & 0. 
\end{eqnarray}
In a topologically trivial superconductor one would account only for the even
parity part, $H_{\alpha}^{p = 0}$, and omit the odd parity part,
$H_{\alpha}^{p = 1}$. The reason is that an odd parity state requires an
additional quasiparticle mode to be occupied, which is associated with an
energy at least as large as the superconducting gap $\Delta$, which is outside
the energy regime we are interested in here.

{\tmem{Majorana bound states.}} The topologically nontrivial
superconducting islands each possess an additional fermionic mode, associated
with the field operators $\op{f}_{1 2} = \left( \op{\gamma}_1 + i
\op{\gamma}_2 \right) / 2$ for the left island and $\op{f}_{3 4} = \left(
\op{\gamma}_3 + i \op{\gamma}_4 \right) / 2$ for the right island,
respectively. When the outer valves in our device are opened and the central
valve is closed, these additional modes derive from pairs of Majorana bound
states (MBS) at zero energy, $\gamma_n$ ($n = 1, \ldots, 4$), localized at the
opposite ends of the wire segments {\cite{AliceaReview,FlensbergReview}} as
sketched in {\color{black} Fig. \ref{fig:model}}(a). The associated
self-conjugate Majorana operators, $\op{\gamma}_n = \op{\gamma}^{\dag}_n$, are
denoted with hats and satisfy anticommutation relations $\{
\op{\gamma}_n, \op{\gamma}_m \} = 2 \delta_{n m}$. We
assume that the two Majorana wave functions on each island do not overlap in
space. As a consequence, occupying the fermionic modes $f_{1 2}$ and $f_{3 4}$
is associated with {\tmem{zero}} ``orbital'' energy. Hence, both fermion
parity sectors are accessible at low energies $< \Delta$, in contrast to a
topologically trivial superconductor. The degeneracy between the even and odd
parity sectors can be lifted either by the charging energy [see {\color{black}
Fig. \ref{fig:model}}(b)] {\cite{MajoranaTransportWithInteractions1}} or by
the tunnel coupling of the central valve, which in both cases fuses MBS.

{\tmem{Phase basis and Majorana operators.}} 
\new{We give an intuitive definition of the Majorana operators first in the phase basis $\ket{\varphi_L,\varphi_R}$, which is} the Fourier transform of the number basis. First considering only the
left island, we define (see \App{app:derh}) \cite{MajoranaTransportWithInteractions1}
\begin{eqnarray}
  | 0_{1 2}, \varphi_L \rangle & = & \frac{1}{\sqrt{2 \pi}} \sum_{n_L \text{
  even}} e^{- i \varphi_L n_L / 2} | n_L \rangle,  \label{eq:phie}\\
  | 1_{1 2}, \varphi_L \rangle & = & \frac{1}{\sqrt{2 \pi}} \sum_{n_L \text{
  odd}} e^{- i \varphi_L n_L / 2} | n_L \rangle,  \label{eq:phio}
\end{eqnarray}
where the variable $\varphi_L
\in [ 0, 2 \pi)$ is continuous and $p_{12}=0_{12}, 1_{12}$ denotes the occupation of fermionic mode
$f_{1 2}=\gamma_1 + i \gamma_2$ and thus the fermion parity of the island {\footnote{We use a different sign
factor in the exponentials of Eqs. (\ref{eq:phie}) and (\ref{eq:phio}) as
compared to {\color{black} Ref. {\cite{MajoranaTransportWithInteractions1}}}
so that the displacement relation (\ref{eq:displ}) becomes compatible with
the commutation relation (\ref{eq:comm}).}}. 
When used as a label, we add a subscript (here $12$) to the numbers $0$ and $1$ to denote the fermionic mode that is meant; however, in mathematical expressions $p_{12}$ should be evaluated as the numbers $0$ and $1$. \new{The wave functions can be represented in phase space as $\psi_{p_{12}}(\varphi_L) = \langle  p_{1 2}, \varphi_L | \psi \rangle$, which obey the periodicity condition 
\begin{eqnarray}
 \psi_{p_{12}} (\varphi_L + 2 \pi ) &=& (-1)^{p_{12}} \psi_{p_{12}}(\varphi_L). \label{eq:boundarycond}
\end{eqnarray}
The action of the Majorana operators
$\op{\gamma}_1$ and $\op{\gamma}_2$ on the wave functions can then be defined through
\begin{eqnarray}
  e^{\pm i \op{\varphi}_L / 2} \op{\gamma}_1 \psi_{p_{1 2}}(\varphi_L) & = & e^{\pm i {\varphi}_L / 2}\psi_{\bar{p}_{1 2}}(\varphi_L),  \label{eq:gamma1}\\
  e^{\pm i \op{\varphi}_L / 2} \op{\gamma}_2 \psi_{p_{1 2}}(\varphi_L) & = & e^{\pm i {\varphi}_L / 2} i^{1 - 2 p_{1 2}} \psi_{\bar{p}_{1 2}}(\varphi_L), \label{eq:gamma2}
\end{eqnarray}
where $\bar{p}_{12}=1-p_{12}$. We added the phase factors here because the Majorana operators appear only in combination with them in the tunneling Hamiltonian \eq{eq:ht} discussed below (which is all we need). Moreover, because of the phase factors $e^{\pm i {\varphi}_L / 2}$, the phase-space wave functions on the right-hand side of Eqs. \eq{eq:gamma1} and \eq{eq:gamma2} obey automatically the boundary conditions \eq{eq:boundarycond}. We give a derivation of the above relations in \App{app:derhi} starting from a standard BCS description of the island.}

\new{Since we express the Hamiltonian in the number basis for our numerics, we further note the following useful relations:
\begin{eqnarray}
  e^{\pm i \op{\varphi}_L / 2} \op{\gamma}_1 | n_L \rangle & = & e^{\pm i
  \op{\varphi}_L / 2} i^{2 n_L - 1} \op{\gamma}_2 | n_L \rangle \\ 
  & = & | n_L \pm 1 \rangle.  \label{eq:fusion-displse}
\end{eqnarray}
}Corresponding relations hold when
replacing $L \rightarrow R$ and $1 2 \rightarrow 34$ for the right island and
a full basis can be formed by tensor-product states. 

\subsection{Majorana-Josephson coupling}\label{sec:tunham}

Without tunnel coupling across the center junction, the even and odd parity
sectors for each of the islands decouple. This changes with a
tunnel coupling, which we model with the following Hamiltonian:
\begin{eqnarray}
  H_T & = & E_{J, C} \left( 1 - \cos \left( \op{\varphi}_L - \op{\varphi}_R
  \right) \right) \nonumber\\
  &  & + E_M \cos \left( \frac{\op{\varphi}_L - \op{\varphi}_R}{2} \right) i
  \op{\gamma}_2 \op{\gamma}_3 .  \label{eq:ht}
\end{eqnarray}
The first term is the ``conventional'' Cooper pair tunneling associated with a
Josephson energy $E_{J, C}$, which conserves the fermion parities of both
islands. The second term, also known as
the fractional Josephson effect {\cite{MajoranaQSHedge}}, involves parity flips.  \new{We derive the Majorana-Josephson term in \App{app:derht}, which also shows that the combination $i \hat{\gamma}_2 \hat{\gamma}_3$ appears naturally since the tunnel coupling is local.
The Majorana-Josephson term can be interpreted most clearly
by comparing with its representation in the number basis [using \Eq{eq:fusion-displse}]:}
\begin{eqnarray}
  H_M & = & - \frac{E_M}{2} \sum_{n_{\alpha}, \eta = \pm 1} | n_L - \eta, n_R
  + \eta \rangle \nobracket \langle n_L n_R \nobracket | . 
  \label{eq:htcharge}
\end{eqnarray}
Equations (\ref{eq:ht}) and (\ref{eq:htcharge}) together show that the
transfer of single electrons across the central junction [described by Eq. (\ref{eq:htcharge})] leads to a
transfer of charge between the islands [through the phase-dependent terms in
Eq. (\ref{eq:ht})] as well as a flip of their fermion parities [through the
Majorana operators in Eq. (\ref{eq:ht})]. 

We note that in combining Eq. (\ref{eq:ht}) with the island Hamiltonians
(\ref{eq:hi}) in the full Hamiltonian (\ref{eq:hfull}), we assume that there
is no phase difference across the two bulk superconductors. This is further
discussed in {\color{black} App. \ref{app:phasedifferences}} and motivated
mainly by the fact that any phase difference would increase the ground-state
energy (and for the protocols we discuss the system should stay mostly in the ground state, at least when phase differences could be relevant).

Even though $E_{J, C}$ and $E_M$ appear as independent parameters in Eq.
(\ref{eq:hfull}), they cannot be controlled individually in an experiment with
a gate at the central junction. We estimate in {\color{black} App.
\ref{app:emejratio}} that they are related by $E_{J, C} \sim E^2_M / \Delta$
for typical parameters. The Josephson energy $E_{J, C}$ and the Majorana
coupling $E_M$ may therefore become of comparable size only when $E_M$ approaches the superconducting gap $\Delta$. Since we assume $E_M\ll \Delta$ during MBS manipulations, we will thus set
$E_{J, C} = 0$ in some parts of our analysis, which simplifies the considerations. However, we
point out that a nonzero central Josephson coupling is not
detrimental to the gate-controlled approach for manipulating MBS.

{\tmem{Total parity conservation.}} The Hamiltonian (\ref{eq:hfull}) conserves
the total fermion parity:
\begin{eqnarray}
  \left[ H, P \right] & = & 0. 
  \label{eq:fusion-parityconserve}
\end{eqnarray}
This is seen from the representation (\ref{eq:htcharge}) of the
Hamiltonian in the number basis. Only the fermion parity of the individual
islands may be changed by the tunneling process. In contrast to the total
charge, the total parity is thus always a good quantum number {\footnote{A state $| \psi \rangle
\tmop{has}$ even (odd) total parity here if the {\tmem{difference}} of the
number of fermionic modes occupied for $| \psi \rangle$ and the reference
state $| n_L, n_R \rangle = | 0, 0 \rangle$ is even (odd). Note that $n_L$ and
$n_R$ are just defined up to arbitrary offset electron numbers whose parities
are not specified.}}. The coupling to
the environment can break fermion parity conservation (so-called quasiparticle poisoning), which is an experimentally relevant issue. Including such
processes is beyond the scope of this paper; a brief discussion of this issue
can be found in {\color{black} Ref. {\cite{Aasen15}}}. Basically, we expect
that such processes happen on time scales long compared to those on which MBS
will be operated in such devices.

\subsection{Hamiltonian in sum and difference variables}\label{sec:sumdiff}

The above terms of the Hamiltonian are expressed in the phase and number
operators referring to the individual islands. For our physical discussion of
the energy spectra, it will be useful to express the Hamiltonian instead in the sums
and differences of the phase and number operators,
\begin{eqnarray}
  \left(\begin{array}{c}
    \op{\Phi}\\
    \Delta \op{\varphi}
  \end{array}\right) & = & \left(\begin{array}{cc}
    1 & 1\\
    1 & - 1
  \end{array}\right) \left(\begin{array}{c}
    \op{\varphi}_L\\
    \op{\varphi}_R
  \end{array}\right), \\
  \left(\begin{array}{c}
    \op{N}\\
    \Delta \op{n}
  \end{array}\right) & = & \left(\begin{array}{cc}
    1 & 1\\
    1 & - 1
  \end{array}\right) \left(\begin{array}{c}
    \op{n}_L\\
    \op{n}_R
  \end{array}\right), 
\end{eqnarray}
which is merely a canonical transformation (up to normalization constants). These
operators form again a set of canonically conjugate operator pairs:
\begin{eqnarray}
  \left[ \op{\Phi} / 2, \op{N} / 2 \right] & = & \left[ \Delta \op{\varphi} /
  2, \Delta \op{n} / 2 \right] \text{ \ = \ } i, \\
  \left[ \op{\Phi}, \Delta \op{n} \right] & = & \left[ \op{N}, \Delta
  \op{\varphi} \right] \text{ \ = \ } 0. 
\end{eqnarray}
In terms of these operators, the Hamiltonian reads
\wideeq{
\begin{eqnarray}
  H & = & 2 \bar{E}_C \left[ ( 1 - \varepsilon_C^2) \left( \tfrac{\op{N} -
  N_g}{2} \right)^2 \right. \left. + \left. \left( \tfrac{\Delta \op{n} -
  \Delta n_g}{2} \right. + \varepsilon_C  \tfrac{\op{N} - N_g}{2} \right)^2
  \right] + 2 \bar{E}_J  \left[ 1 - \cos \left( \tfrac{\op{\Phi}}{2} \right)
  \cos \left( \tfrac{\Delta \op{\varphi}}{2} \right) \right. - \left.
  \varepsilon_J \sin \left( \tfrac{\op{\Phi}}{2} \right) \sin \left(
  \tfrac{\Delta \op{\varphi}}{2} \right) \right] \nonumber\\
  &  & + E_M \cos \left( \tfrac{\Delta \op{\varphi}}{2} \right) i
  \op{\gamma}_2 \op{\gamma}_3 + E_{J, C} \left( 1 - \cos \left( \Delta
  \op{\varphi} \right) \right)  \label{eq:hsumdiff}
\end{eqnarray}}

This contains the averages,
\begin{eqnarray}
  \bar{E}_C \text{ \ = \ } \frac{E_{C, L} + E_{C, R}}{2}, &  & \bar{E}_J
  \text{ \ = \ } \frac{E_{J, L} + E_{J, R}}{2},  \label{eq:ecbar}
\end{eqnarray}
and asymmetry parameters,
\begin{eqnarray}
  \varepsilon_C \text{ \ = \ } \frac{E_{C, L} - E_{C, R}}{E_{C, L} + E_{C,
  R}}, &  & \varepsilon_J \text{ \ = \ } \frac{E_{J, L} - E_{J, R}}{E_{C, L} +
  E_{C, R}} ,  \label{eq:epsj}
\end{eqnarray}
as well as the sum and difference of the gatings:
\begin{eqnarray}
  N_g & = & n_{g,L} + n_{g,R}, \\
  \Delta n_g & = & n_{g,L} - n_{g,R} . 
\end{eqnarray}
This completes our discussion of the model Hamiltonian and we explain next how we diagonalize it numerically.

\section{Numerical diagonalization}\label{sec:numerics}

We compute the spectrum of the model Hamiltonian (\ref{eq:hfull}) by
expressing it as a matrix in the number basis $| n_L, n_R \rangle = | n_L
\rangle \otimes | n_R \rangle$ based on Eqs. (\ref{eq:hc}), (\ref{eq:hj}), and
(\ref{eq:ht}) together with Eqs. (\ref{eq:displ}) and (\ref{eq:htcharge}).
Since the Hamiltonian (\ref{eq:hfull}) conserves the total parity, both the
total parity sectors can be diagonalized individually for all parameters. We
restrict our calculations to the subspace of even total parity unless stated
otherwise.

For the numerical diagonalization, we introduce a cutoff for the maximal
electron number that we include: $| n_L |, | n_R | \leqslant N_{\max}$. This
is sufficient because the charging energy acts like a parabolic potential for
a ``particle'' that is confined in number space. In this analogy, the kinetic
energy of the particle is given by the energy scale $E_T = \max ( E_{J, C},
E_M, E_{J, L}, E_{J, R})$. If the potential energy $\sim E_C ( n_L^2 + n_R^2)$
exceeds the kinetic-energy scale $E_T$, the contributions from the corresponding
number states become exponentially small for the lowest eigenstates. Thus,
one may neglect states for $n_L, n_R \gtrsim N_{\max} = \sqrt{E_T / E_C}$.
Since we keep $E_T \lesssim 100 E_C$ in our simulations, choosing a cutoff
$N_{\max} > 10$ already yields the low-energy spectrum with high accuracy.

Diagonalizing the Hamiltonian in the number basis turns out to be much more convenient in the presence of charging energies
as compared to diagonalizing it in the phase basis as we further substantiate in
{\color{black} App. \ref{app:numerics}}.

\section{Low-energy spectrum}\label{sec:spectrum}

We next discuss the low-energy spectrum of the coupled topological
superconducting islands. In addition to providing an overview of the different
operating regimes, this is also of experimental interest: A thorough
characterization of such devices will be needed as a preparation step before
the protocols presented in {\color{black} Ref. {\cite{Aasen15}}} can be
implemented experimentally.

To keep the analysis simple in this Section, we neglect the Josephson energy
of the central junction ($E_{J, C} = 0$) and assume symmetric islands ($E_{J,
L} = E_{J, R} = E_J$ and $E_{C, L} = E_{C, R} = E_C$). We dedicate
{\color{black} App. \ref{app:low-energy}} to investigating effects of
deviating from these assumptions. Analyzing the spectrum under the simplifying
assumptions has the advantage that it provides a {\tmem{conservative}}
estimate of the time scales for operating MBS in these devices (see {\color{black} Sec.
\ref{sec:fusion}}). 

The characteristics of the spectrum can be divided into
four different parameter regimes as illustrated in {\color{black} Fig.
\ref{fig:regimes}}: (i) If the charging energy dominates all other energy
scales, $E_C \gg E_J, E_M$, the system behaves as two closed, uncoupled
islands (green in {\color{black} Fig. \ref{fig:regimes}}, see {\color{black}
Sec. \ref{sec:closeduncoupled}}). The islands host a definite number of
electrons except for charge-degeneracy points, depending on the gatings $n_{g,
L}$ and $n_{g, R}$. (ii) If the Majorana tunneling dominates, $E_M \gg
E_J, E_C$, the system behaves as a single, larger island at low energies $<
\sqrt{4 E_M E_C}$ (orange in {\color{black} Fig. \ref{fig:regimes}}, see
{\color{black} Sec. \ref{sec:singleisland}}). The superconducting phases of
both islands are then locked to each other. At larger energies $>\sqrt{4 E_M E_C}$, the dynamics of their phase difference has to be taken into account.

(iii) If instead the Josephson
coupling dominates, $E_J \gg E_M, E_C$, the system has to be treated
rather as two separate open islands (blue in {\color{black} Fig.
\ref{fig:regimes}}, see {\color{black} Sec. \ref{sec:twoopen}}). Here, the phases
of the islands have to be treated as individual degrees of freedom even at low
energies. (iv) Finally, there is the four-MBS regime (yellow in {\color{black}
Fig. \ref{fig:regimes}}, see {\color{black} Sec. \ref{sec:majorana}}), which
appears for two open uncoupled islands. Here, both the tunnel coupling of the
MBS across the center junction and the charging-mediated couplings of
the MBS on each island are strongly suppressed.
\begin{center}
  \Bigfigure{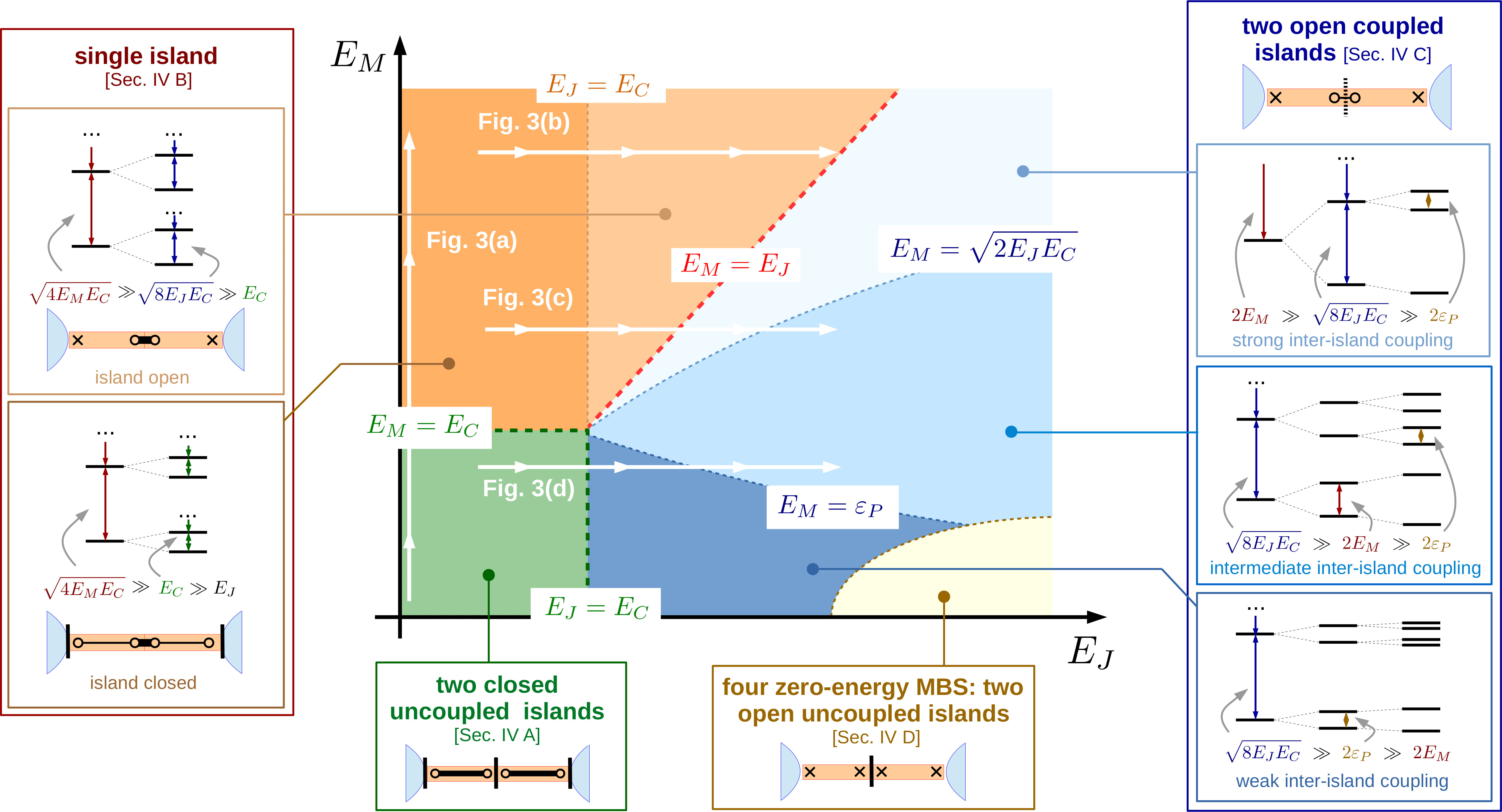}{Sketch
  of the parameter regimes for the energy spectra characteristics of the
  coupled topological superconducting island Hamiltonian (\ref{eq:hfull}). In the boxes, we sketch the energy level spectrum and indicate the different energy scales dominating the level structure. The lower part of the boxes shows the couplings (lines) between the MBS, which are fused when denoted as circles and at close to zero energy when denoted as crosses.
We
  assume symmetric islands, $E_{J, L} = E_{J, R} = E_J$, $E_{C, L} = E_{C, R}
  = E_C$, and have set the central Josephson coupling to zero, $E_{J, C} = 0$.
  The white arrows indicate the paths taken in the $( E_J, E_M)$ plane for the
  plots in {\color{black} Fig. \ref{fig:emejdep}}\label{fig:regimes}}.
\end{center}
Representative plots of the energy spectra along paths in the $( E_J, E_M)$
plane shown in {\color{black} Fig. \ref{fig:regimes}} are shown in
{\color{black} Fig. \ref{fig:emejdep}}. The color of the horizontal axes in
{\color{black} Fig. \ref{fig:emejdep}} corresponds to the regimes shown in
{\color{black} Fig. \ref{fig:regimes}}. From 
{\color{black} Fig. \ref{fig:emejdep}} \ it is clear that the dashed
lines in {\color{black} Fig. \ref{fig:regimes}}, marking the boundaries of the
different regimes, should not be understood as lines of a ``phase
transition''. The transition from one regime to the other is gradual and may
even be shifted for higher-lying excited states. Figure {\color{black}
\ref{fig:regimes}} should thus be understood rather as a rough guide for the
characteristics of the spectra. We next discuss the different regimes in
detail.

\subsection{Two closed uncoupled islands: $E_C \gg E_M,
E_J$}\label{sec:closeduncoupled}

When the charging energy $E_C$ dominates, the eigenstates are close to the
number states $| n_L, n_R \rangle$ except at degeneracy points where $n_{g,
\alpha} = \pm 1/2, \pm 1, \pm 3/2, ... $.
This results in a charge-stability diagram similar to
nonsuperconducting double-dot devices {\cite{vanderWiel03rev}} but with the
constraint that the total parity is conserved (parity-switching processes are
not considered here). The eigenenergies are roughly given by $E = E_{C, L}
n_L^2 + E_{C, R} n_R^2$, which approximately reproduces the low-energy
spectra shown in the left (green) parts of Figs. \ref{fig:emejdep}(a) and (d) for $E_M,
E_J \ll E_C$.

The MBS $( \gamma_1, \gamma_2)$ and $(\gamma_3, \gamma_4)$ are {\tmem{fused}} in this regime, which is a viable way
to initialize and readout MBS (see Sec. {\sec{sec:fusion}). Away from degeneracy points, the eigenstates of
different fermion {\tmem{parity}} possess a different {\tmem{charge}} --
parity and charge are not independent.
\begin{center}
  \Bigfigure{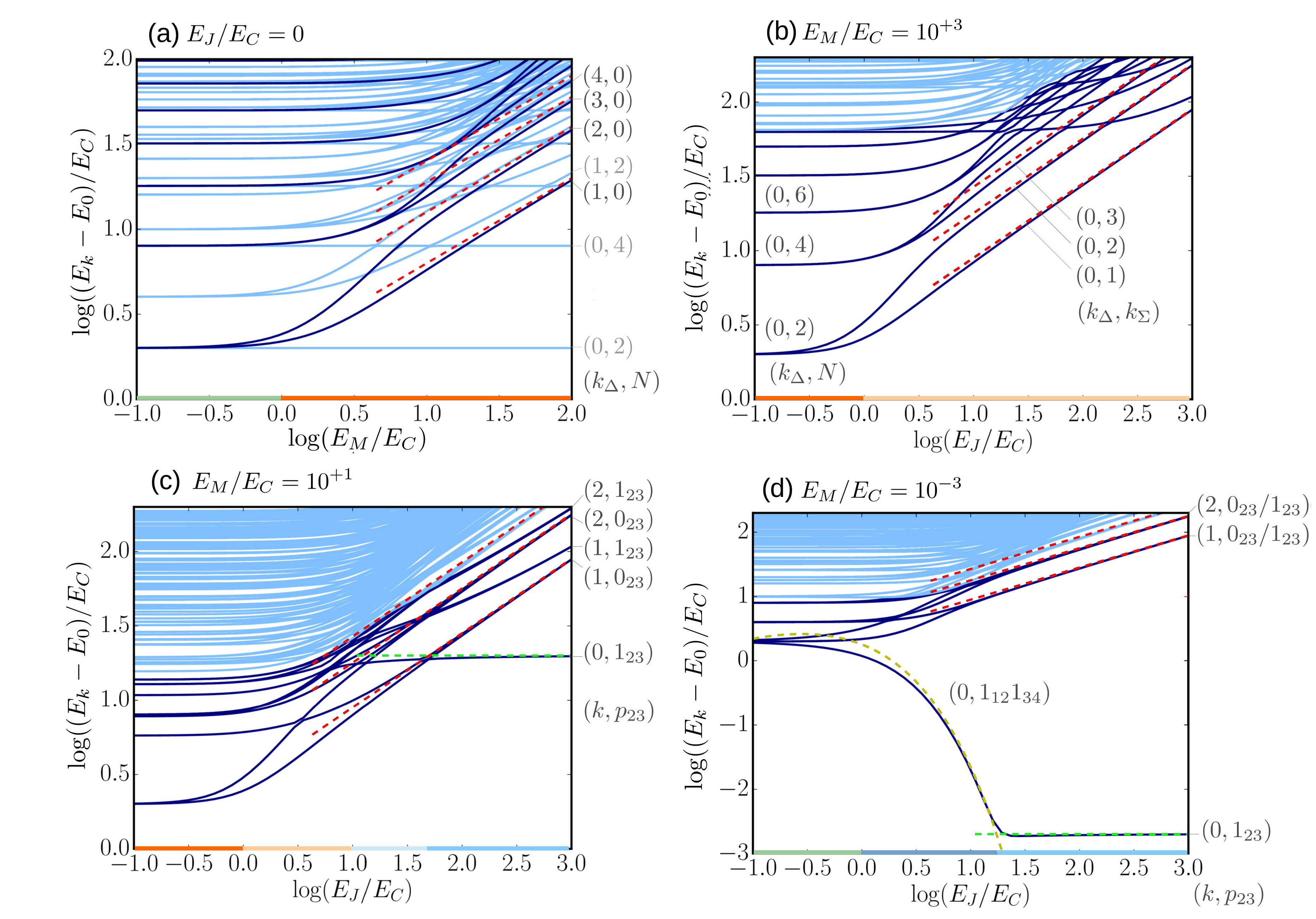}{Low-energy
  spectrum of coupled topological superconducting islands. Solid lines are the
  numerically computed energy splittings (blue) between succeeding 
  excited states (energy $E_k$) and the ground state (energy $E_0$) in
  the even total parity sector. Dashed lines indicate analytic approximations
  to the energy gaps (see below). The parameters are changed along the paths in the $(
  E_J, E_M)$ plane as denoted by the white arrows in {\color{black} Fig.
  \ref{fig:regimes}} and the colors on the horizontal axis correspond to the
  regimes in {\color{black} Fig. \ref{fig:regimes}}. In (a), we show the
  spectrum as a function of a the Majorana coupling $E_M$ for $E_J = 0$. We
  indicate in darker color those states with zero excess charge, $N = 0$,
  while the bright color corresponds to other excess excess charge, $N \neq
  0$. The red-dashed lines indicate the analytic approximations from Eq.
  (\ref{eq:tundomea0}). In (b)--(d), we show the spectra as a function of the
  bulk Josephson coupling $E_J$ for different values of $E_M$ as indicated.
  Here, the darker color highlights the 12 lowest eigenstates and the bright
  color corresponds to higher-lying states. The dashed lines indicate the
  transition energies $E_k - E_0 = k \sqrt{8 E_J E_C}$ related to Josephson
  plasma oscillations (red), the parity splitting $2 \varepsilon_P$, Eq.
  (\ref{eq:epsp}) (yellow), and the Majorana splitting $2 E_M$ (green). We
  assume $E_{C, L} = E_{C, R} = E_C \nocomma$, $E_{J, L} =
  E_{J, R} = E_{J, C} = 0$, $n_{g, L} = n_{g, R} = 0$, $E_{J, C} = 0$, and use a
  number-state cutoff $N_{\max} = 25$ for the numerical
  calculations.\label{fig:emejdep}}
\end{center}
\subsection{Single island: $E_C, E_J \ll E_M$}\label{sec:singleisland}

When the Majorana coupling between the islands is the largest energy scale, the physics can be understood most clearly from the
representation (\ref{eq:hsumdiff}) of the Hamiltonian in the sum and
difference variables. Equation (\ref{eq:hsumdiff}) can here be interpreted
analogous to a pair of strongly coupled oscillators, identifying the phases
with positions and the number of electrons with momentum. For $E_M \gg E_J,
E_C$, the dynamics of the ``relative coordinate'' $\Delta \varphi$ is fast
since it is subject to a strong confining potential, while the dynamics of the
``center coordinate'' $\Phi$ is slow since it is much more weakly confined.

To understand the energy spectrum, we first decompose the Hamiltonian into $H  =  H_{\Delta} + H_{\Sigma}$
with
\begin{eqnarray}
  H_{\Delta} & = & 2 E_C \left. \left( \tfrac{\Delta \op{n} - \Delta n_g}{2}
  \right. \right)^2 + E_M \cos \left( \tfrac{\Delta \op{\varphi}}{2} \right) i
  \op{\gamma}_2 \op{\gamma}_3  \label{eq:hdelta},\\
  H_{\Sigma} & = & 2 E_C \left( \tfrac{\op{N} - N_g}{2} \right)^2 + 2
  E_J  \left[ 1 - \cos \left( \tfrac{\Delta  \op{\varphi}}{2} \right) \cos \left( \tfrac{\op{\Phi}}{2} \right) \right]. \nonumber \\
 & &
  \label{eq:finestructure}
\end{eqnarray}
On a rough energy scale, the spectrum is determined by $H_{\Delta}$, while $H_{\Sigma}$ can be treated as a perturbation for $E_M \gg E_J,E_C$. First setting $H_{\Sigma}=0$, the
Hamiltonian is formally the same as that of a Cooper pair box, with the
physical difference that the Majorana term couples consecutive electron number
states on the islands instead of consecutive Cooper pair number states
{\footnote{The Hamiltonian for a Cooper pair box [Eq. (\ref{eq:hi}) for one
island] reads in electron number states: $H = \sum_n [ E_C | n \rangle \langle
n | + E_J / 2 ( | n + 2 \rangle \langle n | + | n - 2 \rangle \langle n |)]$.
The Hamiltonian $H_{\Delta}$ follows by replacing $n \rightarrow \Delta n$,
$E_C \rightarrow E_C / 2$, $E_J \rightarrow E_M$), and changing the number of electrons by 1 instead of 2 in the Hamiltonian.}}. The eigenenergies are
approximately given by
\begin{eqnarray}
  E_{k_{\Delta}} & = & \sqrt{4 E_C E_M} \left( k_{\Delta} + 1/2 \right) \\
  & & - (E_C/4) \left(k_{\Delta}^2+k_{\Delta} +1/2 \right) + O( \sqrt{E_C/E_M}). \nonumber \label{eq:tundomea0}
\end{eqnarray}
with $k_{\Delta} = 0, 1, ...$. We obtained this result by expanding $H_{\Delta}$ in $\Delta \varphi$ and including anharmonic corrections perturbatively along the lines of Ref. \cite{Koch}.
The resulting lowest-order energy gaps of $\sqrt{4 E_C E_M}k_{\Delta}$ to the ground state are indicated as red-dashed lines in {\color{black}
Fig. \ref{fig:emejdep}}(a). We note that the energy levels are altered when
the central Josephson energy is included, which we discuss further in
{\color{black} App. \ref{app:coupledtopcpb}}.

We next discuss the effect of the ``fine-structure'' term $H_{\Sigma}$, which
gives rise to smaller energy splittings than those induced by $H_{\Delta}$.
Depending on the bulk Josephson coupling $E_J$, we find two subregimes, which
are marked in {\color{black} Fig. \ref{fig:regimes}} in two shades of orange.

{\tmem{(i) Closed islands:}} $E_J \ll E_C ( \ll E_M)$. Let us first consider
the simplest case of $E_J = 0$, the situation shown in the orange part of Fig.~\ref{fig:emejdep}(a): Here, the total number of electrons on the islands is
conserved, $ [ \op{N}, H ] = 0$, and $\op{N}$ can thus be treated as
a number $N$. In {\color{black} Fig. \ref{fig:emejdep}}(a), we highlight the
energy gaps of all states with zero excess charge, $N = 0$, labeled by $(k_{\Delta},0)$ in darker color (corresponding to $H_{\Sigma} = 0$).
The full energy spectrum including all other states with (even) $N \neq 0$ is then
obtained as replicas of the spectrum for $N = 0$ by adding the
charging energies $H_{\Sigma} = 2 E_C ( N / 2)^2$. The corresponding energy
differences to the total ground state are shown in Fig. \ref{fig:emejdep}(a) in brighter color.
Among those, we see a sequence of constant energy gaps. These correspond to
the energetically lowest states ($k_{\Delta} = 0$) for the different values of
$N \neq 0$. Correspondingly shifted ``replicas'' of the states with excited
oscillation quanta $( k_{\Delta} = 1, \ldots)$ can also be identified. 

{\tmem{(ii) Open islands}}: $E_C \ll E_J ( \ll E_M)$. Treating the
``fine-structure'' term (\ref{eq:finestructure}) as a perturbation, we  
replace in leading order $\cos(\Delta \varphi /2)$ by its average:

\begin{eqnarray}
  H_{\Sigma} & \approx & 2 E_C \left( \tfrac{\op{N}}{2} \right)^2 + \tfrac{E_J}{2} \left(1-\rho \right) \left(\tfrac{\op{\Phi}}{2} \right)^2 + \rho ,  \label{eq:heff}
\end{eqnarray}
and the eigenenergies are $E=E_{k_{\Delta}}+\Delta E_{k_{\Sigma}}$ with
\begin{eqnarray}
  \Delta E_{k_{\Sigma}} & = & \sqrt{8 E_C E_J\left(1-\rho\right)} \left( k_{\Sigma} + 1/2 \right) +\rho .  \label{eq:ksigma}
\end{eqnarray}
Here, $k_{\Sigma} = 0, 1, ... $  and $\rho = E_{k_{\Delta}}/2 E_M$.
Thus, each $k_{\Delta}$ is associated with a ladder of states and each of these ladder states is specified by $k_{\Sigma}$. The different ladders are spaced by the large energy $E_{k_\Delta} \approx \sqrt{4 E_C E_M}$ and the states within each ladder are spaced by an energy $\sim \sqrt{8 E_J E_C}$. Note that for $E_J \gg E_C$, anharmonic corrections to $E_{k_{\Delta}}$ can be neglected as compared to the latter splitting.

The crossover from regime (i) to (ii)
is indicated by the transition from the dark orange to the light orange part in Fig.~\ref{fig:emejdep}(b). If we focus on low
energies $E \ll \sqrt{4 E_C E_M}$, all states correspond to $k_{\Delta} = 0$,
i.e., there are no excitations corresponding to oscillations in the difference phase $\Delta \varphi$ involved. 
This physically means that both phases are rigidly coupled at low
energies and the two islands behave as one.

In view of the Majorana physics, this means that the ``nonlocal'' parity
$\langle i \hat{\gamma}_2 \hat{\gamma}_3 \rangle = - 1$, indicating that the inter-island
fermionic mode, associated with the annihilator $\op{f}_{2 3} = \op{\gamma}_2
+ i \op{\gamma}_3$, is empty. The MBS pair $( \gamma_2, \gamma_3)$ is thus
{\tmem{fused}} -- the occupation of $f_{2 3}$ requires a finite energy $E_M$,
which is much larger than all other energy scales. For low energies, the
fermion parity degree of freedom is gapped out (recall that the spectra are
shown here only for {\tmem{even}} total parity).

\subsection{Two open islands: $E_C, E_M \ll E_J$}\label{sec:twoopen}

We next turn to the more intricate case when the Josephson energies of the
junctions with the bulk superconductors dominate. Here, 
$\varphi_L$ and $\varphi_R$ are not rigidly coupled to each
other at low energies (as in the foregoing section) and their individual dynamics becomes
important. The low-energy spectrum depends crucially on how the Majorana
coupling energy $E_M$ compares with (i) the Josephson plasma frequency
$\sqrt{8 E_J E_C}$ and (ii) the charging-induced energy splitting of even and
odd fermion parity states of the individual islands. The corresponding energy
scale {\cite{BraidingWithoutTransport}} is
\begin{eqnarray}
  \varepsilon_P & = & \frac{32}{( 2 \pi^2)^{1 / 4}}  ( E_J^3 E_C)^{1 / 4} e^{-
  \sqrt{8 E_J / E_C}},  \label{eq:epsp}
\end{eqnarray}
which is much smaller than $\sqrt{8 E_J E_C}$. This yields three
possibilities, which are shown in {\color{black} Fig. \ref{fig:regimes}} as
three shades of blue and which we discuss next.

{\tmem{(i) Weak inter-island coupling:}} $2 E_M \ll 2 \varepsilon_P \ll
\sqrt{8 E_J E_C}$: In this regime, the system behaves as two weakly coupled
topological superconducting islands. To discuss the physics, let us first set
$E_M = 0$ so that $H ( E_M = 0) = \sum_{\alpha} H_{\alpha} \left(
\op{n}_{\alpha}, \op{\varphi}_{\alpha} \right)$ simply decomposes into the two
island parts. The eigenstates of the system are trivially the product states
\begin{eqnarray}
  | k_L k_R ; 0_{1 2} 0_{3 4} \rangle & = & | k_L 0_{1 2} \rangle \otimes |
  k_R 0_{3 4} \rangle,  \label{eq:prodeven}\\
  | k_L k_R ; 1_{1 2} 1_{3 4} \rangle & = & | k_L 1_{1 2} \rangle \otimes |
  k_R 1_{3 4} \rangle .  \label{eq:prododd}
\end{eqnarray}
Here, $| k_{\alpha}, p_{n m} \rangle$ are the excitations of the Josephson plasma
oscillations of each island with $k_{\alpha}$ excited quanta and parity $p_{n m}$ of
the fermionic mode $\op{f}_{n m} = \op{\gamma}_n + i \op{\gamma}_m$. The total
energies are approximated by
\begin{eqnarray}
  E & \approx &\sqrt{8 E_J E_C} ( k_L + k_R + 1) \nonumber \\
  & & + ( p_{1 2} + p_{3 4} - 1)
  \varepsilon_P,  \label{eq:ek1k2}
\end{eqnarray}
On a rough energy scale, the spectrum can be grouped into pairs of $( k_L + k_R + 1)$-fold degenerate
states, split by the Josephson plasma frequency $\sqrt{8 E_J E_C}$ (see
{\color{black} Fig.~\ref{fig:regimes}}). 
The corresponding splittings are denoted by red-dashed lines in the right blue parts of Figs. {\color{black} \ref{fig:emejdep}}(c) and (d) with $k = k_L + k_R$.
These are valid not only for the weak inter-island coupling, $E_M \ll \varepsilon_P$, but also for the other subregimes discussed in this section. 

Within each of the ``parity pairs'', the odd-odd parity combination is split
from the lower even-even parity combination \footnote{We assume $n_{g, L} = n_{g, R} = 0$; for other gatings the
energy can be reduced and the order of the levels may reverse.}  by the smaller energy $2
\varepsilon_P$ (we focus now on the case $E_M \ll \varepsilon_P$). For the two lowest lying states with $k_L = k_R =
0$, the excited state $|00; 1_{1 2} 1_{3 4} \rangle$ is therefore split from the
ground state $|00; 0_{1 2} 0_{3 4} \rangle$ by $2 \varepsilon_P$ as given by Eq.
(\ref{eq:epsp}), which is indicated by the yellow-dashed line in the dark blue part of Fig.~\ref{fig:emejdep}(d). We have verified that a nonzero capacitive coupling between the islands, modeled by a term $ E_{C,LR} \hat{n}_L \hat{n}_R$ ($E_{C,LR} \leq E_C)$, influences this parity energy splitting only slightly and does not affect the exponential suppression. 
We also neglected anharmonic corrections of $O(E_C)$ to the Josephson-plasma frequency in \Eq{eq:ek1k2} because these only \emph{shift} levels of the same $k_L+k_R$ by the same amount but do not contribute their \emph{splitting}. Compared to the rough energy scale $\sqrt{8 E_J E_C}$, these anharmonic corrections are negligible.

When $E_M \ll \varepsilon_P$, the Majorana tunneling has little effect on
the spectrum. For the lowest parity pair, a nonzero $E_M$ results in
a small shift of the energy levels. This is different for the
higher-lying excited states, which exhibit degeneracies that may be lifted for
nonzero $E_M$ (not resolved on the scale shown in {\color{black}
Fig. \ref{fig:emejdep}}).

{\tmem{(ii) Intermediate inter-island coupling}}: $2 \varepsilon_P \ll 2 E_M
\ll \sqrt{8 E_J E_C}$. When the Majorana coupling exceeds the charging-induced
parity splitting, the inter-island tunneling strongly mixes the even-even and
odd-odd parity sectors. The energy eigenstates are therefore ``bonding'' and
``antibonding'' combinations of the ``local'' parity states:
\begin{eqnarray}
  | k_L k_R ; 0_{2 3} 0_{1 4} \rangle & = & | k_L k_R \rangle \otimes
  \frac{|0_{12} 0_{34} \rangle + | 1_{12} 1_{34} \rangle }{\sqrt{2}}, 
  \label{eq:sym} \\
   | k_L k_R ; 1_{2 3} 1_{1 4} \rangle & = & | k_L k_R \rangle \otimes
  \frac{|0_{12} 0_{34} \rangle - | 1_{12} 1_{34} \rangle }{\sqrt{2}},  \label{eq:asym}
\end{eqnarray}
split by an energy $2 E_M$ [green dashed in Figs. {\color{black}
\ref{fig:emejdep}}(c) and (d)]. The crossover from regime (i) to (ii) can be
clearly seen for the lowest-lying excited state in {\color{black} Fig.
\ref{fig:emejdep}}(d) from the dark blue to the lighter blue part. As our notation in
Eqs. (\ref{eq:sym}) and (\ref{eq:asym}) suggests, increasing the tunnel
coupling fuses the MBS in a complementary way: The MBS at the central junction
become more strongly {\tmem{fused}} than the pairs on each island. This is a
key ingredient to test the Majorana fusion rules as discussed in
{\color{black} Sec. \ref{sec:fusion}}.

{\tmem{(iii) Strong inter-island coupling:}} $2 \varepsilon_P \ll \sqrt{8 E_J
E_C} \ll 2 E_M ( \ll 2 E_J)$. In this regime, the tunnel coupling between the
two islands is not a ``fine-structure'' effect: Roughly speaking, the
lower end of the spectrum is given by bonding states of the two islands, while
the upper end of the spectrum is given by antibonding states (see
{\color{black} Fig. \ref{fig:regimes}}). This effectively removes the parity
degree of freedom from the low-energy spectrum, similar to the single-island
regime. Accordingly, the Josephson plasmon excitations do not appear
in parity pairs. 

This can be seen in Fig.~\ref{fig:emejdep}(c)
in the light-blue part: Here, the lowest state of flipped nonlocal parity, labeled by $(0,1_{23})$, is at a much higher energy than other states with the same nonlocal parity $(k>0,0_{23})$ as the ground state. This contrasts with the situation for the darker blue parts shown in the Fig.~\ref{fig:emejdep}(c) and in Fig.~\ref{fig:emejdep}(d), where the state $(0,1_{23})$ is closest to the ground state.

Finally, we emphasize that there is still a difference between the regime of
strongly coupled individual islands ($E_J \gg E_M,E_C$) and the regime of a single
island ($E_M \gg E_J,E_C$). In the former case $\varphi_L$ and
$\varphi_R$ are {\tmem{not}} locked to each other and one may still use $k_L$,
$k_R$ as independent quantum numbers to give a rough construction of the
low-energy spectrum. In contrast to the single-island regime, the
oscillator levels are degenerate here [compare Eq. (\ref{eq:ksigma}) and Eq.
(\ref{eq:ek1k2})]. It is only the parity degree of freedom that is ``gapped
out'' in both regimes.

\subsection{Regime of four zero-energy MBS $( E_M, \varepsilon_P \lll
E_C)$}\label{sec:majorana}

The system hosts four zero-energy MBS $\gamma_1$,..,$\gamma_4$ as sketched
{\color{black} Fig. \ref{fig:model}} when both $E_M$ and $\varepsilon_P$
become negligibly small. Equation (\ref{eq:epsp}) shows that the
charging-mediated energy splitting $\varepsilon_P$ can become exponentially
small in $E_J / E_C$. The lowest-energy parity states $|00; 0_{1 2}, 0_{3 4}
\rangle$ and $|00; 1_{1 2}, 1_{3 4} \rangle$ ($k_L = k_R = 0$) are then
degenerate up to exponential accuracy. Including also states of odd total
parity, the ground state becomes four-fold degenerate and is spanned by $|00;
p_{1 2}, p_{3 4} \rangle$ ($p_{n m} = 0, 1$).

\section{Time scales for gate-controlled fusion-rule testing
protocol}\label{sec:fusion}

The segmented nanowire structure shown in {\color{black} Fig. \ref{fig:model}}
has recently been proposed as en experimental testbed for Majorana physics
that could be realized in the near future {\cite{Aasen15}}. MBS may be
manipulated by opening and closing junctions through gate control. In
this Section, we derive the time-scale conditions (\ref{eq:timescalelower})
and (\ref{eq:timescaleupper}) stated in Sec. \ref{sec:intro}, which are required
to perform the fusion-rule test suggested in {\color{black} Ref.
{\cite{Aasen15}}}.

The fusion rules of MBS can lead to nontrivial parity correlations by fusing
four MBS in complementary pairs. This is rooted in the nonlocal character of
the MBS and not possible for local fermions. To prepare and probe such parity
correlations, one goes through the steps sketched in {\color{black} Fig.
\ref{fig:fusion-plane}}(a), changing the parameters along the path in the $(
E_J, E_M)$ plane as shown in {\color{black} Fig. \ref{fig:fusion-plane}}(b).
The starting point (A) is to initialize the system in the ground state when
the MBS are fused in pairs ($\gamma_2$,$\gamma_3$) and
($\gamma_1$,$\gamma_4$). This corresponds to a superposition of even-even and
odd-odd fermion parities in the complementary pairs ($\gamma_1$,$\gamma_2$)
and ($\gamma_3$,$\gamma_4$). To detect these parity correlations, one first
forms all four zero-energy MBS (C) and and then converts them into charge
states for the islands (D). Subsequent charge detection then probes the
prepared parity correlations. To repeat the experiment by going back from point D to A, a resetting step is needed, in which the system has to relax into the ground state again.

We now go step by step through the protocol and verify the time-scale criteria
(\ref{eq:timescalelower}) and (\ref{eq:timescaleupper}). Our considerations
here concern the cycle sketched in {\color{black} Fig.
\ref{fig:fusion-plane}}; the time-scale conditions on the readout are discussed in {\color{black} App. \ref{app:readout}}.

\begin{center}
  \Figure{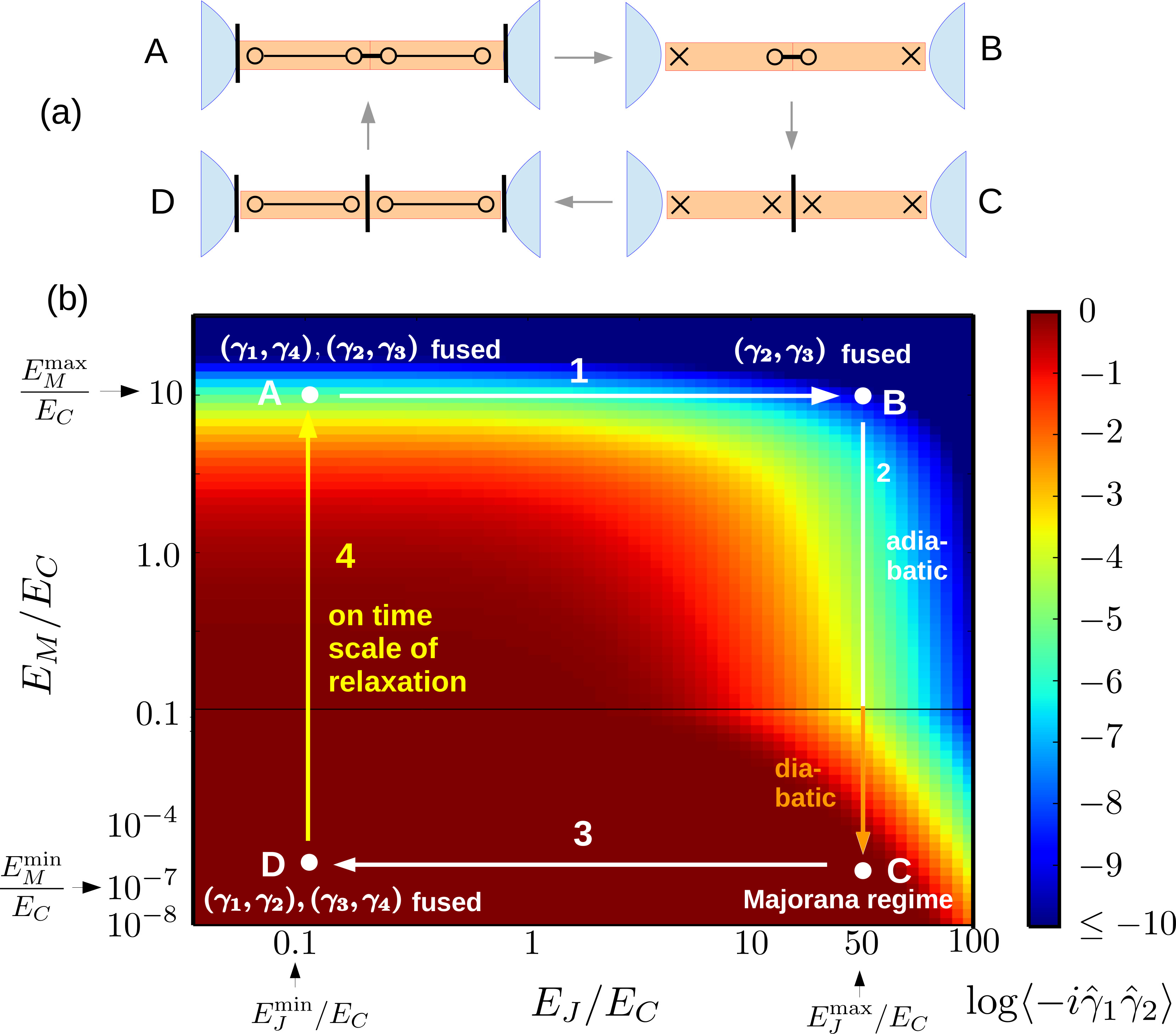}{
  Protocol for testing the Majorana fusion rules. The steps of the protocol are sketched in (a) and
  the corresponding path taken in the $( E_J, E_M)$ parameter space in (b)
  (compare with {\color{black} Fig.~\ref{fig:regimes}}). White arrows indicate
  that these processes have to be adiabatic, while the orange arrow indicates
  a diabatic step, in which the system should not follow the ground state evolution. The yellow arrow indicated the resetting step, which should be done on the time scale of the charge relaxation to the ground state.
  The color scale gives the ground-state expectation value of
  $\left\langle - i \op{\gamma}_1 \op{\gamma}_2 \right\rangle$ [see Eq.
  (\ref{eq:sigmaz})], indicating the preferred parity combination of the ground
  state. In the charge-dominated regime ($E_J, E_M \ll E_C$), the Majorana
  pairs ($\gamma_1, \gamma_2$) and ($\gamma_3, \gamma_4$) are fused
  [$\left\langle - i \op{\gamma}_1 \op{\gamma}_2 \right\rangle \rightarrow
  1$], while in the tunneling-dominated regime ($E_M \gg \varepsilon_P$), the
  Majoranas ($\gamma_2, \gamma_3$) are fused [$\left\langle - i \op{\gamma}_1
  \op{\gamma}_2 \right\rangle \rightarrow 0$]. To bring all the parity states
  of the two islands as close to zero energy as possible, the path should
  cover the Majorana regime ($E_M, E_J \gg E_C$ and $E_M, E_J$ should be, in principle, as large as possible). We
  use $E_{J, C} = 5 E_M^2 / \Delta$, $E_{J, L} = E_{J, R} = E_J$, $\Delta = 100
  E_C$, $n_{g, R} = - n_{g, L} = 0.3$, and $N_{\max} =
  21$.\label{fig:fusion-plane}}
\end{center}

\subsection{Time-scale conditions}\label{sec:fusion-timescales}

To derive the time-scale conditions for each step of the fusion-rule protocol,
one has to demand the following {\tmem{adiabaticity condition}}
{\cite{SchiffBook}}:
\begin{eqnarray}
  \max_{k, t} f_k ( t) & \ll & 1,  \label{eq:fusion-adiabatic}
\end{eqnarray}
with
\begin{eqnarray}
  f_k ( t) & = & \left| \frac{\langle \psi_k ( t) | \partial H ( t) / \partial t | \psi_0 (
  t) \rangle}{[ E_k ( t) - E_0 ( t)]^2} \right| . 
\end{eqnarray}
Here, $k$ labels all excited states $| \psi_k \rangle$ at energy $E_k$ \ and 0
labels the ground state $| \psi_0 \rangle$ at energy $E_0$ [except for step 2, in which the first excited state
has to be excluded from condition (\ref{eq:fusion-adiabatic})].

Our first goal is to turn Eq. (\ref{eq:fusion-adiabatic}) into a condition
for the entire time interval $\Delta t$ when changing the parameters, labeled
in the following by $\vecg{\lambda} = ( \lambda_1, \ldots)$, from
$\vecg{\lambda} ( 0)$ at time $t = 0$ to $\vecg{\lambda} ( \Delta t)$
at time $t = \Delta t$. In our case, the parameters are given by
$\vecg{\lambda} = ( E_J, E_M)$. The optimal way to change the parameters
is to keep the function $f_k ( t)$ constant during the parameter sweep because
that minimizes the sweeping time $\Delta t$ for a given value of $\max_{k, t}
f_k ( t)$. We can then turn Eq. (\ref{eq:fusion-adiabatic}) into the following
condition by integrating over time:
\begin{eqnarray}
  \Delta t & \gg & \int_0^{\Delta t} d t \left| \frac{\langle \psi_k ( t) |
  \partial H ( t) / \partial t | \psi_0 ( t) \rangle}{[ E_k ( t) - E_0 (
  t)]^2} \right| \nonumber\\
  & = & \int_C d \vecg{\lambda} \cdot \left| \frac{\langle \psi_k (
  \vecg{\lambda}) | \partial H ( \vecg{\lambda}) / \partial
  \vecg{\lambda} | \psi_0 ( \vecg{\lambda}) \rangle}{[ E_k (
  \vecg{\lambda}) - E_0 ( \vecg{\lambda})]^2} \right| . 
  \label{eq:fusion-adiabatic2}
\end{eqnarray}
Here, $C$ is the path in the parameter space connecting $\vecg{\lambda} (
0)$ and $\vecg{\lambda} ( \Delta t)$. We emphasize that Eq.
(\ref{eq:fusion-adiabatic}) is sufficient only if $f_k ( t)$ is constant (in
practice remaining of the same order of magnitude), which might correspond to a
rather complicated time dependence for $\vecg{\lambda} ( t)$. This means
that one has to know the properties of the system quite well to design the
best gate pulse (in that respect the estimate (\ref{eq:fusion-adiabatic2}) is
optimistic).

Our estimates for the energy gaps $E_k ( t) - E_0 ( t)$ are based on the
numerically computed energy spectrum of Hamiltonian (\ref{eq:hfull}) and the
analytic approximations worked out in {\color{black} Sec.
\ref{sec:spectrum}}. The gaps to the lowest and further selected excited states along the path shown in Fig. \ref{fig:fusion-plane} are shown in Fig. \ref{fig:fusion-splitting}. We motivate our parameter choices for Figs. \ref{fig:fusion-plane} and  \ref{fig:fusion-splitting} in App.
\ref{app:fusion-parameters}. In the following, we work out Eq.
(\ref{eq:fusion-adiabatic2}) only for the lowest accessible excited state and
show in {\color{black} App. \ref{app:higherexcited}} that the time scale
derived from this is not modified if the effect of transitions into the entire
spectrum of higher-lying excited states is included.

Provided optimal pulsing shapes can be achieved, our time-scale estimates are
{\tmem{conservative}} in four respects: (i) We implement the cosine
approximation for the Josephson energies in Eq. (\ref{eq:hfull}). As we
explained in the Appendix of {\color{black} Ref. {\cite{Aasen15}}}, we expect
corrections due to higher harmonics to {\tmem{enhance}} energy gaps between
the ground state(s) and the excited states and to {\tmem{reduce}} splittings
within the ideally degenerate ground-state manifold. The time-scale window we estimate here is
therefore narrower than what one could expect including these corrections.
(ii) While our numerical results in {\color{black} Figs.
\ref{fig:fusion-plane}} and {\color{black} \ref{fig:fusion-splitting}}
include a nonzero Josephson energy $E_{J, C}$ for the central junction, we use
$E_{J, C} = 0$ for our analytic estimates below. Since nonzero $E_{J, C}$ also
enhances the relevant gaps, this again tends to underestimate the actual
time-scale window. (iii) For the estimates, we take the islands to be
symmetric, $E_{J, L} = E_{J, 2} = E_J$, and $E_{C, 1} = E_{C, 2} = E_C$. In
practice, they will be asymmetric and therefore the energy gaps associated
with one of the islands will be less constraining for the time scales than
those associated with the other island. (iv) We estimate all matrix elements
in Eq. (\ref{eq:fusion-adiabatic2}) by maximal values (if nonzero) even though
they can be smaller in practice.
We next go through each of the steps of the protocol sketched in Fig. \ref{fig:fusion-plane} in detail.

\begin{center}
  \Bigfigure{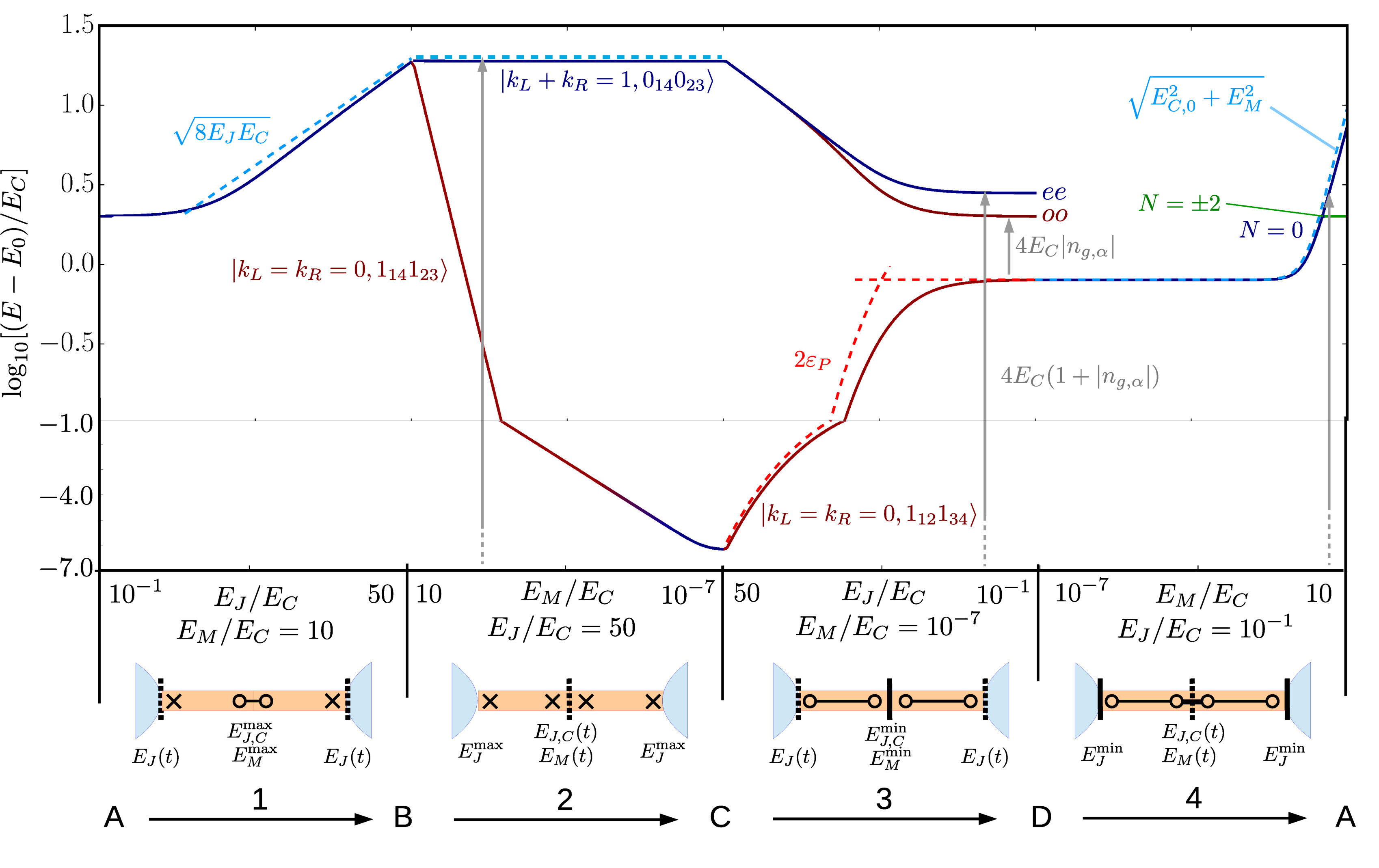}
  {Energy gap between the ground state and first excited state for even total parity 
  (in some intervals the gap to other excited states is shown as well). The parameters $E_M$ (and accordingly $E_{J, C}
  = 5 E_M^2 / \Delta$), as well as $E_J$ on the horizontal axis are changed
  logarithmically along the path $A \rightarrow B \rightarrow C \rightarrow D
  \rightarrow A$ marked in {\color{black} Fig. \ref{fig:fusion-plane}}(b). The
  solid lines correspond to the numerically computed splittings while the
  dashed lines are the approximation formulas for the splittings (see text).
  Gray arrows indicate allowed transitions and we mark the lowest accessible excited state
  from the ground state in blue. 
  In several steps transitions into the lowest excited states (red,
  green) are prohibited by parity or charge conservation as explained next: In step 2, the lowest excited state $| 1_{2 3}, 1_{1 4} \rangle$ is decoupled from the
  ground state $| 0_{2 3}, 0_{1 4} \rangle$ for a large energy range because the tunneling Hamiltonian
  $H_T$ conserves the nonlocal parities $p_{2 3}$, $p_{1 4}$. However, when $\varepsilon_P$ becomes dominant, the parity character of the ground state changes to $| 0_{12}0_{34} \rangle$. Thus, when $E_M$ approaches $\varepsilon_P$, transitions to the lowest excited state become possible. By tuning fast the system stays in the desired state  $| 0_{2 3}, 0_{1 4} \rangle = ( | 0_{12}0_{34} \rangle + | 1_{12}1_{34} \rangle ) /\sqrt{2}$, i.e., the system can be both in the ground and first excited state after step 2. Since the islands are decoupled then, the local
  parities $p_{12},p_{34}$ are conserved in step 3 and transitions from even-even to odd-odd parity or vice versa are prohibited.
From the ground state, one can therefore only reach the third-lowest excited state [blue, parity $(0_{12},0_{34})$]. In addition, transitions from the first to the second excited state are possible [red, parity $(1_{1 2},1_{34})$], which determines the adiabaticity condition.
  Finally, in step 4, the islands host a different total charge $N = \pm 2$ in
  the lowest excited state, while transitions are only possible into the
  lowest $N = 0$ state ($N = n_L + n_R$ is the total number of electrons). The
  parameters are $n_{g, R} = - n_{g, L} = 0.3$, $\Delta = 100 E_C$, $E_{C,0}= 2 E_C (1+n_{g,L} -n_{g,R})$ (see text), and
  the number-state cutoff is $N_{\max} = 25$. Note that we chose a rather large value for $\Delta$ to be consistent with the assumption $E_C \ll E_M^{\max} \ll \Delta$ for our numerical approach. We expect that the protocol should also work for smaller values of $\Delta$ as we further explain in App.~\ref{app:parameter-estimation}. \label{fig:fusion-splitting}}
\end{center}

\subsection{Initial point}

We start at point A in the parameter space in {\color{black} Fig.
\ref{fig:fusion-plane}}(b). Here, the central valve is open $( E_M =
E_M^{\max})$, while the valves to the bulk superconductors are closed $( E_J =
E_J^{\min})$. In this configuration, the MBS pairs $( \gamma_2, \gamma_3)$ and
$ ( \gamma_1, \gamma_4)$ are fused. In the ground state, $| N=0,k_{\Delta} = 0 ; 0_{1 4}
0_{2 3} \rangle$, the corresponding fermionic modes are empty and no
Josephson-plasma oscillations are excited. We will from hereon suppress quantization indices for charge state / plasma oscillations whenever we refer to the lowest state with respect to these degrees of freedom,  i.e.,
$|p_{1 4}p_{2 3} \rangle =| N=0,k_{\Delta} = 0 ; p_{1 4}p_{2 3} \rangle $.

To illustrate the evolution of the parity of the ground state along the protocol, we show with the color scale in
{\color{black} Fig. \ref{fig:fusion-plane}}(b) the numerically computed
ground-state expectation value of the operator
\begin{eqnarray}
  &  & - i \op{\gamma}_1 \op{\gamma}_2 \text{ \ = \ } \op{P}_{e e} -
  \op{P}_{o o} \nonumber\\
  &  & \text{ \ = \ } \left( \sum_{n_L, n_R \text{ even}} - \sum_{n_L, n_R
  \text{ odd}} \right) | n_L n_R \rangle \langle n_L n_R |.  \label{eq:sigmaz}
\end{eqnarray}
If $\left\langle - i \op{\gamma}_1 \op{\gamma}_2 \right\rangle = + 1 ( - 1)$,
a state has even-even (odd-odd) parity, while if $\left\langle - i \op{\gamma}_1 \op{\gamma}_2 \right\rangle = 0$, the state is
a linear combination of an even-even and an odd-odd state with equal probability (we restrict ourselves here to the subspace of even total parity).
Figure {\color{black} \ref{fig:fusion-plane}}(b) demonstrates that, at point A,
$| \left\langle - i \op{\gamma}_1 \op{\gamma}_2 \right\rangle| \ll 1 $ in the
ground state, which is consistent with the form $| 0_{1 4} 0_{2 3} \rangle =
[ | 0_{1 2} 0_{3 4} \rangle + | 1_{1 2} 1_{3 4} \rangle] / \sqrt{2}$.

\subsection{Step 1: Initialization of $\gamma_1$ and
$\gamma_4$}\label{sec:step1}

The first step of the protocol brings the MBS $\gamma_1$ and $\gamma_4$
located at the two outer ends of the island close to zero
energy {\footnote{The ground state becomes two-fold degenerate. We checked
numerically that the splitting between the corresponding states $| 1_{1 4}
0_{2 3} \rangle$ and $| 0_{1 4} 0_{2 3} \rangle$ is given by $\varepsilon_P$
[Eq. (\ref{eq:epsp})] and therefore exponentially small in $\sqrt{8 E_J /
E_C}$. We recall that we show in {\color{black} Fig.
\ref{fig:fusion-splitting}} only results for even total parity (including $|
0_{1 4} 0_{2 3} \rangle$) since transitions into states of odd total parity
(including $| 1_{1 4} 0_{2 3} \rangle$) are forbidden and therefore irrelevant
for the time-scale considerations.}}. This is achieved by opening the valves to the
bulk superconductors [$\lambda = E_J \rightarrow E_J^{\max}$ at point B in
{\color{black} Fig. \ref{fig:fusion-plane}}(b)], which suppresses charging
effects on both islands. Yet, the Majorana pair ($\gamma_2, \gamma_3$) remains
fused, i.e., $\left\langle - i \op{\gamma}_1 \op{\gamma}_2 \right\rangle$ remains essentially zero along the curve $A \rightarrow B$ in
{\color{black} Fig. \ref{fig:fusion-plane}}(b). The numerically computed
energy gap $E_1 - E_0$ between the first excited state and the ground state $|
0_{1 4} 0_{2 3} \rangle$ is shown in {\color{black} Fig.
\ref{fig:fusion-splitting}} as a function of $E_J$.

Concerning the time-scale $T_1$ for this step, we can use $| \langle \psi_1 |
\partial H \text{/} \partial E_J | \psi_0 \rangle | \leqslant 1$ and express
Eq. (\ref{eq:fusion-adiabatic2}) as
\begin{eqnarray}
  T_1 & \gg & \int_0^{E_C} \frac{d E_J }{| E_1 - E_0 |^2} +
  \int_{E_C}^{E_J^{\max}} \frac{d E_J }{| E_1 - E_0 |^2}.  \label{eq:t1cond}
\end{eqnarray}
We decomposed the integral into two parts according to the two parameter
regimes crossed during the $E_J$ sweep. For $E_J \lesssim E_C$, the gap is
dominated by the charging energy and we can estimate $E_1 - E_0 \geq 2 E_C$.
For $E_C \lesssim E_J < E_J^{\max} < ( E_M^{\max})^2 / 2 E_C$, the gap is dominated by
Josephson plasma oscillations and given roughly by $E_1 - E_0 \approx \sqrt{8
E_J E_C}$ as the light blue dashed line in {\color{black} Fig.
\ref{fig:fusion-splitting}} illustrates. Inserting these two estimates into
Eq. (\ref{eq:t1cond}), we obtain
\begin{eqnarray}
  T_1 & \gg & \frac{\ln ( E_J^{\max} / E_C)}{E_C} .  \label{eq:t1adiabatic}
\end{eqnarray}
where we used $E_J^{\max} / E_C \gg 1$. We see that the time-scale condition
scales only logarithmically with $E_J^{\max}$. Accounting for the dependence
of the matrix elements $| \langle \psi_1 | \partial H \text{/} \partial E_J |
\psi_0 \rangle |$ on $E_J$ might result in even less restrictive conditions
than Eq. (\ref{eq:t1adiabatic}) {\footnote{The matrix element $A_{k} = |
\langle \psi_k | \partial H \text{/} \partial E_J | \psi_0 \rangle | = \left|
\langle \psi_k | \cos \left( \op{\varphi}_1 \right) + \cos \left(
\op{\varphi}_2 \right) | \psi_0 \rangle \right|$ in Eq.
(\ref{eq:fusion-adiabatic2}) can actually become much smaller than 1 in the
limit $E_J \gg E_M \gg E_C$. If one approximates $| \psi_k \rangle \approx |
k_1, k_2 ; p_{1 4}, p_{2 3} \rangle$ in that limit, one first finds $A_{k_1=1,k_2=0} =A_{k_1=0,k_2=1} =A_{k_1=1,k_2=1}=0$, that is,
transitions into the first excited states vanish. The reason for this is that $\cos \left(
\op{\varphi}_{\alpha} \right)$ contains only an even power of
$\op{\varphi}_{\alpha} = \op{a}_{\alpha} + \op{a}^{\dag}_{\alpha}$ and
therefore a product of an even number of ladder operators $\op{a}_{\alpha},
\op{a}^{\dag}_{\alpha}$, each changing $k_{\alpha}$ by $- 1, + 1$,
respectively. However, taking the next excited states in condition
(\ref{eq:t1cond}) would just increase the gap by a factor of $\sqrt{2}$. What
is more important is that the matrix element is suppressed as $A_{k_2=2, k_1=0},A_{k_2=0, k_1=2} \propto
\sqrt{E_C/E_J}$. Inserting this into Eq. (\ref{eq:fusion-adiabatic2}),
integrand in Eq. (\ref{eq:t1cond}) scales as $\propto 1 / \sqrt{E_J}$
instead of $\propto 1 / E_J$. This would remove logarithmic dependence in the
condition (\ref{eq:t1adiabatic}). However, to give a solid quantitative
prediction, one should also include corrections for deviations of $| \psi_k
\rangle$ from $|k_1, k_2 ; p_{1 4}, p_{2 3} \rangle$ for $E_J \sim E_C$. In the same order,
higher-harmonic corrections to the Josephson energy might become relevant. We
limit our discussion therefore to our conservative estimate. }}. However, for
the experimental situation we have in mind, the ratio $E_J^{\max} / E_C$ will
be a few tens and the logarithmic term is less important.

We finally mention that condition (\ref{eq:t1adiabatic}) has to be
reconsidered if $E_J^{\max} \gg ( E_M^{\max})^2 / 2 E_C$: The gap to the first
excited state is then constant and given by $2 E^{\max}_M$. However, transitions into this lowest state are forbidden (see below) and
for the
parameters used {\color{black} Fig. \ref{fig:fusion-splitting}}, one hardly
enters into this regime.

\subsection{Step 2: Initialization of $\gamma_2$ and
$\gamma_3$}\label{sec:step2}

Closing the central valve brings the Majoranas $\gamma_2$ and $\gamma_3$ to
zero energy [$E_M \rightarrow E_M^{\min}$ at point C in {\color{black} Fig.
\ref{fig:fusion-plane}}(b)]. This separates the system into two decoupled
islands and the two states $| 0_{1 4} 0_{2 3} \rangle, | 1_{1 4} 1_{2 3}
\rangle = [ | 0_{1 2} 0_{3 4} \rangle \pm | 1_{1 2} 1_{3 4} \rangle] /
\sqrt{2}$ become  degenerate, at least ideally. As in any other topological setup, the ground-state degeneracy is in practice slightly broken
due to the charging-induced
parity splitting $\varepsilon_P$ on each of the islands (see {\color{black} Sec.
\ref{sec:twoopen}}). This splitting is suppressed with exponential accuracy but renders the even-even parity state $| 0_{1 2} 0_{3
4} \rangle$ the ground state for $E_M \ll \varepsilon_P$. 
Therefore, this step has to be
performed {\tmem{diabatically}} regarding the two lowest states when $E_M$
becomes of the order of $\varepsilon_P$. 
The goal is here
that the system remains in the prepared superposition $| \psi_0 \rangle = ( |
0_{1 2} 0_{3 4} \rangle + | 1_{1 2} 1_{3 4} \rangle) / \sqrt{2}$. Moreover,
this step should be {\tmem{adiabatic}} regarding all other excited states, which is
the condition we discuss first.

{\tmem{Adiabaticity condition.}} For $E_M \gg \varepsilon_P$, the
tunnel splitting $2 E_M$ sets the gap $E_1 - E_0$ between the ground and first
excited state (red line in Fig. \ref{fig:fusion-splitting}). Importantly, the
tunneling Hamiltonian $H_T$ does not allow for transitions between the two
lowest states \ $| 0_{1 4} 0_{2 3} \rangle$ and $| 1_{1 4} 1_{2 3} \rangle$ because
$H_T$ cannot flip the nonlocal parities of the fermionic modes $\op{f}_{1 4} =
\op{\gamma}_1 + i \op{\gamma}_4$ and $\op{f}_{2 3} = \op{\gamma}_3 + i
\op{\gamma}_3$. The lowest accessible excited state is therefore $| \psi_2
\rangle \approx | k_L + k_R = 1 ; 0_{1 4} 0_{2 3} \rangle$ with an excited
Josephson plasma oscillation on one of the islands as explained in
{\color{black} Sec. \ref{sec:twoopen}} (ii). The energy gap is constant here:
$E_2 - E_0 \sim \sqrt{8 E^{\max}_J E_C}$ (blue dashed line in {\color{black}
Fig. \ref{fig:fusion-splitting}}). Inserting this into Eq.
(\ref{eq:fusion-adiabatic2}) and using $| \langle \psi_k | \partial H \text{/} \partial E_M
| \psi_0 \rangle | \leqslant 1$ yields the condition:
\begin{eqnarray}
  T_2 & \gg & \frac{E_M^{\max}}{8 E_J^{\max} E_C}, 
\end{eqnarray}
If $E_M^{\max} / E_J^{\max} < 1$, as assumed here, this condition is fulfilled
if $T_2 \gg 1 / E_C$, which is a looser condition than Eq.
(\ref{eq:t1adiabatic}).

{\tmem{Diabaticity condition}}. 
If $E_M \ll \varepsilon_P$, the two lowest
eigenstates are $| 0_{1 2} 0_{3 4} \rangle$ and $| 1_{1 2} 1_{3 4} \rangle$
and split by $2 \varepsilon_P$ as discussed in {\color{black} Sec.
\ref{sec:twoopen}} (i). 
The system has to evolve {\tmem{diabatically}} when
tuning through the crossover point at $E_M = \varepsilon_P$, i.e., step 2 has
be carried out {\tmem{fast}} in the sense that
\begin{eqnarray}
  T_2 & \ll & \frac{1}{\varepsilon_P^{\min}},  \label{eq:fusion-t4t3}
\end{eqnarray}
where $\varepsilon_P^{\min}$ is the value for $\varepsilon_P$ at point C, where it is minimal because $E_J/E_C$ is maximal [see \Eq{eq:epsp}].
Roughly speaking, the above condition ensures that the time-evolution operator with the
ground-state manifold can be approximated as 

\begin{eqnarray}
 U ( T_2) &=& e^{- i \int_0^{T_2} d
t H ( t)} = e^{- i \int_0^{T_2} d t \varepsilon_P ( t)} | 1_{1 2} 1_{3 4}
\rangle \langle 1_{1 2} 1_{3 4} | \nonumber \\
& & + e^{+ i \int_0^{T_2} d t \varepsilon_P (
t)} | 0_{1 2} 0_{3 4} \rangle \langle 0_{1 2} 0_{3 4} | \approx 1 .
\end{eqnarray}

\subsection{Step 3: Parity-to-charge conversion}\label{sec:step3}

The next step is to read out the fermion parities of each of the disconnected
islands, which necessitates closing of the valves to the bulk superconductors again [$E_J
\rightarrow E_J^{\min}$ at point D in {\color{black} Fig.
\ref{fig:fusion-plane}}(b)]. The two decoupled parity components in the
prepared state $| 0_{1 4} 0_{2 3} \rangle = ( | 0_{1 2} 0_{3 4} \rangle + |
1_{1 2} 1_{3 4} \rangle) / \sqrt{2}$ are thereby mapped onto different charge
states. We assume for the following estimations that the gate voltages are
adjusted such that
\begin{eqnarray}
  - 1 / 2 \text{ \ } < \text{ \ } n_{g, L} \text{ \ } < & 0 & < \text{ \ }
  n_{g, R} \text{ \ } < \text{ \ } 1 / 2,  \label{eq:ngcond}
\end{eqnarray}
and that the $n_{g, \alpha}$ are not close to either of the boundaries. Under
this assumption, the parity states are transferred to {\footnote{Note that one
of the two charge states is an excited state of the system (here $| n_L = - 1,
n_R = + 1 \rangle$) but transitions to the ground state (here $| n_L = 0, n_L
= 0 \rangle$) are not possible because the central valve is closed.}}
\begin{eqnarray}
  | 0_{1 2} 0_{3 4} \rangle & \rightarrow & | n_L = 0, n_R = 0 \rangle, \\
  | 1_{1 2} 1_{3 4} \rangle & \rightarrow & | n_L = - 1, n_R = + 1 \rangle . 
\end{eqnarray}
These charge states are subsequently detected either by proximal charge
sensors or by a charge pumping scheme. The details of these two readout
schemes are explained in {\color{black} Ref. {\cite{Aasen15}}} and the
time-scales for the charge pumping are considered further in {\color{black}
App. \ref{app:readout}}. Note that the above scheme also works for asymmetric charging energies or in the presence of cross-capacitive couplings between the islands; however, the conditions on the gatings may be altered.

{\tmem{Adiabaticity condition.}} The adiabaticity criteria for step 3 are closely
related to those of step 1. The only difference is that we have to consider
the subspaces of even-even and odd-odd parity separately. They are decoupled
because the two wire segments are decoupled ($E_M = E_M^{\text{min}}$). For $E_J \gg E_C$, the system behaves as two
decoupled superconducting islands and the gaps in both parity sectors are the
same: $E_{1, \text{ee}} - E_{0, \text{ee}} = E_{1, \text{oo}} - E_{0,
\text{oo}} \approx \sqrt{8 E_J E_C}$. For $E_J \ll E_C$ the system behaves as
two decoupled topological Cooper pair boxes. Here the eigenstates are close to
charge states and the gaps between them depend sensitively on the gating: Under
condition (\ref{eq:ngcond}), we get $E_{1, \tmop{ee}} - E_{0, \tmop{ee}} = 4
E_C ( 1 + n_{g, L})$, while $E_{1, \text{oo}} - E_{0, \text{oo}} = 4 E_C \min
( | n_{g, L} |, | n_{g, R} |)$. Provided $| n_{g, \alpha} | = O ( 1)$, both
gaps are on the order of the charging energy. We can thus follow the
argumentation of step 1 and obtain the following adiabaticity condition:
\begin{eqnarray}
  T_3 & \gg & \frac{\ln ( E_J^{\max} / E_C)}{E_C} . 
\end{eqnarray}
{\tmem{No-relaxation condition.}} Clearly, this operation has to be done fast
enough so that no electrons can be exchanged between the two islands, i.e.,
\begin{eqnarray}
  T_3 & \ll & \frac{1}{E_M^{\min}} . 
\end{eqnarray}
Otherwise a leakage into the even-even ground state may spoil the readout.

\subsection{Step 4: Reset}\label{sec:step4}
The final step is to close the cycle in parameter space in order to repeat the
protocol again. Opening the central valve ($E_M \rightarrow E_M^{\max}$ at
point A) fuses the MBS pair $( \gamma_2, \gamma_3)$ and brings the parameters of the system back
to the initial point.

{\tmem{Relaxation condition}}. Depending on the measurement outcome, the reset requires a relaxation process because the odd-odd parity configuration is an excited state of the device. Before repeating the cycle, one thus has to wait until the system has relaxed to the ground state (which is possible if the central valve is opened):
\begin{eqnarray}
  T_4 & \gg & \tau_{\text{relax}}.
\end{eqnarray}
 We estimate the involved charge relaxation processes by typical charge relaxation times for GaAs double quantum dots, which are on the order of 10~ns \cite{Fujisawa02,Petta04} due to phonon emission
\footnote{Note that relaxation rates in the other steps of the protocol should be slower: Josephson plasma oscillations in nontopological nanowire setups relax typically on the microsecond scale \cite{Gatemon}.}.  
However, the relaxation time scales for nanowire setups in the presence of a screening superconductor might be different and an interesting future task to investigate.
Since such a relaxation process is needed anyway, there is also no general reason to perform the parameter sweep adiabatically, at least in the charge-readout scheme. For the charge readout, one can simply wait long enough after opening the central valve until the relaxation has certainly happened. 

{\tmem{Adiabaticity condition}}.
For the pumping readout scheme, it would be desirable to perform the resetting step fast as possible in order to maximize the pumping current.  Besides waiting for charge relaxation, it is therefore favorable to open the central valve adiabatically to make sure that the system is not driven to even higher excited states (necessitating possibly multiple relaxation steps). 
For this reason, and because we use it later in \Sec{sec:sweepleftright}, we investigate here the conditions needed to perform the opening of the central valve adiabatically.

As mentioned in {\color{black} Sec.
\ref{sec:singleisland}}, the Hamiltonian conserves the total number of
electrons $N = n_L + n_R$ on the islands when the junctions to the bulk
superconductors are closed. Therefore, transitions are possible only within
the sectors for fixed $N$. 
Using the inequality $| \langle \psi_{1, N = 0}
| \partial H \text{/} \partial E_M | \psi_0 \rangle | \leqslant 1$, we obtain from Eq.
(\ref{eq:fusion-adiabatic2}) the sufficient condition:
\begin{eqnarray}
  T_4  \gg  \left( \int_0^{E_C} + \int_{E_C}^{E_M^{\max}} \right) \frac{d
  E_M }{| E_{1, N = 0} - E_0 |^2},  \label{eq:t4cond}
\end{eqnarray} 
where we have split the integral again into two parts analogous to our
considerations in step 1.
For $E_M \lesssim E_C$, the gap is dominated by the charging energy
(blue-dashed line in {\color{black} Fig. \ref{fig:fusion-splitting}}): $E_1 -
E_0 \approx \sqrt{E_{C, 0}^2 + E_M^2} \geqslant E_{C, 0}:= 2 E_C ( 1 + n_{g,
L} - n_{g, R}) \sim E_C$ under condition (\ref{eq:ngcond}). For $E_M \gg
E_C$, the system is tuned into the single island regime considered in
{\color{black} Sec. \ref{sec:singleisland}} and the gap can be estimated by
$E_{1, N = 0} - E_0 \approx \sqrt{4 E_C E_M}$ for $E_{J, C} = 0$. The latter regime is not reached for the parameters used in Fig. \ref{fig:fusion-splitting}. Inserting
these two gap estimates into the above integrals, we get the condition
\begin{eqnarray}
  T_4 & \gg & \frac{\ln ( E_M^{\max} / E_C)}{E_C} ,\label{eq:t4adiabatic} 
\end{eqnarray}
for $E_M^{\max} / E_C \gg 1$ and assuming that $n_{g, L} - n_{g, R}$ does not come close to 1.

This finally completes the account of the time-scale conditions
(\ref{eq:timescalelower}) and (\ref{eq:timescaleupper}) for the fusion-rule
testing protocol.

\subsection{Time-scale estimate for entire cycle}\label{sec:estimate}
According to the above estimates, all steps are adiabatic if carried out on a time scale $\sim 1/E_C$. For charging energies of a few hundred mK, which is much smaller than the typical superconducting gap in Al (around 2~K \cite{Marcus15_NatureNano_10_232}), we obtain $1/E_C \sim 0.1$~ns. Taking all the steps and logarithmic correction factors into account, steps 1-3 have to be carried out on a time scale of 10~ns to be adiabatic. This is a time scale on the same order as typical times for charge relaxation needed in the resetting step \cite{Fujisawa02,Petta04}. To be sure the system has completely relaxed into the ground state, 100~ns for the entire cycle in the pumping scheme seems reasonable. This results in a pumping current of a few pA.

We have so far not considered the time-scale
conditions for the {\tmem{readout}}, which we postpone to {\color{black} App.
\ref{app:readout}}. We discuss there that a parity-selective pumping process
may be implemented on a time scale faster than $\sim 1 / E_C$, i.e., the steps for the pumping play a minor role in estimating the minimal cycle period.

\section{Time scales for gate-controlled Majorana
manipulations in nanowire networks}\label{sec:braiding}

In this final section, we estimate time scales for performing gate-controlled
exchanges of MBS, which requires going from single-wire structures as sketched in {\color{black}
Fig. \ref{fig:operations}}(a) to branched structures such as trijunction
setups as depicted in {\color{black} Fig. \ref{fig:operations}}(b). We
envisage the three topological wire segments (orange) to be connected via a
nontopological region (blue), which may be either normal or superconducting.

We specifically work out two basic operations that are needed for realizing
braiding: We first consider in {\color{black} Sec. \ref{sec:majtransfer}} the
transfer of a MBS from one nanowire segment to another as sketched in
{\color{black} Fig. \ref{fig:operations}}(c). The second operation, considered
in {\color{black} Sec. \ref{sec:majcentral}}, is the transfer of a MBS across
a trijunction as sketched in {\color{black} Fig. \ref{fig:operations}}(d).
While the first operation can be analyzed from simulations of the
segmented nanowire device, the second operation would, in general, require a
numerical simulation of a trijunction geometry, which we do not pursue here. However, with
our insights from {\color{black} Sec. \ref{sec:spectrum}} for the single-wire
geometry, we may identify the relevant energy scales for the trijunction
geometry, allowing us to give a conservative time-scale estimate also for the
operations in nanowire networks. By concatenating operations of 
{\color{black} Figs. \ref{fig:operations}}(c) and (d), it is possible to
exchange two MBS as shown in {\color{black} Fig. \ref{fig:operations}}(e).
This realizes the braiding protocol discussed in {\color{black} Ref.
{\cite{Aasen15}}}, closely related to Refs.
{\cite{AliceaBraiding,ClarkeBraiding,SauBraiding,HalperinBraiding,BondersonBraiding}} but using electrical control over the Josephson couplings as an alternative to magnetic flux manipulation schemes.

\begin{center}
  \Bigfigure{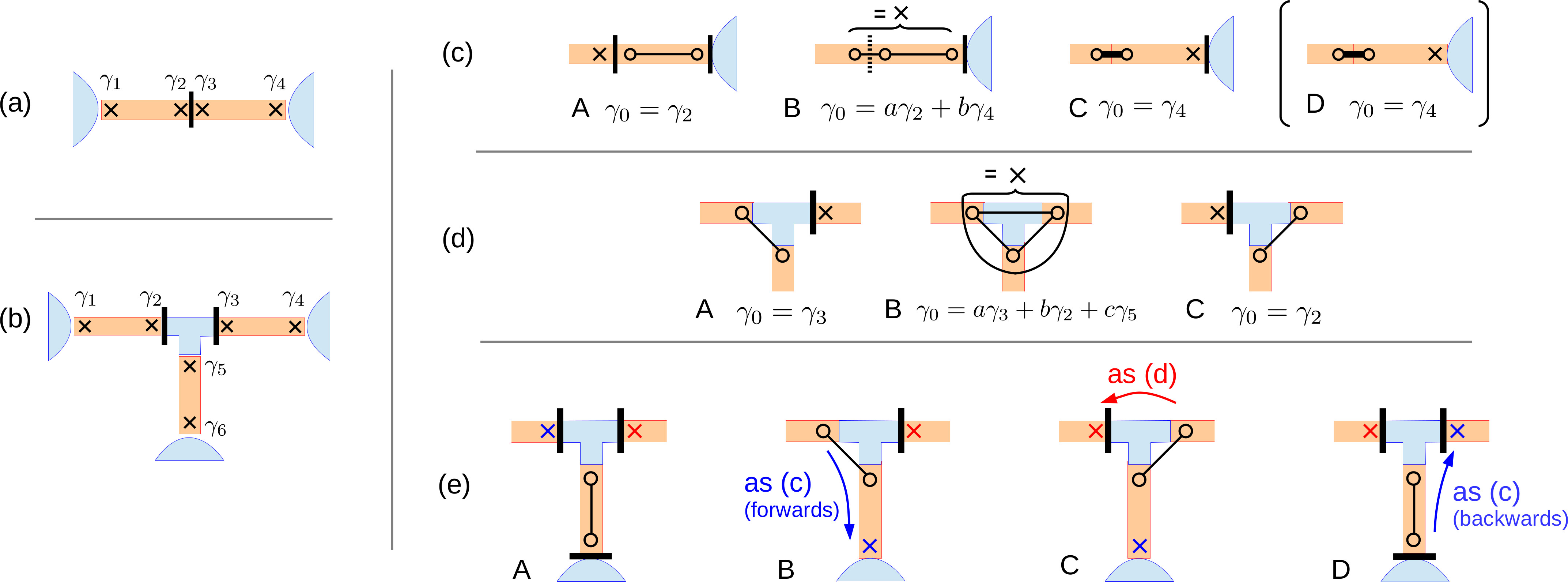}{Basic
  operations for gate-controlled manipulations of MBS in nanowire networks.
  (a) and (b) show the positions of the four and six MBS in the segmented and
  trijunction nanowire geometry, respectively. In (c), we show the steps of
  transferring a MBS from one end of the left wire segment to the opposite end
  of the right segment. During the entire transfer process, a MBS $\gamma_0$ remains at zero energy (see \Eq{eq:gamma0} and the explanation below), which is a linear combination with changing weights of $\gamma_2$ and $\gamma_4$ as indicated in the figure.
We put the last step in parenthesis since opening the
  valve to the right superconductor is not necessary, provided the coupling of
  the central junction can be made large enough (as specified in
  {\color{black} Sec. \ref{sec:majtransfer}}). In (d), we show the steps to
  transfer a MBS from the right side of the trijunction to the left side. (e)
  Elementary braid operation, composed of operations (c) and (d). Analogous to
  (c), the vertical island needs not be connected to a bulk superconductor for
  large couplings across the trijunction. We emphasize that all operations in
  (c) [(d) and (e)] must be performed such that the ground-state degeneracy is not
  changed during the operation, meaning that during the braid operation the system should never be in the
  configurations shown in (a) [(b)] with 4 (6) uncoupled MBS.\label{fig:operations}}
\end{center}

\subsection{Transfer of Majorana bound states across coupled nanowire
segments}\label{sec:majtransfer}

We first discuss the transfer of a MBS from one nanowire segment to the other
as sketched in {\color{black} Fig. \ref{fig:operations}}(c). To be specific,
let us first consider the situation when (A) the valve to the left bulk
superconductor is maximally open ($E_{J, L} \gg E_{C, L}$) and all other
valves are closed ($E_M = E_{J, C} = E_{J, R} = 0$). Then both MBS $\gamma_1$
and $\gamma_2$ on the left island are very close to zero energy (the energy
splitting is exponentially small in $\sqrt{E_{J, L} / E_{C, L}}$). By
contrast, the MBS $\gamma_3$ and $\gamma_4$ on the right island are fused due to
their coupling through the charging energy $E_{C, R}$. To transfer the MBS
$\gamma_2$ at the central junction to the right wire segment, one first opens
the central valve, going through the configuration (B) to (C) and then
opens the valve to the right bulk superconductor resulting in (D).

We emphasize that the order of opening the valves is important: Opening the
right valve first would increase the ground-state degeneracy by bringing
$\gamma_3$ and $\gamma_4$ to zero energy resulting in the configuration shown in
{\color{black} Fig. \ref{fig:operations}}(a). Opening the central valve could
then lead to uncontrolled rotations in the ground-state subspace, which must
be avoided. If instead the central valve is opened first, the ground state
degeneracy remains two-fold (including both even and odd total parity). To
illustrate this point, let us consider the following Majorana Hamiltonian,
\begin{eqnarray}
  H & = & i E_M \op{\gamma}_2 \op{\gamma}_3 + i \varepsilon_{P, R} 
  \op{\gamma}_3 \op{\gamma}_4,  \label{eq:hmaj1}
\end{eqnarray}
accounting for the charging-induced coupling $\varepsilon_{P, R}$ of MBS
$\gamma_3$ and $\gamma_4$ and the tunnel coupling $E_M$ of MBS $\gamma_2$ and
$\gamma_3$. Diagonalization shows that fusing three
MBS yields one fermionic mode at finite energy and one MBS, $\gamma_0$, that remains at
zero energy {\cite{BeenakkerBraiding}}:
\begin{eqnarray}
  \op{\gamma}_0 & = & a \op{\gamma}_2 + b \op{\gamma}_4 .  \label{eq:gamma0}
\end{eqnarray}
The MBS operator $\op{\gamma}_3$ does not appear in Eq. (\ref{eq:gamma0})
because MBS $\gamma_2$ and $\gamma_4$ are not directly coupled. Equation
(\ref{eq:gamma0}) implies that irrespective of the values for $E_M$ and $E_{J,
R}$, which control the coefficients $a$ and $b$, the ground state is
always two-fold degenerate with respect to the fermionic mode $\hat{f} = \hat{\gamma}_1 +
i \hat{\gamma}_0$. To verify this, we numerically computed the energy spectrum of
Hamiltonian (\ref{eq:hfull}) for both total even parity as well as total odd
parity. We have verified that the energy splitting between the lowest states in both
sectors is given by Eq. (\ref{eq:epsp}) for $E_{J, L} / E_C \gg 1$, that is,
it is exponentially small in $\sqrt{E_{J, L} / E_C}$.

We note that the Majorana Hamiltonian (\ref{eq:hmaj1}) is a good
effective description only for the regime $E_C \ll E_M \ll E_{J, L}$, as it otherwise
fails to describe the relevant energy gaps otherwise because it ignores the
Cooper pair condensate. (The
validity is further discussed in {\color{black} App. \ref{app:eps-extract}}.)
The parameter $\varepsilon_{P, R}$ depends, in
principle, on all other parameters and can be suppressed with increasing
$E_M$. In general, the energy gaps have to be inferred from the full
Hamiltonian.

\subsubsection{Time-scale conditions}\label{sec:sweepleftright}

Based on the adiabaticity criterion (\ref{eq:fusion-adiabatic2}), we next show
that the transfer proceeds adiabatically if it is carried out on a time scale
\begin{eqnarray}
  T & \gg & \frac{\ln^{} [ \max ( E_{J, R}^{\max}, E_M^{\max}) / E_C]}{E_C}, 
\end{eqnarray}
provided $E_M^{\max}, E_{J, R}^{\max} \leq E^{\max}_{J, L}$. This is a conservative
estimate derived by assuming $E_{J, C} = 0$. In {\color{black} Fig.
\ref{fig:transfer-splitting}}, we show the energy gap between the lowest
excited and the ground state for both steps of the MBS transfer, both
including and excluding the central Josephson coupling $E_{J, C}$. Our results
are again restricted to even total parity (transitions into states of
different total parity are forbidden). We next discuss the time scales for
each of the two steps.

\begin{center}
  \Figure{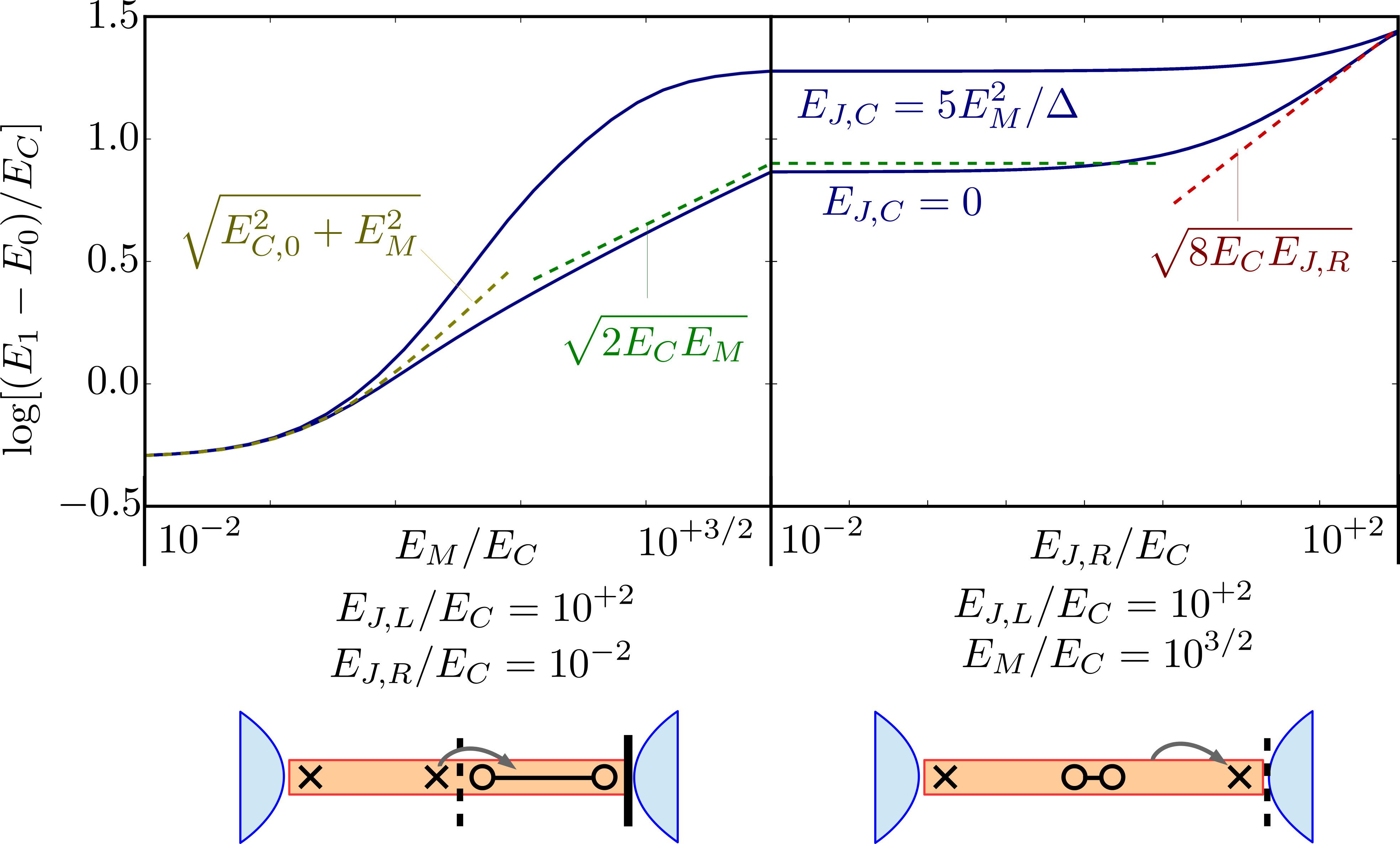}{Transfer
  of MBS between two nanowire segments. Numerically computed energy gap (solid
  blue) between the ground state and the first excited state. The two curves
  are for two different values of $E_{J, C}$.  
  We change
  $E_M$ in the left part and $E_{J, R}$ 
  in the right part as illustrated in the sketches below
  the plots. We also show analytic approximations for the energy splitting
  (dashed lines) as denoted and further discussed in the text. Due to the
  gating of the right island, the relevant charging energy is given by $E_{C,
  0} = E_C ( 1 - 2 n_{g, R})$. The other parameters are given by $E_{C, L} =
  E_{C, R} = E_C$, $n_{g, L} = 0$, $n_{g, R} = 1 / 4$, and 
  $N_{\max} = 20$.\label{fig:transfer-splitting}}
\end{center}

{\tmem{Step 1: Opening the central valve $( E_M \rightarrow E_M^{\max})$.}}
This step is related to the parameter sweep performed in step 4 in the fusion-rule
protocol. Here, we are, however, only interested in the adiabatic evolution of the ground state and 
the total charge on the two islands is not
conserved because $E_{J, L} \neq 0$. Thus, transitions are also possible
into the {\tmem{lowest}} excited state. However, the condition $E_{J, L} \gg
E_C$ also implies that the smallest energy gap is not independent of $E_M$ as
in the case of $E_{J, L} = 0$. In fact, by comparing {\color{black} Fig.
\ref{fig:transfer-splitting}} with {\color{black} Fig.
\ref{fig:fusion-splitting}} we can see that the lowest excited state for
$E_{J, L} \gg E_C$ ({\color{black} Fig. \ref{fig:transfer-splitting}}) follows
the same $E_M$ dependence as the lowest state in the $N = 0$ sector for $E_{J,
L} = 0$ (step 4 in {\color{black} Fig. \ref{fig:fusion-splitting}}). We can
thus follow the same considerations as for step 4 in the fusion-rule
protocol and obtain the following sufficient condition for adiabatic
evolution:
\begin{eqnarray}
  T_1 & \gg & \frac{\ln ( E_M^{\max} / E_C)}{E_C} .  \label{eq:t1transfer}
\end{eqnarray}
We note that an underlying assumption here is that $E_M^{\max} < E_{J, L}$.
Otherwise the two islands are strongly hybridized and the lowest excited state
is the first excited state of the Josephson plasma oscillations [see
{\color{black} Sec. \ref{sec:singleisland}} (B)]. This regime is not
reached for the parameters we assume in {\color{black} Fig.
\ref{fig:transfer-splitting}}.

{\tmem{Step 2: Opening the right valve $( E_{J, R} \rightarrow E_{J,
R}^{\max})$.}} As the right half of {\color{black} Fig.
\ref{fig:transfer-splitting}} shows, the energy gap $E_1 - E_0 \sim \sqrt{2
E_J E_M}$ remains constant over a larger range and increases only for larger
$E_{J, R}$. If the Josephson coupling to right bulk superconductor in the final
stage does not exceed that to left bulk superconductor (i.e., $E_{J,
R}^{\max} \leqslant E_{J, L}^{\max}$), the gap $\Delta E$ satisfies
\begin{eqnarray*}
  \Delta E ( E_{J, L}^{\max}, E_{J, R}) & \geqslant & \Delta E ( E_{J, L} =
  E_{J, R}),
\end{eqnarray*}
i.e., the gap is always larger than that given when symmetrically increasing
both Josephson couplings. The symmetric increase is studied in {\color{black}
Sec. \ref{sec:step1}} for step 1 of the fusion-rule protocol and the
adiabaticity criterion (\ref{eq:fusion-adiabatic}) should be satisfied here
even better due to the larger gap. Employing the result (\ref{eq:t1adiabatic})
given there, the evolution proceeds adiabatically if the time for this step satisfies
\begin{eqnarray}
  T_2 & \gg & \frac{\ln ( E_{J, R}^{\max} / E_C)}{E_C} . 
  \label{eq:t2transfer}
\end{eqnarray}
Conditions (\ref{eq:t1transfer}) and (\ref{eq:t2transfer}) show that
transferring MBS between different wire segments can be performed on the same
time scale as initializing them for, e.g., testing the fusion rules as
discussed in {\color{black} Sec. \ref{sec:fusion}}. \

\subsubsection{Suppression of charging effects through the center junction}\label{sec:majeps}

We next show that the above transfer of the MBS between two segments can
actually be achieved even {\tmem{without}} the second step of opening the
right valve to the bulk superconductor. A prerequisite is that the condition
$E_C \ll E_M, E_{J, L}$ can be satisfied.

Intuitively, one could expect that the connection to the right bulk is not
needed once the system behaves as larger superconducting island for large
Majorana coupling $E_M > E_J$. We verify this expectation in {\color{black}
App. \ref{app:eps-extract}}, where we show that asymmetries of the bulk
Josephson couplings $E_{J, \alpha}$ become irrelevant for $E_M \gg E_J$.
However, it turns out that the condition $E_M > E_J$ is, in fact, not
necessary at all.

\begin{center}
  \Figure{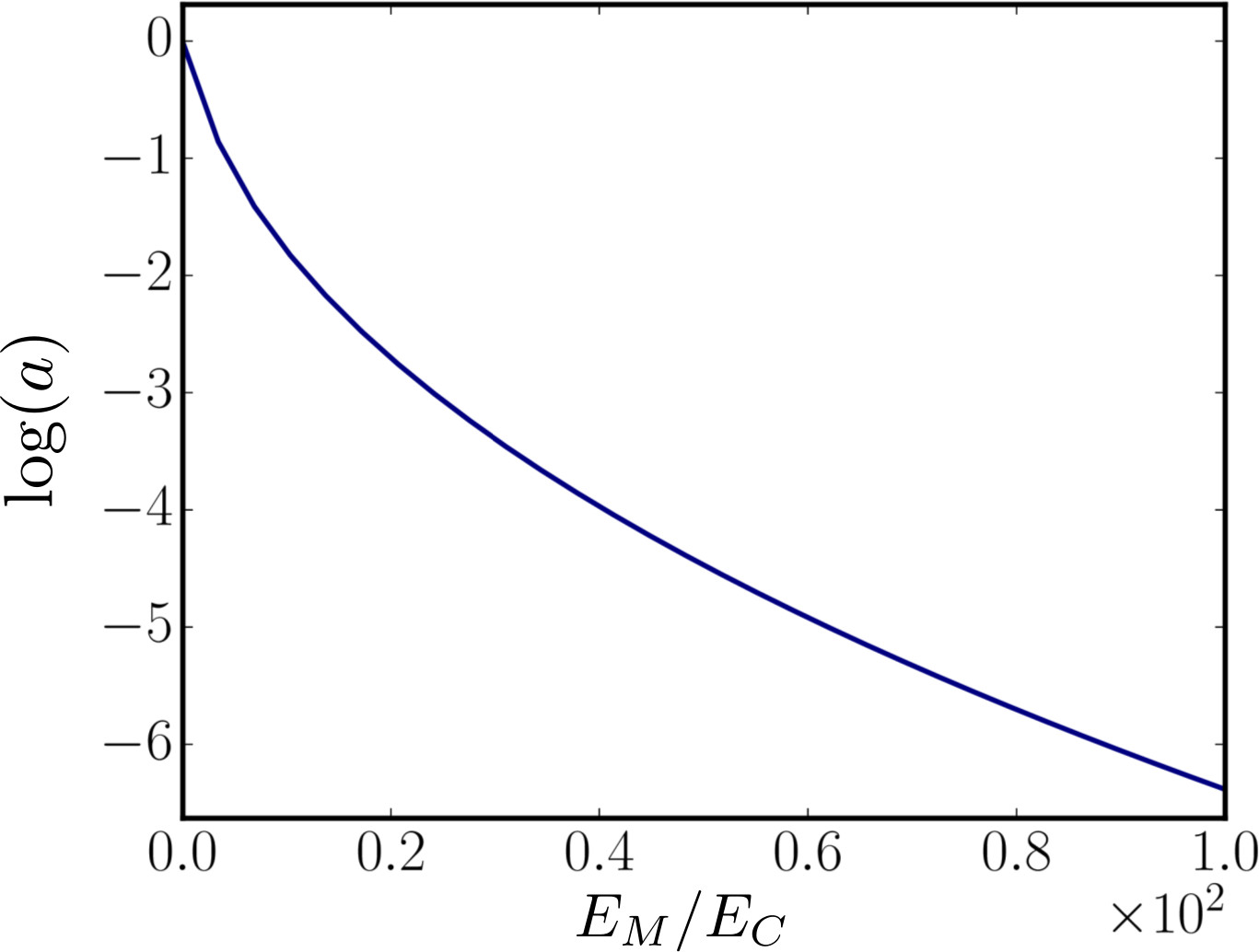}{Gate-controlled
  transfer of a MBS between two nanowire segments. We show the numerically
  computed coefficient $a$ in Eq. (\ref{eq:gamma0}) as a function of the
  Majorana coupling $E_M$. The calculation of the coefficient $a$ is explained in
  {\color{black} App. \ref{app:eps-extract}}. The parameters used here are
  $E_{J, L} = 100 E_C$, $E_{J, R} = E_{J, C} = 0$, $n_{g, L} = 0$, $n_{g, R} =
  0.25$, and $N_{\max} = 10$.\label{fig:alpha}}
\end{center}

To demonstrate this, we show the dependence of the coefficient $a$ in the
linear combination $\hat{\gamma}_0 = a \hat{\gamma}_2 + b \hat{\gamma}_4$ [Eq.
(\ref{eq:gamma0})] in {\color{black} Fig. \ref{fig:alpha}}. Indeed, we find
that $a$ is strongly suppressed as a function of the Majorana
tunneling $E_M$ (exponentially for very large ratios $E_M/E_C$). The technical details how we extract $a$ and $b$ from our
numerical simulations are postponed to {\color{black} App.
\ref{app:eps-extract}}. The exponential suppression of $a$ crucially
relies on a suppression of the coupling $\varepsilon_{P, R}$ between the two MBS
on the right island. In other words, tuning the Majorana coupling changes not only
the parameter $E_M$ in the effective model (\ref{eq:hmaj1}), but also the
coupling $\varepsilon_{P, R}$ . This demonstrates that it is important to
include the dynamics of the Cooper pair condensates on the islands to arrive
at an effective description of the Majorana physics.

\subsection{Transfer of Majorana bound states across
trijunctions}\label{sec:majcentral}

The second operation we consider is the transfer of a MBS across a trijunction
as sketched in {\color{black} Fig. \ref{fig:operations}}(d). At first, the
left valve at the trijunction is open, connecting the left and the vertical
segment, while the right valve at the trijunction is closed as shown in (A). A
MBS is thus located at the right side of the trijunction. Then the right valve
is opened and all three MBS located near the branching point
are fused as depicted in (B). Analogous to the MBS transfer discussed above, this
leaves behind a finite-energy fermionic mode and one zero-energy MBS, which is
a linear combination of all three coupled MBS. Following the notation
introduced in {\color{black} Fig. \ref{fig:operations}}(b), this mode can be expressed as
\begin{eqnarray}
  \hat{\gamma}_0 & = & a \hat{\gamma}_2 + b \hat{\gamma}_3 + c \hat{\gamma}_5. 
\end{eqnarray}
Finally, the left valve is closed and the MBS is located at the left of the
trijunction, while now the right and vertical segments are connected as shown
in (C).

To estimate time scales for this transfer process, we first need to specify a
model for the trijunction geometry. We assume that the Hamiltonian is a simple
extension of the two-fold segmented nanowire structure
{\cite{BeenakkerBraiding}}:
\begin{eqnarray}
  H & = & \sum_{\alpha = L, R, V} E_{C, \alpha} \left( \op{n}_{\alpha} - n_{g,
  \alpha} \right)^2 + E_{J, \alpha} \left( 1 - \cos \left(
  \op{\varphi}_{\alpha} \right) \right) \nonumber\\
  &  & + \sum_{\langle \alpha, \beta \rangle} E_{M, \alpha \beta} \cos \left(
  \frac{\op{\varphi}_{\alpha} - \op{\varphi}_{\beta}}{2} \right) i
  \op{\gamma}_{\alpha} \op{\gamma}_{\beta} . 
\end{eqnarray}
In the second term above, we sum over the pairs $\langle \alpha,\beta \rangle = \langle L, R \rangle, \langle R, V \rangle, \langle V,L \rangle $ and the Majorana operators refer to those MBS at the trijunction as indicated in \Fig{fig:alpha}.
We neglect in our considerations the Josephson coupling between the islands
since it tends to increase all energy gaps. We will argue in the following
that transfer of the MBS is adiabatic if it is carried out on a time scale
\begin{eqnarray}
  T & \gg & \frac{\tmop{arsinh} ( E_M^{\max} / E_M^{\min})}{E_C}, 
  \label{eq:trijunctiontransfer}
\end{eqnarray}
in the parameter regime
\begin{eqnarray}
  E_C & \ll & E^{\min}_M \lesssim E_{J,L/R}\left( \ \lesssim \frac{(
  E_M^{\min})^2}{E_C}  \ < \ \frac{\Delta^2}{E_C} \right) . 
  \label{eq:braidcond}
\end{eqnarray}
Here, $E^{\min ( \max)}_M$ denotes the minimal (maximal) value of
\begin{eqnarray*}
  E_M & = & \sqrt{E_{M, L R}^2 + E_{M, R V}^2 + E_{M, V L}^2},
\end{eqnarray*}
which is the energy of the fermionic mode created by fusing the three MBS at
the trijunction.

Let us discuss the physics behind the different conditions in Eq.
(\ref{eq:braidcond}). First of all, the condition $E_{J,L/R} \gg E_C$
implies that the charging energies on both the left and right island are
strongly suppressed and the MBS pairs $( \gamma_1, \gamma_2)$ and $( \gamma_3,
\gamma_4)$ are decoupled to exponential accuracy. Second, since $E^{\min}_M
\gg E_C$, the charging energy on the vertical island is also exponentially
small in $E_M / E_C$ during the entire process (see {\color{black} App.
\ref{app:eps-extract}}): The vertical island is then strongly coupled either
to the left or the right island. Under this condition, the Josephson coupling
$E_{J, V}$ to the bulk superconductor attached to the vertical island would
not be necessary at all. Connecting the vertical island to the bulk may,
however, be useful in experiments to suppress charging effects on the
vertical island even further if $E_M$ cannot be made sufficiently large in
practice.

The condition $E^{\min}_M \lesssim E_{J, L}, E_{J, R}$ implies that the lowest
energy scale of the system is set by \ $\Delta E \sim \sqrt{E_M E_C}$, similar
to the Majorana-tunneling dominated regime discussed in Sec.
\ref{sec:singleisland}. The Josephson plasma frequencies of the islands $\sim
\sqrt{E_J E_C}$ or the Majorana couplings $\sim E_M$ are much larger. We
therefore conclude that the gap of the lowest excited state $| \psi_1 \rangle$
to the ground-state manifold is at least on the order of $\sqrt{E_M E_C}$.
Using the adiabaticity criterion (\ref{eq:fusion-adiabatic2}), we can write
down the condition
\begin{eqnarray}
  T & \gg & \int_C \frac{\sum_{\langle \alpha, \beta \rangle} | d E_{M, \beta
  \alpha} ( \vecg{\lambda}) |}{E_M E_C}  \label{eq:tcentral}
\end{eqnarray}
where $C$ denotes the curve in parameter space given by the three
Majorana couplings $\vecg{\lambda} = ( E_{M,L R}, E_{M, RV}, E_{M, VL})$. The
integral on the right-hand side of Eq. (\ref{eq:tcentral}) scales roughly as
stated by Eq. (\ref{eq:trijunctiontransfer}), which depends logarithmically on
$E_M^{\max} / E_M^{\text{min}}$.

We emphasize that it is important that the junctions at the branching point
are not closed at the same time (then $E_M \approx$0 and the denominator in
Eq. (\ref{eq:tcentral}) diverges). If the vertical island is connected to the
bulk, this would result in the situation shown in {\color{black} Fig.
\ref{fig:operations}}(b): Then the number of zero-energy MBS would be increased
from four to six, which doubles the ground-state degeneracy. Since relevant
gaps become small, the evolution would not necessarily be adiabatic any more and
uncontrolled rotations of the state may result.

We finally note that the transfer would in principle \ also work if the
rightmost condition in Eq. (\ref{eq:braidcond}), $E_{J, L / R} \lesssim (
E_M^{\min})^2 / E_C$, was not be satisfied. We added it here because the entire
protocol may involve more steps where this condition may be needed, for
example, to {\tmem{initialize}} the system.

\section{Summary and conclusions}\label{sec:summary}
  
  In this paper, we have analyzed the energy spectrum of two tunnel-coupled
  topological nanowire segments, which is a candidate setup for testing the
  topological nature of Majorana bound states.
  
  We have computed the energy spectrum of this device at low energies numerically
  and identified four different operating regimes with several fine-structure
  subregimes as depicted in {\color{black} Fig. \ref{fig:regimes}}: (i) We find
  a charge-dominated regime $E_C \gg E_M, E_{J, \alpha}$ utilizable for
  initialization and readout of Majorana bound states. (ii) We identify a
  single-island regime for $E_M \gg E_C, E_{J, \alpha}$. (iii) We find a double-island regime for $E_{J, \alpha} \gg E_M, E_C$, 
  where charging
  effects are strongly suppressed with $ \varepsilon_{P, \alpha} \sim (
  E_C E_{J, \alpha}^3)^{1 / 4} \exp \left( - \sqrt{8 E_{J, \alpha} / E_C}
  \right)$. (iv) Finally, one can enter a regime with four MBS near zero energy, a regime
  allowing for topologically protected Majorana manipulations.
  
  Based on the low-energy spectrum, we have analyzed the time scales for preparation,
  manipulation, and readout of MBS in this setup. We discussed specifically
  each step of the fusion-rule test suggested in {\color{black}
  Ref. {\cite{Aasen15}}} and also basic operations for manipulating MBS in
  nanowire networks as needed for braiding. It turns out that the time scale
  is limited from below by $\Delta t \gg A / E_C \sim$0.1~ns, where $A$ depends on
  logarithmic terms in $E_M / E_C$ and $E_{J, \alpha} / E_C$ and from
  above by $\Delta t \ll 1 / \varepsilon_{P, \alpha},1/ E_M^{\min}$, which is
  the minimal ground-state energy splitting.  In addition, the resetting step of the fusion-rule protocol includes a charge-relaxation process, which is not included in the above criteria. Charge-relaxation times of semiconductor quantum dots are typically on the order of 10~ns.
  
  Since all the time-scale estimates we give are rather conservative, it would be
  interesting to see to which extent one could do better than the time-scale
  window given here. This would require a more accurate description of the
  Josephson energy for the nanowire junctions, both for the connection to the
  bulk superconductors and between the two islands. Since the charging
  energy might not be very small compared to the superconducting gap in the
  experimental devices {\cite{Marcus15_NatureNano_10_232}}, future studies
  should also include the effect of quasiparticles. Quasiparticle poisoning has been studied in particular for nontopological transmon qubits \cite{Catelani11,Riste13,Catelani14,Wang14} and also aiming at topological devices \cite{RainisNoise,PoisoningTime,Woerkom15}, but especially the interplay with charging energy has, to our knowledge, not
  yet been explored.

\begin{acknowledgments}
  We acknowledge stimulating discussions with  David Aasen, Jason Alicea, Joshua A. Folk, Fabian Hassler, Andrew Higginbotham, Thomas S. Jespersen, Ferdinand Kuemmeth, Charles M. Marcus, and Ryan V. Mishmash. We acknowledge support from the Crafoord Foundation (M. L. and M. H.), the Swedish Research Council (M. L.), and The Danish National Research Foundation (K. F.).
\end{acknowledgments}

\appendix

\section{Derivation of the Hamiltonian}\label{app:derh}
\new{In this Appendix, we derive the island Hamiltonian \eq{eq:hc} and tunneling Hamiltonian \eq{eq:ht} presented in \Sec{sec:model} of the main
part starting from a BCS description. Our effective
low-energy Hamiltonian is valid at energies much smaller than the
superconducting gap ($E \ll \Delta$). Aside from this, we define the
phase-basis states we introduced in Eqs. \eq{eq:phie} and \eq{eq:phio} in the main part and give the connection to the
definition of the Majorana operators.}
\new{\subsection{Island Hamiltonian}\label{app:derhi}}
\new{The island Hamiltonian $H_I =H_F + H_C $ is given by two contributions.
The first, single-particle part $H_F$ describes the fermionic modes of the
islands and takes the form of a BCS Hamiltonian:
\begin{eqnarray}
  H_F ( \varphi) & = & \int d^3 x \left\{ \sum_{\sigma}
  \hat{\psi}^{\dag}_{\sigma} ( x) H_0 ( x) \hat{\psi}_{\sigma} ( x) \right.
  \nonumber\\
  &  & \nobracket + ( \Delta ( x) e^{- i \varphi}
  \hat{\psi}^{\dag}_{\uparrow} ( x) \hat{\psi}^{\dag}_{\downarrow} ( x) +
  \text{H.c.}) \} \label{eq:hisland}. 
\end{eqnarray}
Here, $\hat{\psi}_{\sigma} ( x)$ denotes field operators for electrons with spin $\sigma$. \newer{The first part, $H_0(x)$, contains all effects not related to superconductivity and especially the spin-orbit coupling and Zeeman energy, which are needed to drive the island into a topologically nontrivial state. The second line of \Eq{eq:hisland} incorporates the superconductivity in a BCS description. The superconducting gap is given by $\Delta ( x)$, which is nonzero in the
metal part of the island, and the superconducting phase of the
island is given by $\varphi$.}}

\new{The second contribution to the island Hamiltonian is the Coulomb
interaction energy,
\begin{eqnarray}
  H_C ( \varphi) & = & E_C \left( \frac{2}{i}  \frac{\partial}{\partial
  \varphi} + \hat{n}_e - n_g \right)^2, \label{eq:hcapp}
\end{eqnarray}
which incorporates the total number of electrons, given by twice the number
$N_C$ of Cooper pairs in the condensate ($2 \hat{N}_C = - 2 i \partial / \partial
\varphi$) plus unpaired electrons occupying other fermionic modes:
\begin{eqnarray}
  \hat{n}_e & = & \sum_{\sigma} \int d^3 x \hat{\psi}_{\sigma}^{\dag} ( x)
  \hat{\psi}_{\sigma} ( x) . 
\end{eqnarray}
\newer{The gating $n_g$ in \Eq{eq:hcapp} accounts for the effect of the applied gate voltages, which change the number of electrons on the island that minimizes the energy}}

\new{The Hamiltonian is expressed here in the phase basis, $| \varphi,
\mathbf{n}_e \rangle$, which characterizes the \newer{Cooper-pair condensate} and 
$|\mathbf{n}_e \rangle = |n_{e,0},n_{e,1},... \rangle $ contains the occupation numbers for modes of unpaired
electrons. The phase-basis states can be expressed as
\begin{eqnarray}
  | \varphi, \mathbf{n}_e \rangle & = & \frac{1}{\sqrt{2 \pi}} \sum_{N_C}
  e^{- i \varphi N_C} | N_C, \mathbf{n}_e \rangle, \label{eq:phibefore}
\end{eqnarray}
where $| N_C, \mathbf{n}_e \rangle$ is a state with a well-defined number of
$N_C$ Cooper pairs. Since $N_C$ is an integer, the basis states are $2
\pi$-periodic in $\varphi$, $| \varphi + 2 \pi, \mathbf{n}_e \rangle = |
\varphi, \mathbf{n}_e \rangle$, and so are the wave functions in phase-space
representation, $\psi_{\mathbf{n}_e} ( \varphi) = \langle \varphi,
\mathbf{n}_e | \nobracket \psi \rangle$.}

\new{To proceed, we remove the superconducting phase from the single-particle part along the lines of, e.g., \Cite{Barkeshli15} by a
unitary transformation of the Hamiltonian and the states,
\begin{eqnarray}
  H' \ = \ U H U^{\dag}, & &
  | \psi \rangle ' \ = \ U | \psi \rangle,\label{eq:psitransform} 
\end{eqnarray}
choosing (different from  \Cite{Barkeshli15})
\begin{eqnarray}
  U & = & e^{- i \varphi \hat{n}_e / 2} .  \label{eq:unitary}
\end{eqnarray}
\newer{The transformed Hamiltonians and states introduced in this Appendix are the ones used in the main part (where we leave out the prime for simplicity). 
Applying the unitary transformation to the single-particle part yields}
\begin{eqnarray}
  H_F' & = & U H_F ( \varphi) U^{\dag} \ = \ H_F ( 0) \ = \ \sum_{i > 0} E_i
  \hat{\chi}_i^{\dag} \hat{\chi}_i,  \label{eq:hfprime}
\end{eqnarray}
where in the last step we diagonalized $H_F$ \newer{with eigenenergies $E_{i} \geq 0$} and Bogoliubov field operators
\begin{eqnarray}
  \hat{\chi}_i & = & \sum_{\sigma} \int d x \{ a_{i, \sigma} ( x)
  \hat{\psi}_{\sigma} ( x) + b_{i, \sigma} ( x) \hat{\psi}^{\dag}_{\sigma} (
  x) \} .  \label{eq:chizerophase}
\end{eqnarray}
\newer{When expressing $H_F'$ in its eigenmodes, we assumed that the wire is infinitely long. The single-particle part thus possesses a mid-gap mode $i=0$ at energy $E_0 =0$ with field operator
$\hat{\chi}_0$, which does not appear in the sum of Eq. (\ref{eq:hfprime}). Here and in the rest of this paper we refer to this mode as the 'zero-energy mode' or 'Majorana mode', even though this mode may move away from zero energy when adding the effect of the charging energy or when the wire has finite length.}
We next exclude all modes with finite energy $E_{i>0} \sim \Delta$, which
means that we may simply drop $H'_F$ when projecting on the low-energy
subspace.
We decompose the only remaining field operator $\hat{\chi}_0$ into the Majorana operators
$\hat{\gamma}_1, \hat{\gamma}_2$ by
\begin{eqnarray}
  \hat{\chi}_0 & = & \hat{\gamma}_1 + i \hat{\gamma}_2.  \label{eq:maj}
\end{eqnarray}
Exploiting Eq. (\ref{eq:chizerophase}), the Majorana operators read
\begin{eqnarray}
   \hat{\gamma}_n & = & \sum_{\sigma} \int d^3 x \{ c_{n i \sigma} ( x)
  \hat{\psi}_{\sigma} ( x) + c^*_{n \nocomma \sigma i} ( x)
  \hat{\psi}^{\dag}_{\sigma} ( x) \},  \nonumber \\
  & & \label{eq:gammadef}
\end{eqnarray}
\newer{where we chose the functions $c_{n i \sigma} ( x)$ such that they are exponentially localized either at the left end (for $n=1$) or the right end (for $n=2$) of the wire}.\\
We next transform the charging energy,
\begin{eqnarray}
  H'_C \ = \ U H_C U^{\dag} & = & E_C \left( \frac{2}{i} 
  \frac{\partial}{\partial \varphi} - n_g \right)^2, 
\end{eqnarray}
in which the term from the unpaired fermionic modes has been removed. The
information about the parity of the island is now contained in the boundary
conditions of the wave function in phase space. The low-energy subspace is
spanned by the basis states
\begin{eqnarray}
  | \varphi, p_{1 2} \rangle' & = & e^{- i \varphi p_{1 2} / 2}  | \varphi,
  p_{1 2} \rangle, 
\end{eqnarray}
where $p_{12} = 0, 1$ is the occupation number of the zero-energy mode. We
have left out here the reference to all other unpaired fermionic modes, which
we assume for simplicity to be occupied by an even number of
electrons {\footnote{We assumed that the number $n'_e$ of occupied
unpaired modes excluding the zero mode is even. Then the parity of the island
$p_{12} = n_0$ equals the occupation $p_{1 2}$ of the zero-energy mode and the
unitary transformation (\ref{eq:psitransform}) contributes an irrelevant $2
\pi$-periodic phase factor $e^{i n_e' \varphi / 2}$ to the wave function
(since it does not change). If $n_e'$ is odd, we could proceed in the same way
by connecting the parity via $p_{1 2} = 1 + n_0$ to the occupation $n_0$ of
the zero-energy mode.}}. The phase basis states can now be written as in the main
part,  Eqs. \eq{eq:phie} and \eq{eq:phio}:
\begin{eqnarray}
  | \varphi, p_{1 2} \rangle' & = & \frac{1}{\sqrt{2 \pi}}  \sum_{N_C} e^{- i
  \varphi ( 2 N_C + p_{12}) / 2} | N_C, p_{1 2} \rangle,  \label{eq:phip0}\\
  & = & \frac{1}{\sqrt{2 \pi}} \sum_{n\text{ with parity } p_{1 2}} e^{- i
  \varphi n / 2} | n \rangle . 
\end{eqnarray}
Moreover, physical wave functions have to obey parity-dependent boundary
conditions:
\begin{eqnarray}
  \psi'_{p_{12}} ( \varphi + 2 \pi) & = & ( - 1)^{p_{12}} \psi'_{p_{12}} ( \varphi) . 
  \label{eq:aperbc}
\end{eqnarray}
\newer{The above boundary conditions are closely related to the phase-space representation of the number operator: One possibility is to use periodic boundary conditions (as before the unitary transformation), then $ - 2 i \partial / \partial \varphi$ counts the number of fermions in the Cooper-pair condensate. The other possibility is to use (anti)periodic boundary conditions and then $ - 2 i \partial / \partial \varphi$ counts the number of fermions contained in the Cooper-pair condensate plus the Majorana mode. This can be easily seen by comparing Eqs. \eq{eq:phibefore} and \eq{eq:phip0}. }}

\new{For an island without charging energy and a well-defined, fixed phase
$\varphi$, the Hilbert space can be further reduced to that of a two-level system $|
p_{1 2} \rangle' = | \varphi, p_{1 2} \rangle'$. In this case, the zero-mode field
operator acts as
\begin{eqnarray}
  \hat{\chi}_0 | p_{1 2} = 1 \rangle ' & = & | p_{1 2} = 0 \rangle ', \\
  \hat{\chi}_0 | p_{1 2} = 0 \rangle ' & = & 0, 
\end{eqnarray}
and using Eq. (\ref{eq:maj}), the Majorana operators flip the parity:
\begin{eqnarray}
  \hat{\gamma}_1 | p_{1 2} \rangle ' & = & | \bar{p}_{1 2} \rangle ', \label{eq:majdef1} \\
  \hat{\gamma}_2 | p_{1 2} \rangle ' & = & i^{2 p_{12} - 1} | \bar{p}_{1 2}
  \rangle '. \label{eq:majdef2}
\end{eqnarray}
Such a simplified description of the Hilbert space is reasonable when the
tunneling energies in our setup are large compared to the charging energy.}

\new{\newer{When the charging energy is increased, phase fluctuations become important and phase is no longer a good quantum number.
In this case, Eqs. \eq{eq:majdef1} and \eq{eq:majdef2} should only be understood as formal definitions of the Majorana operators and one has to be careful when trying to assign a physical meaning to them. Applying the Majorana operators to a general state $\psi'_{p_{12}}(\varphi)$ that obeys the boundary condition \eq{eq:aperbc}, produces a new wave function that violates the periodicity condition and is thus not physical. Specifically, the Majorana operators interchange the wave function components in the even and odd parity sector. Thus, the odd (even) component of the new wave function obeys (anti)periodic boundary conditions, i.e., the opposite of what \Eq{eq:aperbc} requires.
Therefore, any observable, including the Hamiltonian, can in general not contain `lonesome' Majorana operators $\hat \gamma_n$.
Rather, the Majorana operators always appear in combination with phase operators, such as in Eqs. \eq{eq:gamma1} and \eq{eq:gamma2},
\begin{eqnarray}
  e^{\pm i \hat{\varphi} / 2} \hat{\gamma}_1 | \varphi, p_{1 2} \rangle' & = &
  e^{\pm i \varphi / 2} | \varphi, \bar{p}_{1 2} \rangle' . 
  \label{eq:gammaphi}
\end{eqnarray}
These operators, when applied to a general wave function $\psi'_{p_{12}}(\varphi)$, produce a state that does obey the boundary condition \eq{eq:aperbc},
\begin{equation}
e^{\pm i \hat \varphi /2}\hat \gamma_1 \psi'_{p_{12}}(\varphi) = e^{\pm i\varphi /2} \psi'_{\bar p_{12}}(\varphi),
\end{equation}
and therefore can have a physical meaning.
In addition, the operators $e^{\pm i \hat \varphi /2}\hat \gamma_1$ also have a simple interpretation in the number basis:
\begin{equation}
e^{\pm i \hat \varphi /2}\hat \gamma_1 = \sum_n |{n\pm 1}\rangle\langle {n} |, \label{eq:fusion-displseapp}
\end{equation}
it simply shifts the charge state by one.}
Corresponding relations hold for $\hat{\gamma}_2$ as stated in the main part
in Eqs. \eq{eq:gamma2} and \eq{eq:fusion-displse}. }
\new{\subsection{Tunneling Hamiltonian}\label{app:derht}}
\new{For the derivation of the Majorana-Josephson coupling we start from a standard bilinear coupling of the two wire segments:
\begin{eqnarray}
  H_T & = & \int  d^3 x [ t ( x) \hat{\psi}_L ( x) \hat{\psi}_R^{\dag} ( x) +
  \text{H.c.}] .
\end{eqnarray}
Applying the above unitary transformation (\ref{eq:unitary}), the tunneling
Hamiltonian is transformed to
\begin{eqnarray}
  H_T' & = & \int  d^3 x [ t ( x) e^{- i ( \hat{\varphi}_L -
  \hat{\varphi}_R) / 2}  \hat{\psi}_L ( x) \hat{\psi}_R^{\dag} ( x) +
  \text{H.c.}]. \nonumber
  \\
  & & \label{eq:htprime}
\end{eqnarray}
To express this Hamiltonian in terms of the Majorana operators, we proceed similar to
Refs. \cite{Barkeshli15,ZeroBiasAnomaly4,MajoranaTransportWithInteractions2} and invert
the relation (\ref{eq:chizerophase}) (re-introducing the island index
$\alpha$),
\begin{eqnarray}
  \hat{\psi}_{\alpha} ( x) & = & f_{\alpha} ( x) \hat{\chi}_{\alpha 0} +
  f^*_{\alpha} ( x) \hat{\chi}^{\dag}_{\alpha 0} + \ldots, \\
  & = & k_{\alpha} ( x) \hat{\gamma}_{\alpha 1} + i l_{\alpha} ( x)
  \hat{\gamma}_{\alpha 2} + \ldots,  \label{eq:psiamaj}
\end{eqnarray}
where $\ldots$ denotes the contributions from finite-energy modes, which we
drop in the low-energy description. In the last step, we have introduced
Majorana operators according to Eq. (\ref{eq:maj}), where $\hat{\gamma}_{L 1}
= \hat{\gamma}_1, \hat{\gamma}_{L 2} = \hat{\gamma}_2, \hat{\gamma}_{R 1} =
\hat{\gamma}_3$, and $\hat{\gamma}_{R 2} = \hat{\gamma}_4$ consistent with our
labeling in \Fig{fig:model} used in the main part. When inserting Eq. (\ref{eq:psiamaj})
into the expression (\ref{eq:htprime}) for the tunneling Hamiltonian, we can
simplify the expression by assuming that $t ( x)$ is a function localized
in the vicinity of the junction, i.e., $k_L ( x) t ( x), l_R ( x) t ( x)
\approx 0$. We may therefore neglect $\hat{\gamma}_1$ and $\hat{\gamma}_4$
from the integral and find
\begin{eqnarray}
  H'_T & \approx & \frac{E_M}{2} e^{- i ( \hat{\varphi}_L - \hat{\varphi}_R) /
  2} i \hat{\gamma}_2 \hat{\gamma}_3 + \text{H.c.}
\end{eqnarray}
with $E_M = 2 \int d^3 x ~  t ( x) l_R ( x) k_L ( x)$. Assuming finally $E_M$ to be
real, we obtain the expression \eq{eq:ht} for the Majorana-Josephson coupling in the
main part. 
\newer{Inserting \Eq{eq:gammaphi} and \Eq{eq:fusion-displseapp} into the tunneling Hamiltonian, we finally obtain an operational definition for the tunneling Hamiltonian both in the phase- and the number-basis representation as used in our numerical calculations.}}\\

\section{Parameter choices and
estimations}\label{app:parameter-estimation}

This Appendix is dedicated to discussing the parameter choices and relations
that we employ in our study. We first argue in App. \ref{app:phasedifferences} why we neglect phase
differences between the two bulk superconductors connected to the nanowire.
We then give in App. \ref{app:emejratio} a heuristic derivation of how the Majorana coupling $E_M$
and the Josephson coupling $E_{J, C}$ of the junction connecting the two
topological superconducting islands are related. In {\color{black} App.
\ref{app:fusion-parameters}}, we finally motivate our parameter choices for the
time-scale estimations of the fusion-rule protocol ({\color{black} Sec.
\ref{sec:fusion}}) regarding also their experimental feasibility.

\subsection{Phase differences across the junction}\label{app:phasedifferences}
We have set the phase difference between the two bulk superconductors
(in the following denoted by $\chi$) to zero in the Hamiltonian (\ref{eq:hj}).
However, in contrast to superconducting circuits with a loop, the phase differences cannot be adjusted by an external magnetic flux for the device we consider. 
In this Appendix, we explain why this is nevertheless a permissible simplification in our analysis of the fusion-rule protocol.

First, phase differences between the bulk superconductors matter only if all three valves are at least partially open (step 1 and 2 of the fusion-rule protocol).  Otherwise the disconnected parts of the semiconductor wire structure can be treated independently and a nonzero phase difference $\chi$ can simply be absorbed into a redefinition of the phases.
This is not possible if all junctions are partially open. However, we have explicitly verified from our numerical simulations that the ground state of the system is assumed for zero phase difference $\chi = 0$ irrespective of the values of all parameter values. Since the system adiabatically follows the ground state in step 1 and 2, there is no reason why a nonzero offset phase should develop.

Second, in addition to such an offset phase, one should also investigate the effect of phase fluctuations, which are associated with current fluctuations. While these fluctuations do not affect the measurement outcomes for the charge sensing, they can affect the pumping current. For small phase differences $\chi$, one can write the energy of the $\chi$-dependent part of the energy as $E \sim  E_{\min} \chi^2/2$, where $E_{\min}$ is roughly given by the smallest of the Majorana / Josephson couplings $E_{\min} = \min(E_{J,\alpha} (t),E_M (t))$ \footnote{The phases across the junctions with larger couplings $E_T > E_{\text{min}}$ are more strongly locked and phase fluctuations are more strongly suppressed as compared to the phase difference across the junction with coupling energy $E_{\text{min}}$.}. We neglect in our considerations the effect of a nonzero charging energy.
Fluctuations in $\chi(t)$ cause a fluctuating current $ I(t) = \partial E/ \partial \chi(t) = E_{\min} \chi$ and the number of electrons to the left and right of the junction therefore fluctuates. This does not affect the measurement outcomes on average because $ \chi$ fluctuates symmetrically around zero.

However, the fluctuations can affect the current noise, which means that one should repeat the cycle sufficiently many times to suppress noise effects. One may wonder whether the number of cycles needed to obtain an acceptable signal-to-noise ratio would not lead to measurement times that are long compared to, e.g., quasiparticle poisoning times. 
 To this end, we estimate the number $\Delta N_{\text{cycle}}$ of electrons that are transferred in a typical cycle due to a fluctuating phase $\chi(t)$ (in addition to the pumped charge). For simplicity, we assume that $\chi$ is constant for each cycle and changes from one cycle to the other randomly. The typical number for $\Delta N_{\text{cycle}}$ thus reads $\Delta N_{\text{cycle}} = I \Delta t = E_{\min} \sqrt{\langle \chi^2 \rangle} \Delta t$ where $\Delta t$ is the time duration of step 1 and 2 of the fusion-rule protocol. Treating $\chi$ as a classical variable, the average energy $E \sim E_{\min} \langle \chi^2 \rangle/2 $ for the phase degree of freedom is given by half the temperature, $T/2$, according to the equipartition theorem in thermal equilibrium. It follows
\begin{eqnarray}
  \Delta N_{\text{cycle}} = \sqrt{E_{\min} T} \Delta t.
\end{eqnarray}
To estimate $\Delta N_{\text{cycle}}$, we insert in the above expression $E_{\min}(t) = E_M^{\text{max}} = 10 E_C$ and $\Delta t \sim 10/E_C$ and $T=E_C/10$ from our considerations in the main part, which yields
\begin{eqnarray}
  \Delta N_{\text{cycle}} \lesssim 10.\label{eq:estimate}
\end{eqnarray}
This is an upper bound for $\Delta N_{\text{cycle}}$ for several reasons: (i) The adiabaticity condition derived in the main part used for the value of $\Delta t$ has been estimated conservatively. (ii) The Majorana coupling $E_M^{\max}$ is rather smaller than $10 E_C$ and, in addition, $E_{\min} \leq E_M^{\max}$ during the cycle. (iii) The presence of a charging energy suppresses charge fluctuations. (iv) The assumption of a constant phase during each cycle leads to a maximal value for $\Delta N_{\text{cycle}}$. If the correlation time of the phase fluctuations is smaller, i.e., phase fluctuations are much faster than the cycle period, the current fluctuations would average out. This means that the relevant $\Delta t$ would be much shorter than the time needed for the steps in the protocol. The fluctuating charge $ \Delta N_{\text{cycle}}$ should thus be rather smaller than the above estimate. Since any charge-pumping experiment will be carried out with a large number of cycles in practice, we thus expect these experiments to have an acceptable signal-to-noise ratio.

\subsection{Relation between $E_M$ and $E_{J, C}$}\label{app:emejratio}

The Majorana tunneling energy $E_M$ and the Josephson energy $E_{J, C}$ cannot
be controlled individually since both are tuned through the gate near the
central valve. We here give a crude estimate of how they are related. The Josephson energy
$E_{J, C}$ can be connected with the junction properties through the
Ambegaokar-Baratoff formula {\cite{TinkhamBook}}: $I_c R_n = \pi \Delta / 2
e$. Here, $I_c = 2 \pi E_{J, C} / \Phi_0$ is the critical supercurrent through
the junction with flux quantum $\Phi_0 = h / 2 e$. Furthermore, $R_n$ denotes
the normal resistance of the junction with $1 / R_n \sim ( e^2 / \hbar) \pi |
t |^2 \nu_L \nu_R$ with tunneling amplitudes $t$ and densities of states
$\nu_{L / R}$ of the left and right wire segments, respectively. This yields
the relation
\begin{eqnarray}
  E_{J, C} & = & \frac{\pi^2}{4} \Delta | t |^2 \nu_L \nu_R .  \label{eq:ej1}
\end{eqnarray}
Note that the tunneling amplitude $t$ in this expression is related to the
overlap of the extended {\tmem{bulk}} wave functions. The
Majorana-Josephson energy $E_M$ is also a tunneling amplitude but it depends on the overlap of the
{\tmem{Majorana}} bound states {\tmem{localized}} at the junction
{\cite{Flensberg10a}}. For simplicity, we take $E_M / t \sim \langle
\psi_L^{\tmop{maj}} | \nobracket \psi^{\tmop{maj}}_R \rangle / \langle \psi_L
| \nobracket \psi_R \rangle \sim L / \xi$. Here, we used that the bulk wave
functions $\psi_{L / R} ( x) \sim 1 / \sqrt{L}$ scale with the length $L$ of
the segments, while the Majorana wave functions $\psi_{L / R}^{\tmop{maj}} (
x) \sim 1 / \sqrt{\xi}$ rather scale with the coherence length $\xi$ of the
superconducting islands. To estimate the density of states, we model the wire
segments as effectively free one-dimensional electron gases, which yields
$\nu_{L / R} = \sqrt{2 m^{\ast} L^2 / E_F}$ with effective mass $m^{\ast}$ and
Fermi energy $E_F$. Inserting these relations into Eq. (\ref{eq:ej1}), we
obtain
\begin{eqnarray}
  E_{J, C} & = & \frac{\pi^2}{4} \Delta \frac{2 m^{\ast} \xi^2}{E_F} E_M^2 . 
  \label{eq:ej2}
\end{eqnarray}
Note that this relation is independent of the wire length as expected for a
property of the junction. In the topological regime, the Fermi energy lies in
the gap and can be estimated as $E_F \sim m^{\ast} \alpha^2$
{\cite{AliceaReview}} with the effective spin-orbit velocity $\alpha$.
Furthermore, in the strong spin-orbit regime, we can estimate $\xi \sim \alpha
/ \Delta$ as the ratio of the Fermi velocity and the gap (see Ref. \cite{SauSmokingGun} and supplemental material),
which gives the simple result
\begin{eqnarray}
  E_{J, C} & = & k \frac{E_M^2}{\Delta}
\end{eqnarray}
where $k$ is a constant of close to 1; for the above values we obtain $k \approx 5$, which we use for our numerical results. 

\subsection{Parameters choices and neglect of quasiparticle
excitations}\label{app:fusion-parameters}

Here we discuss the feasibility of the parameter ratios used in our time-scale estimations of the fusion-rule protocol and the braiding steps.
Since our model does not include quasiparticle excitations, it is applicable
only as long as the Majorana coupling satisfies $E_C,E_M^{\max} \lesssim \Delta$.
Here, $\Delta$ is the superconducting
gap, which we assume to be the same in the nanowire as in the bulk
superconductor. This is not unrealistic in view of recent experiments
{\cite{Marcus15_NatureNano_10_232,PoisoningTime}} measuring
$\Delta$ on the order of a few K (e.g. for aluminum). The charging energy
should thus be a few hundreds of mK. While charging energies in prototypical
nanowire structures are on the same order as $\Delta$ for wires of $\sim 1 - 2~
\mu$m length {\cite{Marcus15_NatureNano_10_232}}, lower charging energies
could be achieved by using longer nanowires. This would also be advantageous
for suppressing the overlap of the Majorana bound states within each wire
segment. Alternatively,
the ratio $E_C / \Delta$ could be reduced by using niobium as a proximal
superconductor instead, which has a larger gap reaching 8 K {\cite{Pronin98}}.
To initialize the MBS, it is furthermore important that $E_C, E_M^{\max} \gg T$, the
thermal energy of the nanowire. This seems possible for fridge
temperatures $\sim 20$ mK {\cite{Marcus15_NatureNano_10_232}}.
Furthermore, one can control the maximal Majorana energy $E_M^{\max}$ by the gates and keeping it between $T$ and $\Delta$ should be no problem.

For our numerical results, we employ a large value of $\Delta / E_C =100$ since we assume $E_C \ll E_M^{\max} \ll \Delta$ in our analysis. This allows us to
set $E_M^{\max} / E_C = 10$ in our numerical calculations. In practice, $E_M^{\max} \sim E_C $ should be sufficient (replacing $\log(E_M^{\max}/E_C) \rightarrow 1$ in the time-scale estimates) and smaller ratios $\Delta / E_C$ should thus work as well. Moreover, only the $E_{J,C}$ depends directly on $\Delta$ in our analysis and as long as $E_M^{\max} \ll \Delta$, $E_{J,C}$ is a rather small correction. 
 We explicitly checked from our numerical analysis that a smaller ratio $\Delta = 10 E_C$ does not lead to a suppression of the energy gaps in the spectrum.

The Josephson couplings to the bulk superconductors, $E_{J,\alpha}$, depend of course also on $\Delta$. However, the only important aspect is here that $E_{J, \alpha} / E_C$ is large enough to
achieve a good suppression of charging effects. At least $E_{J, \alpha}^{\max} / E_C > 10$ is
needed [at $E_{J, \alpha}^{\max} / E_C = 10$, one obtains $\varepsilon_P / E_C \sim
0.01$ from Eq. (\ref{eq:epsp})]. We take $E_{J,\alpha}^{\max} / E_C \sim 50$ in our numerical calculations to make sure that $E_J^{\max} \gg E_M^{\max}$.
We note that the energy splittings arising from the Josephson
energies scale as $\sim \sqrt{8 E_{J, C} E_C}, \sqrt{8 E_{J, \alpha}
E_C}$, which does not
exceed $\Delta$ for the values we assume. It is therefore reasonable to
exclude quasiparticle excitations from our numerical calculations.

\section{Numerical
diagonalization}\label{app:numerics}

In this appendix, we discuss the advantages of numerically diagonalizing the Hamiltonian in the number basis as compared to the phase basis.

\begin{center}
  \Figure{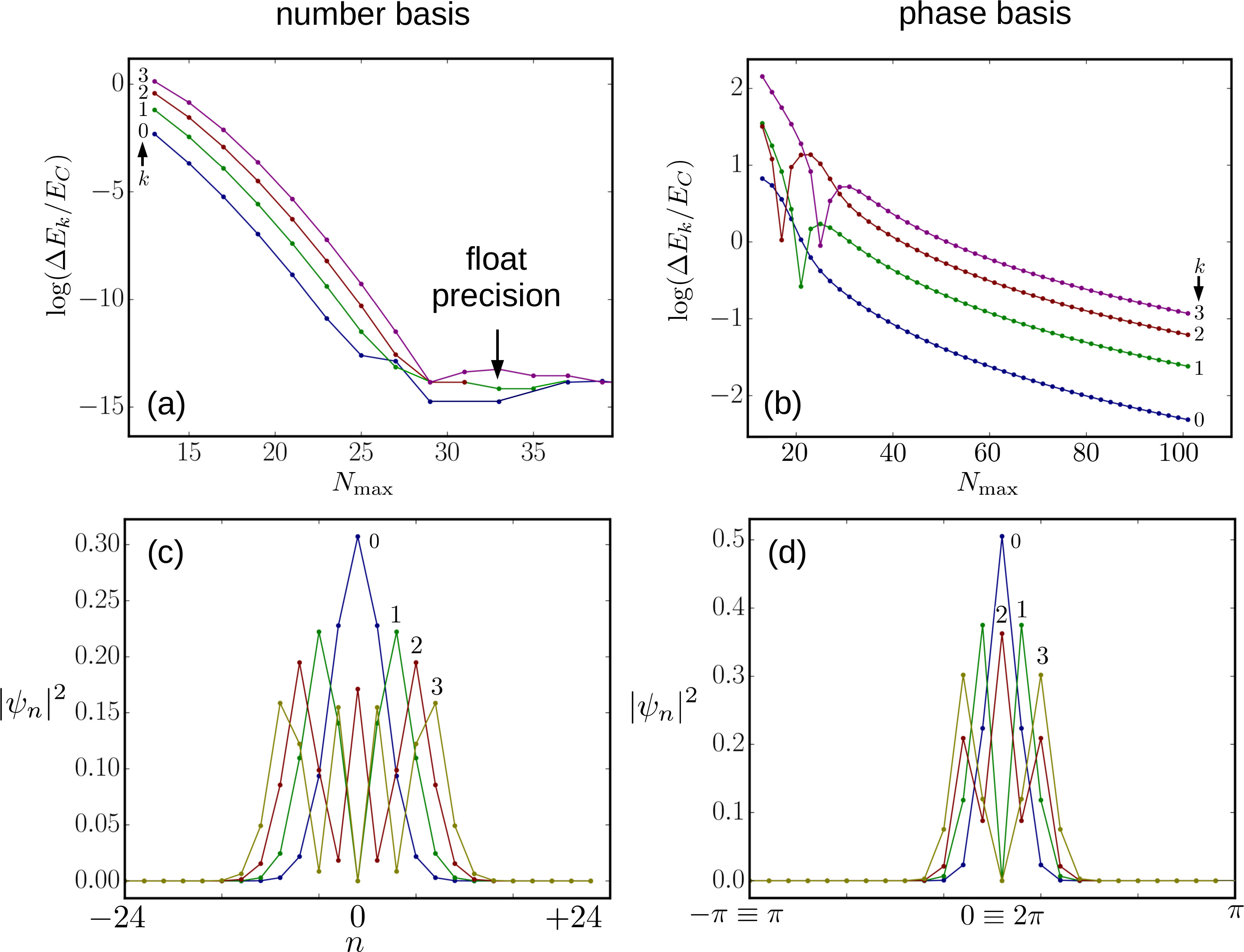}{Comparison
  of the convergence of the numerical approach in number space and 
  phase space. (a) and (b) Energy difference of the four lowest eigenenergies
  $\Delta E_k ( N_{\max}) = E_k ( N = N_{\max}) - E_k ( N = N_{\max} - 2)$ as
  a function of $N$ included basis states. The states are labeled in ascending
  order ($k = 0, 1, 2, 3$). We show the result both in number space (a) and in
  discretized phase space (b). The eigenenergies are obtained from a
  single-island Hamiltonian, i.e., $H = E_C \left( \op{n} - n_g \right)^2 +
  E_J \left( 1 - \cos \left( \op{\varphi} \right) \right)$, see Eq.
  (\ref{eq:hi}). In (c) and (d), we show the corresponding probability
  densities in number and discretized phase space for the case $N_{\max} =
  25$. \ Note that we only include even electron numbers (i.e., the points in
  (c) differ by $\Delta n = 2$). To compute the eigenenergies and
  eigenfunctions in phase space, we express the corresponding Schr{\"o}dinger
  equation in phase space with the replacement $\op{n}^2 \rightarrow -
  \partial^2 / \partial \varphi^2$ and discretize the phase variable, $\psi (
  \varphi) \rightarrow \psi ( \varphi_j)$, $j = 1, \ldots, N$, $\varphi_j = j
  \cdot 2 \pi / N_{\max}$ and the second derivative $\partial^2 \psi ( \varphi) /
  \partial \varphi^2 = \psi [ ( \varphi_{j + 1}) + \psi ( \varphi_{j - 1}) - 2
  \psi ( \varphi_j)] / ( 2 \pi / N_{\max})^2$. For the even-parity sector, we
  impose periodic boundary conditions, i.e., $\psi ( \varphi_{N_{\max} + 1}) =
  \psi ( \varphi_1)$. The results are shown for $E_J / E_C = 100$ and $n_g =
  0$.\label{fig:numerics}}
\end{center}

The number basis has a natural advantage because it is a countable basis due to the
discreteness of the electron charge. By contrast, the phase basis is generated
by a continuous variable, which has to be discretized in a numerical
procedure. One
then maps the Hamiltonian on a lattice model (see caption of Fig. \ref{fig:numerics} and Ref.~\cite{Boykin04}). Taking a uniform mesh in phase space, we compare in
{\color{black} Fig. \ref{fig:numerics}} the convergence of the lowest energy
eigenstates of a single-island Hamiltonian when increasing the number of basis
states both in number and in phase basis. We can clearly see that the number-basis
approach converges exponentially fast, whereas the phase-basis
approach yields only algebraic convergence.

The reason is that the wave function in phase space is sharply localized
near $\varphi = 0$ and therefore one needs a very fine resolution
to describe the wave function properly. When taking $N_{\max} = 25$ for the
parameters used in {\color{black} Fig. \ref{fig:numerics}}(d), the wave
function is localized around a few points around
$\varphi = 0$, which is equivalent to $\varphi = 2 \pi$. To improve the performance in phase space, one could use a nonuniform mesh for the discretization. However, a grid with a fine resolution around zero phase would only work optimal for energetically low-lying states, so it might be useful to adapt the grid for different parts of the spectrum.

One may of course
argue that limiting the maximally included number state to $| n | \leqslant
N_{\max}$ also introduces an approximation similar to using a finite number of
points in phase space. However, as we can see in {\color{black} Fig.
\ref{fig:numerics}}(c), the probabilities for different number states are
still localized only to a number of $n \lesssim 10$ states and they drop
exponentially for $n \gtrsim 10$. This is in accordance with our estimate
$N_{\max} \sim \sqrt{E_J / E_C} \sim 10$ given in {\color{black} Sec.
\ref{sec:numerics}} for $E_J / E_C = 100$ as used in Fig.
\ref{fig:numerics}. Moreover, the accuracy for energetically higher-lying states can be improved by including more charge states depending on its energy $E/E_C$. Thus, choosing a finite set of states for the numerical diagonalization is much easier in the number basis as compared to the phase basis.

Another advantage of the number basis approach is that the interpretation of
the wave functions is simpler. We are also interested in the regime when
the charging energy is the dominant energy scale (discussed in Sec.
\ref{sec:closeduncoupled}) which is useful for initialization and readout of parity
states. In this regime, the wave function is not localized in phase space but
in charge space. Also, number-space probability distributions are directly
related to the measurement outcomes. All these points are different from the
situations studied in other works
{\cite{BeenakkerBraiding,BraidingWithoutTransport}}, where charging effects
are always one of the smallest energy scales. 
In these cases, it may be advantageous to use the phase basis.

\section{Effect of central Josephson coupling and
asymmetries of bulk Josephson couplings}\label{app:low-energy}

In our discussion of the parameter regimes of the coupled topological Cooper pair boxes in \Sec{sec:spectrum} of the main part, we assumed (i) zero central Josephson coupling, $E_{J,C}=0$ and (ii) symmetric Josephson couplings to the bulk superconductors, $E_{J,L}=E_{J,R}$. We next extend this analysis by investigating the effects of loosening each of the assumptions separately. 

\subsection{Effect of central Josephson coupling}\label{app:coupledtopcpb}

\begin{center}
  \Figure{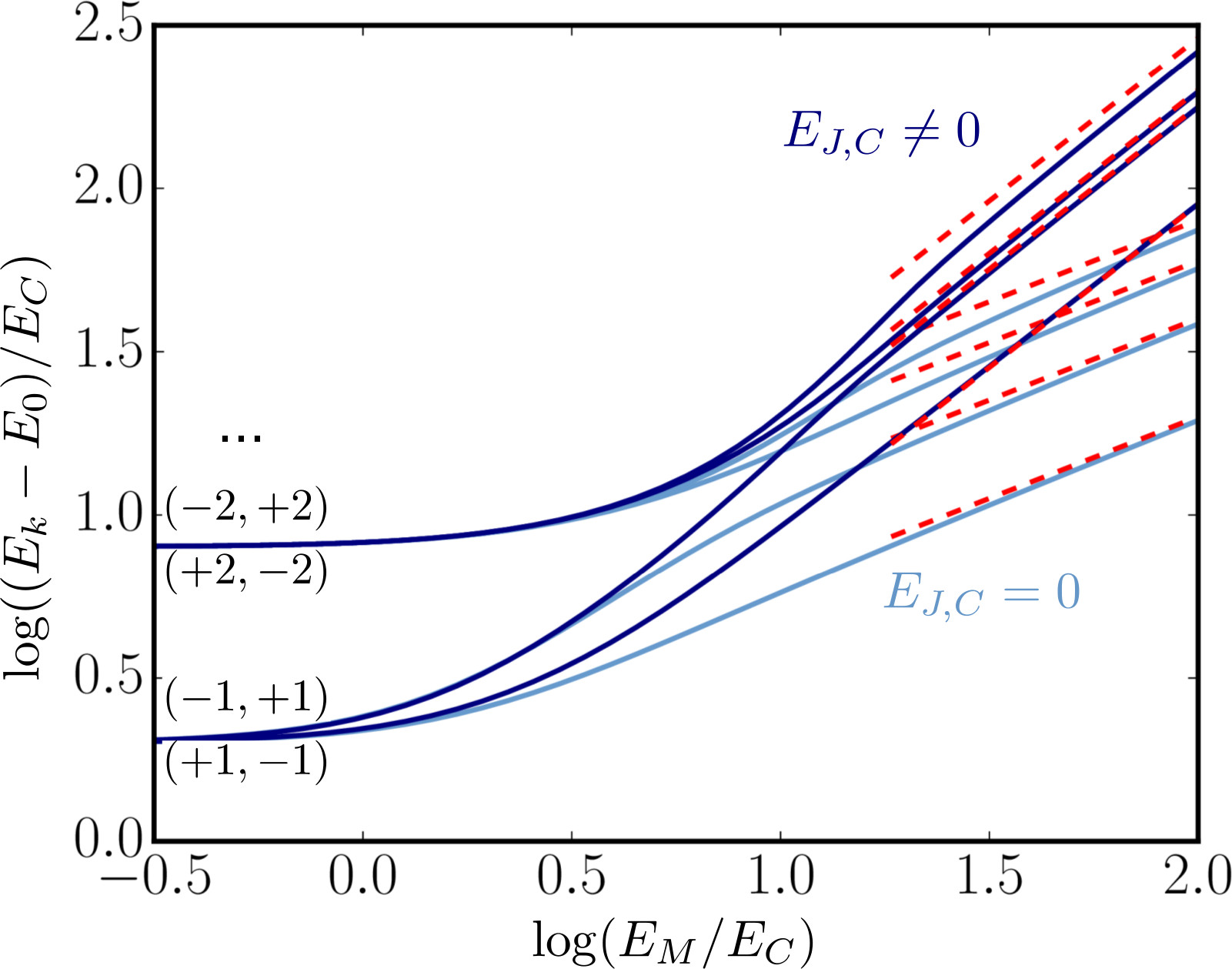}{Effect
  of the conventional Josephson coupling of the central junction $( E_{J, C})$
  on the low-energy spectrum. Solid lines show the numerically computed energy
  splitting between the first four excited states with zero excess charge, $N =
  0$, (at energies $E_k$) and the ground state (at energy $E_0$) as a function
  of the Majorana-Josephson coupling $E_M$. The center Josephson coupling is
  given by $E_{J, C} = 5 \cdot E^2_M / \Delta$, which we exclude $(
  E_{J, C} = 0)$ and include $( E_{J, C} \neq 0)$ from the computation as
  indicated. The red-dashed lines indicate the approximations for the
  splittings from Eq. (\ref{eq:tundomea0app}) for $E_{J, C} = 0$ and from Eq.
  (\ref{eq:tundomeapp}) for $E_{J, C} \neq 0$. All the states are continuously
  connected to charge states $( n_L, n_R)$ for $E_M = 0$ as indicated at
  curves on the left. The other parameters are $E_{C, L} = E_{C, R} = E_C
  \nocomma$, $E_{J, L} = E_{J, R} = 0$, $\Delta = 100 E_C$, and $n_{g, L} =
  n_{g, R} = 0$.\label{fig:appemdep}}
\end{center}

We first investigate the effect on the low-energy spectrum of including the central Josephson coupling
$E_{J, C}$. Any tunnel coupling leading to a
nonzero $E_M$ will also involve a nonzero $E_{J, C}$, i.e., these Josephson
couplings cannot be switched off individually. Our crude
estimate in {\color{black} App. \ref{app:emejratio}} yielded the relation
$E_{J, C} \sim E_M^2 / \Delta$, that is, the ratio $E_{J, C} / E_M$ is tunable
with $E_M$. Our analysis here shows that the relevant energy
splittings increase when $E_{J, C}$ is included (see Fig. \ref{fig:appemdep}). This implies that the time
scale estimates given in the main part are conservative since they assume the
``worst-case scenario'' of $E_{J, C} = 0$.

\begin{center}
  \Bigfigure{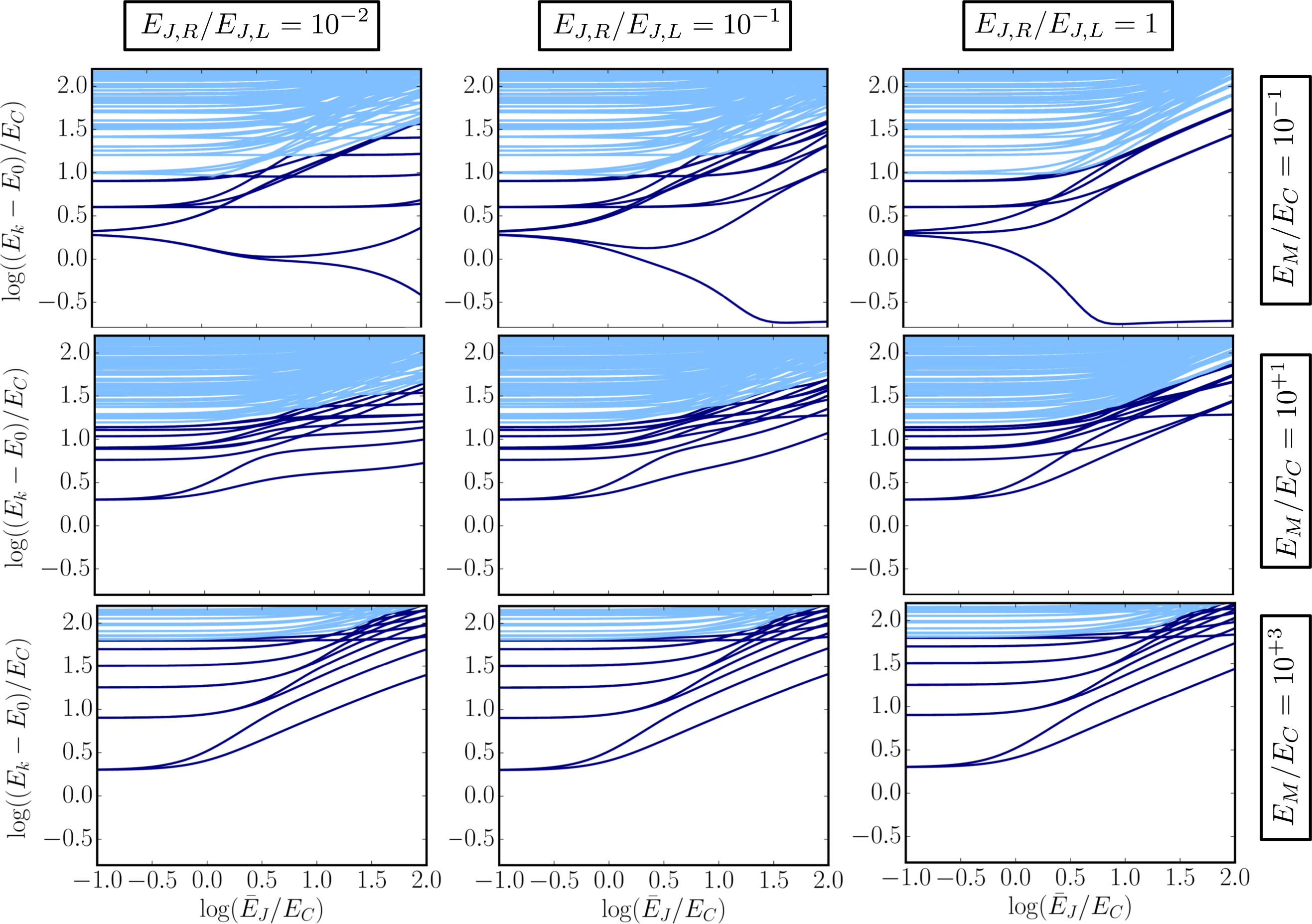}{Low-energy
  spectrum of two hybridized topological superconducting islands
  asymmetrically coupled to bulk superconductors. The figure shows the numerically
  computed energy splittings between succeeding excited states (at energy
  $E_k$) and the ground state (at energy $E_0$) as a function of the average
  Josephson coupling $\bar{E}_J = ( E_{J, L} + E_{J, R}) / 2$ to the
  bulk superconductors. We highlight the 12 lowest states in darker color
  for comparison with {\color{black} Fig. \ref{fig:emejdep}} and show all
  higher-lying states in brighter color. The Majorana coupling $E_M$ is
  changed from one row to the other while the asymmetry $E_{J, R} / E_{J, L}$
  of the Josephson couplings is changed from one column to the other as
  indicated. We use $n_{g, L} = n_{g, R} = 0$, $E_{J, C} = 0$, and $N_{\max} =
  20$.\label{fig:asymdep}}
\end{center}

For simplicity, we focus our discussion here on zero bulk Josephson
couplings, $E_{J, L} = E_{J, R} = 0$, and symmetric charging energies, $E_{C,
L} = E_{C, R} = E_C$. Using the notation of {\color{black} Sec.
\ref{sec:sumdiff}}, the Hamiltonian (\ref{eq:hsumdiff}) reads
\begin{eqnarray}
  H & = & 2 E_C \left[ \left( \tfrac{\op{N}}{2} \right)^2 + \left. \left(
  \tfrac{\Delta \op{n}}{2} \right. \right)^2 \right]  \label{eq:apph}\\
  &  & + E_M \cos \left( \tfrac{\Delta \op{\varphi}}{2} \right) i
  \op{\gamma}_2 \op{\gamma}_3 + E_{J, C} \left( 1 - \cos \left( \Delta
  \op{\varphi} \right) \right), \nonumber
\end{eqnarray}
As discussed in {\color{black} Sec. \ref{sec:singleisland}}, the total number
of electrons is conserved and it is sufficient to focus only on the $N = 0$
subspace. Equation (\ref{eq:apph}) is similar to the Hamiltonian of a
single Cooper pair box with the only difference that the phase-dependent term
includes two harmonics. We compare the resulting $E_M$-dependence of the
low-energy spectrum in {\color{black} Fig. \ref{fig:appemdep}} for $E_{J, C} =
0$ and $E_{J, C} = 5 E_M^2 / \Delta$ as indicated. The spectrum for $E_{J, C} = 0$
has been analyzed in {\color{black} Sec. \ref{sec:singleisland}} and for $E_M
\gg E_C$, the energy levels are well-approximated by
\begin{eqnarray}
  E_k & \approx & \sqrt{4 E_C E_M} ( k + 1 / 2) .  \label{eq:tundomea0app}
\end{eqnarray}
The simple physical picture is here that the Majorana term couples adjacent
charge states of the islands $( n_L, n_R) \leftrightarrow ( n_L \pm 1, n_R \mp
1)$, which leads to {\tmem{single-electron}} plasma oscillations.

This changes when the
Josephson coupling becomes nonneglible for $E_{J,C} \gtrsim E_M$. Let us derive
an approximation of the eigenenergies in the limit
$E_{J, C} / E_M \gg 1$ to contrast it with the physics of the case of $E_{J, C} = 0$.
In the former case, the system behaves as two copies of {\tmem{Cooper \
pair}}-plasma oscillators corresponding to the two possible parity degrees of
freedom of the island (even-even and odd-odd parity; note that we restrict our
considerations to even total parity). The ordinary Cooper pair tunneling
across the junction can only couple states within the even-even or odd-odd
parity sector, changing the number of electrons on both islands by two.
Neglecting the fractional Josephson term $\propto \cos \left( \Delta
\op{\varphi} / 2 \right)$, the eigenstates are given by the degenerate
Josephson plasmon states $| k, 0_{12} 0_{34} \rangle$ and $| k, 1_{12} 1_{34}
\rangle$ with energies $E_k = \sqrt{8 E_{J, C} E_C} ( k + 1 / 2)$.

The two parity states then become mixed due to the fractional Josephson term
$\propto \cos \left( \Delta \op{\varphi} / 2 \right)$, yielding bonding and
antibonding combinations for the eigenstates {\footnote{This is correct for
real $E_M$ assumed throughout the paper. For complex $E_M = e^{i \kappa} | E_M
|$, the eigenstates are given by $( | k, 0_{1 2} 0_{3 4} \rangle \pm e^{i
\kappa} | k, 1_{1 2} 1_{3 4} \rangle) / \sqrt{2}$.}}:
\begin{eqnarray}
  | k, 0_{2 3} 0_{1 4} \rangle & = & \tfrac{1}{\sqrt{2}} [ | k, 0_{12} 0_{34}
  \rangle + | k, 1_{12} 1_{34} \rangle],  \label{eq:symapp}\\
  | k, 1_{2 3} 1_{1 4} \rangle & = & \tfrac{1}{\sqrt{2}} [ | k, 0_{12} 0_{34}
  \rangle - | k, 1_{12} 1_{34} \rangle].  \label{eq:asymapp}
\end{eqnarray}
This corresponds to fusing the MBS pair $( \gamma_2, \gamma_3)$ localized at the
central junction. The eigenenergies are split according to the nonlocal
parity $\langle i \gamma_2 \gamma_3 \rangle = \pm 1$, yielding
\begin{eqnarray}
  E_{k, \pm} & = & \sqrt{8 E_{J, C} E_C} ( k + 1 / 2) \mp E_M, 
  \label{eq:tundomeapp}
\end{eqnarray}
where the minus (plus) sign corresponds to the (anti)symmetric combination, respectively. The associated energy
gaps to the ground state with energy $E_{0, +}$ are shown as red-dashed lined in
{\color{black} Fig. \ref{fig:appemdep}}. Provided the charging energy is
known, one could actually determine both $E_{J, C}$ and $E_M$ independently
from the low-energy spectrum in an easy way using Eq. (\ref{eq:tundomeapp}).

This shows that the physics and also the energy spectrum is different depending on 
whether the Josephson Cooper pair coupling or the single-electron tunneling dominates.

\subsection{Asymmetric Josephson couplings to bulk superconductors}\label{app:asymmetries}

In this Appendix, we turn to the more general situation when the Josephson
energies with the two bulk superconductors are asymmetric $( E_{J, L} \neq
E_{J, R})$. We show that the importance of asymmetries is suppressed the
larger the Majorana coupling $E_M$ becomes.

The effect of asymmetric Josephson energies on the low-energy spectrum is
illustrated in {\color{black} Fig. \ref{fig:asymdep}}: We change the ratio
$E_{J, R} / E_{J, L}$ in the horizontal direction and $E_M$ in the vertical
direction. The figure shows that asymmetries have a large effect on the
spectrum when the coupling $E_M$ between the islands is small ($E_M /
\bar{E}_J \ll 1$ with $\bar{E}_J = ( E_{J, L} + E_{J, R}) / 2$, upper row),
whereas the effect is rather small if the coupling $E_M$ is large ($E_M /
\bar{E}_J \gg 1$, lower row). This finding can be readily understood from the
representation (\ref{eq:hsumdiff}) in sum and difference variables. It takes
the form
\begin{eqnarray}
  H & = & \tilde{H} + ( E_{J, L} - E_{J, R}) \sin \left( \frac{\op{\Phi}}{2}
  \right) \sin \left( \frac{\Delta \op{\varphi}}{2} \right) 
\end{eqnarray}
with a part $\tilde{H}$ that is independent of the asymmetry, that is, it
depends only on the average $\bar{E}_J$. As soon as the phase dynamics are locked,
$\Delta \varphi = \varphi_L - \varphi_R \approx 0$ for $E_M \gg \bar{E}_J$,
the asymmetry-dependent term can be dropped. The simple physical picture is
that the system behaves as a single superconducting island connected to
the bulk superconductors by two Josephson junctions in parallel (with zero
flux enclosed).

This indicates that charging effects can be efficiently suppressed on
{\tmem{both}} segments even if only {\tmem{one}} of them is coupled to a bulk superconductor. This finding is relevant for transferring MBS in nanowire
networks as we discuss in {\color{black} Sec. \ref{sec:majeps}}. Interestingly, it turns out that the requirement $E_M \gg
\bar{E}_J$ is, in fact, not even needed for shuffling MBS between the
segments as we show in {\color{black} Sec. \ref{sec:majeps}}.

\section{Time-scale conditions for readout}\label{app:readout}

Sections {\sec{sec:fusion}} and {\sec{sec:braiding}} only focused so far on
the time scales required for preparing and manipulating MBS in nanowire structures.
In this Appendix, we complement these considerations by discussing the time
scales for performing the {\tmem{readout}} of the MBS.
 
\subsection{Charge sensing}

By tuning the islands to the charge-dominated regime 
different parity states of the island are mapped onto different charge states. The charge
state is then read out by simple charge sensors well-known from quantum-dot
physics. This technique works as long as the readout time, $\tau_D \ll 1 /
E_M^{\min}, 1 / E_{J, \alpha}^{\min}$, is much shorter than all time scales
related to tunneling of electrons or Cooper pairs, respectively. As discussed in {\color{black} Ref.
{\cite{Aasen15}}}, one expects $\tau_D < 1~\mu$s with standard charge readout techniques. 

\subsection{Charge-pumping readout}

An alternative readout method is to pump charge
conditioned upon the fermion parities of the islands. 
The steps of the pumping
procedure, explained in detail in {\color{black} Ref. {\cite{Aasen15}}}, are
sketched in {\color{black} Fig. \ref{fig:pumping}}(a). 
Before we derive
time-scale conditions for the pumping procedure, we briefly review the steps.

{\tmem{Parity-dependent charge pumping.}} 
The simplest way to understand this pumping
procedure is to follow the steps sketched in Fig. \ref{fig:pumping}(a). In step 1, one reduces the Josephson energies $E_{J, \alpha}$ from $E_{J, \alpha}^{\text{max}}$ after step 2 of the fusion-rule protocol (see Fig. \ref{fig:fusion-plane}) until they reach an energy $E_{J,
\alpha}^{\tmop{sw}}$ on the order of the charging energy $E_C$. In fact, reducing $E_{J, \alpha}$ is not needed at all as we explain at the end of this section but it is helpful for the illustration of the mechanism. We assume that the gate voltages are adjusted such that the lowest state of even-even parity is $| n_L, n_R
\rangle = | 0, 0 \rangle$, while the lowest state of odd-odd parity is given
by $| - 1, + 1 \rangle$. As indicated in Fig. \ref{fig:pumping}(b), one then in step 2 adiabatically sweeps the gate voltages from $n_{g,
\alpha}^{( 1)} \rightarrow n_{g, \alpha}^{( 2)}$. This is done such that one crosses a
charge-degeneracy point for odd-odd parity [mapping $| - 1, + 1 \rangle
\rightarrow | + 1, - 1 \rangle$] but not for even-even parity [leaving $| 0, 0
\rangle$ unchanged]. Consequently, a Cooper pair is transferred between the
islands and the bulk superconductors dependent on the island parities. To
complete the pumping, one closes the junctions to the bulk superconductors (step 3),
$E_{J, \alpha} \rightarrow E_{J, \alpha}^{\text{min}}$, opens the central valve $E_M \rightarrow E_M^{\max}$ (step 4), and sweeps the
gate voltages back to their original values: $n_{g, \alpha}^{( 2)} \rightarrow
n_{g, \alpha}^{( 1)}$ (step 5). Finally, one lets the system relax to the unique ground
state again.

Proceeding in this way, one expects for the fusion-rule experiment discussed in Sec.~\ref{sec:fusion} a pumping current of $I= 1/2 \times 2e \times f$, where $f$ is the pumping frequency. The factor 1/2 accounts for the 50\% probability to project the prepared state $ (\ket{0_{12}0_{34}} +\ket{1_{12}1_{34}} ) /\sqrt{2}$ on the odd-odd parity combination. This contrasts with the nontopological case: If there are no subgap states close to zero energy, the islands always remain in the even-even parity configuration and no charge is pumped. In the presence of notopological subgap states, it is further unlikely that the odd-odd combination appears with 50\% probability because these states are unlikely to be at exactly at zero energy. The probability should also depend sensitively on the parameters. We emphasize again that we assumed in our discussion that the total parity of the islands does not change even over many cycles.

We now come back to our initial comment that it is in fact not needed to lower the Josephson energies down
to $E_{J, \alpha} \sim E_C$ in step 1. The gate-voltage sweep in step 2 could
be carried out for {\tmem{any}} $E_{J, \alpha}^{\tmop{sw}}$ as long as $E_{J,
\alpha}^{\tmop{sw}} \gtrsim E_C$. The only important aspect is that
the gate voltage is swept over a point that would become a
charge-degeneracy point for $E_{J, \alpha}=0$ for the odd-odd
parity subspace [$n_{g, \alpha} = 0$ in {\color{black} Fig.
\ref{fig:pumping}}(b)]. By contrast, one must not cross over points that would become charge-degenerate for $E_{J, \alpha}=0$ for the even-even subspace [$n_{g, \alpha} =
\pm 1$ in {\color{black} Fig. \ref{fig:pumping}}(b)]. The projection on a
particular charge state happens when closing the outer valves $( E_{J, \alpha}\rightarrow E_{J, \alpha}^{\text{min}})$.
The time-scale estimation in the next paragraph shows that it is even favorable to perform the gate-voltage sweep at $E^{\tmop{sw}}_{J, \alpha} = E_{J, \alpha}^{\max}$.

\begin{center}
  \Bigfigure{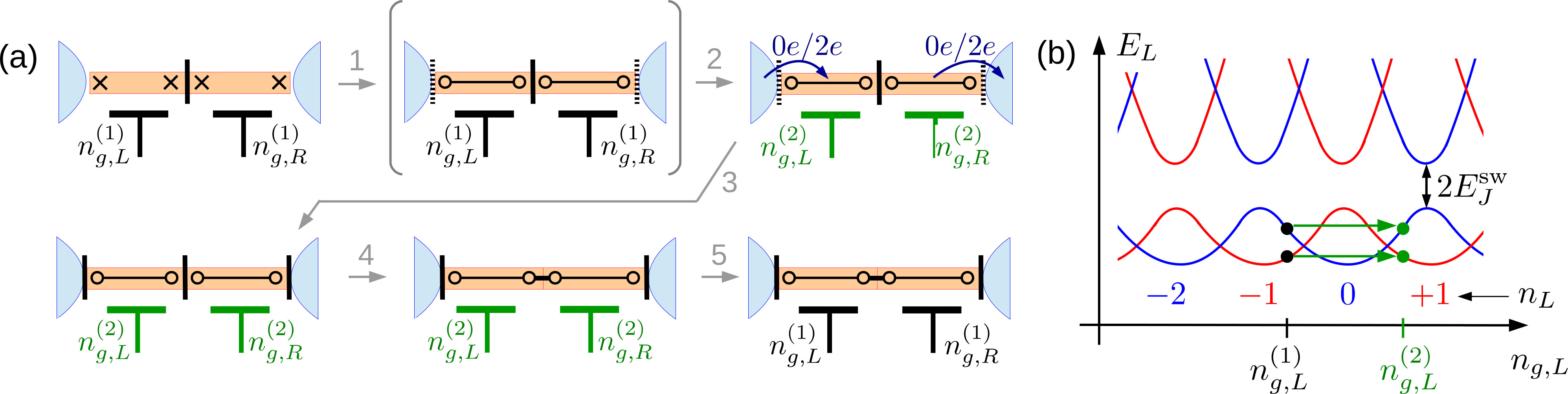}
  {Parity-dependent
  charge pumping. (a) Sketch of the steps of the pumping process. (b) Sketch
  of the energy spectrum for the left island as a function of the gating
  $n_{g, L}$. Sweeping the gating from $n_{g, L}^{( 1)} \rightarrow n^{(
  2)}_{g, L}$ changes the charge on the island by 0 if the parity is even
  (blue) and by $2 e$ if the parity is odd (red). This works oppositely for the right island with $n_{g, R}^{( 1)}=n_{g, L}^{( 2)}$ and $n_{g, R}^{( 2)}=n_{g, L}^{( 1)}$. Note that the labeling of
  the charge states is not valid near the charge-degeneracy points, where
  states that differ by one Cooper pair mix with each
  other.\label{fig:pumping}}
\end{center}

{\tmem{Time-scale condition.}} We next show that this charge pumping can
proceed adiabatically if the gate voltage sweep in step 2 is performed on a
time scale
\begin{eqnarray}
  T_2 & \gg & \frac{1}{E_{J, \alpha}^{\tmop{sw}}}  \text{$\text{ \ \ \ } (
  E^{\tmop{sw}}_{J, \alpha} \gtrsim E_C)$} .  \label{eq:tpump}
\end{eqnarray}
It is thus most favorable to perform the gate sweep at $E^{\tmop{sw}}_{J, \alpha}=E_{J,\alpha}^{\text{max}}$.
The manipulations of the Josephson and Majorana tunneling energies have to be
performed on time scales $T_i \gg 1 / E_C$ as worked out earlier in the main part.

To verify Eq. (\ref{eq:tpump}), we again start from the adiabaticity condition
(\ref{eq:fusion-adiabatic2}). Ideally, the gate voltages should be manipulated
such that the Josephson and Majorana energies are kept constant and only the
charging energy is manipulated through the gatings $n_{g, \alpha}$. In this
case, we can express Eq. (\ref{eq:fusion-adiabatic2}) as
\begin{eqnarray}
  T_2 & \gg & \int_{n_{g, \alpha}^{( 1)}}^{n_{g, \alpha}^{( 2)}} d n_{g,
  \alpha} \left| \frac{\langle \psi_k | \partial H / \partial n_{g, \alpha} | \psi_0
  \rangle}{( E_k - E_0)^2} \right| .
\end{eqnarray}
We check the above condition for the lowest excited state ($k = 1$). Using $| \langle \psi_1 ( t) | \partial H ( t) / \partial n_{g, \alpha} | \psi_0 ( t)
\rangle | \sim E_C$, $n_{g, \alpha}^{( 2)} - n_{g, \alpha}^{( 1)} \sim 1$, and
estimating the minimal gap as $E_1 - E_0 \geqslant \sqrt{E_{J,
\alpha}^{\tmop{sw}} E_C}$ {\footnote{This holds both when $E_{J,
\alpha}^{\tmop{sw}} \sim E_C$, because $E_1 - E_0 \geqslant 2
E^{\tmop{sw}}_{J, \alpha}$ as {\color{black} Fig. \ref{fig:pumping}}(b)
illustrates, but also when $E_{J, \alpha}^{\tmop{sw}} \gg E_C$ when $E_1 - E_0
\approx \sqrt{8 E_{J, \alpha}^{\tmop{sw}} E_C}$.}}, one obtains Eq.
(\ref{eq:tpump}).

The steps of changing the Majorana coupling $E_M$ and the Josephson coupling
$E_{J, \alpha}$ (steps 1, 3, and 4) have to be performed on the same time
scales given by Eqs. (\ref{eq:timescalelower}) and (\ref{eq:timescaleupper})
as derived in {\color{black} Sec. \ref{sec:fusion}}. Sweeping the gate
voltages back to their original values (step 5) does not need to be done
particularly slowly because when we open the central valve again, the system
must have time to relax into the even-even parity ground state anyway. However, it is again favorable to perform this step adiabatically in other to avoid that the system is further excited (see discussion of step 4 in the fusion-rule protocol in \Sec{sec:step4})

This, in sum, illustrates that the additional pumping steps in the current-readout
scheme are in fact {\tmem{fast}} compared to the Majorana manipulation time
scales and also usual charge sensing times (see {\color{black} Ref.
{\cite{Aasen15}}}). The disadvantage is that the current-readout scheme only
allows for an ensemble measurement. Moreover, the topological Majorana case can be distinguished from nontopological Andreev states on the islands just by comparing with the predicted magnitude of the current ($I=e f$). Trivial subgap states close to zero energy would lead in general to different, parameter-dependent but still nonzero pumping current since these states most probably lead to different probabilities for the even-even and odd-odd parity outcome.

\section{Effect of transitions into higher excited states}\label{app:higherexcited}

In this Appendix, we argue that the adiabaticity criteria we derived in Secs.
{\sec{sec:fusion}}, {\sec{sec:braiding}}, and {\color{black} App.
\ref{app:readout}} are valid even if we account for the possibility of
transitions into higher excited states. The criteria worked out there only
account for transitions to the lowest accessible state while transitions into
higher-lying excited states have been disregarded. However, they may be
important, in principle, because there are {\tmem{many}} excited states and
their cumulative effect might lead to stricter adiabaticity criteria than 
those derived in the foregoing section. Note that we do not take
quasiparticle states into account here; we merely include the
excited states within our Hamiltonian (\ref{eq:hfull}).

We argue in this Appendix that even if we sum condition
(\ref{eq:fusion-adiabatic}) over all excited states, one can still satisfy the
condition
\begin{eqnarray}
  \sum_k g_k & \ll & \Delta t,  \label{eq:sumad}
\end{eqnarray}
with the contributions
\begin{eqnarray}
  g_k & = & \int_C d \vecg{\lambda} \cdot \left| \frac{\langle \psi_k (
  \vecg{\lambda}) | \partial H ( \vecg{\lambda}) / \partial
  \vecg{\lambda} | \psi_0 ( \vecg{\lambda}) \rangle}{[ E_k (
  \vecg{\lambda}) - E_0 ( \vecg{\lambda})]^2} \right|, 
\end{eqnarray}
without major corrections to the time scales estimated in Secs.
{\sec{sec:fusion}}, {\sec{sec:braiding}}. In all the steps we considered in these sections, we
change only a single parameter $\lambda = E_M, E_{J, \alpha}$.
Our analysis in the main part showed that one traverses (maximally) two regimes
when changing $\lambda$: the charge-dominated regime, $\lambda \lesssim E_C$, 
and the tunneling / Josephson dominated regime, $\lambda \gtrsim
E_C$. We next discuss these regimes separately.

\subsection{Charge-dominated part}

To make a conservative estimate for condition (\ref{eq:sumad}) in the
charge-dominated regime, we use that the eigenenergies of the Hamiltonian are
roughly given by $E_{n_L n_R} = E_C ( n_L^2 + n_R^2)$. 
Here, we label the
states by the number of electrons $n_L$ and $n_R$ on the left and right
island, respectively, and assume that the number of electrons in the ground
state $| \psi_0 ( t) \rangle$ is close to $n_L = n_R = 0$. In particular, we assumed here that the gatings $n_{g,L}=n_{g,R}=0$ but this assumption is uncritical because the quadratic increase of the energies with $n_L,n_R$ is also present for other gatings. The quadratic increase is important for the suppression of the effect of higher-lying states.
Furthermore, the
resulting energy gaps are larger when corrections due to the Majorana and
/ or Josephson couplings are included so that condition (\ref{eq:sumad})
should be fulfilled even more easily. 
\begin{eqnarray}
  \sum_k g_k & \leqslant & g_1 \times \left( \sum_{n_L, n_R \geqslant 1} \frac{1}{[
  n_L^2 + n_R^2]^2} \right), 
\end{eqnarray}
with $g_{10} = \int_{\lambda_{\min}}^{\lambda_{\max}} d \lambda / E_C^2$. Replacing the sum by 1, we would recover the time-scale estimate given in the main part \footnote{Note that $g_{10}$ refers here to an excited state with odd total parity ($n_L=1,n_R=0$ or $n_L=0,n_R=1$), which cannot be reached from the ground state. However, in the main part we use $E_1 -E_0 =2 E_C \sim E_C$, which is why  $g_{10} = \int_{\lambda_{\min}}^{\lambda_{\max}} d \lambda / E_C^2$ reproduces the result from the main part.}.
In writing down the above equation, we have also exploited that for all our estimates in Secs. {\sec{sec:fusion}},
{\sec{sec:braiding}}, and {\color{black} App. \ref{app:readout}}, we never
compute the matrix element $| \langle \psi_k | \partial H ( t) / \partial
\vecg{\lambda} | \psi_0 \rangle |$ (unless it is zero) but instead
estimate it by its maximal value.

One can easily check that the sum on the right-hand side just gives a
correction factor close to 1: Using the inequality $n_L^2 + n_R^2 \geqslant 2
n_L n_R$, we get
\begin{eqnarray*}
  \sum_{|n_L|, |n_R| \geqslant 1} \frac{1}{[ n_L^2 + n_R^2]^2} & \leqslant &
  \left( \sum_{n_L \geqslant 1} \frac{1}{n_L^2} \right) \left( \sum_{n_R \geqslant 1} \frac{1}{n_R^2}
  \right) \\ &\approx& 2.7,
\end{eqnarray*}
where we used $\sum_{n \geqslant 1} 1 / n^2 \approx 1.64$. This shows that the
adiabaticity criterion can indeed be verified by just considering the lowest
excited state.

\subsection{Tunneling-dominated part}

Let us next turn to the tunneling-dominated regime. Here, the energy gaps of
the excited states to the ground state read $E_k - E_0 = k E_{1 0}$ with the
gap $E_{1 0} = \sqrt{\alpha \lambda E_C}$ ($\lambda = E_M, E_J$) of the first excited state to the
ground state ($\alpha$ is a proportionality constant) {\footnote{Note that
this formula does not hold in step 2 of the fusion-rule protocol discussed in
{\color{black} Sec. \ref{sec:step2}}; however, we can use $E_{k 0} \geqslant
k \sqrt{\alpha E_C E_M}$ there and therefore apply our considerations in this
section. }}. The transition matrix elements are given by $| \langle
\psi_k ( t) | \partial H / \partial \lambda | \psi_0 ( t) \rangle | \sim |
\langle k | A | 0 \rangle |$ with $A = \cos \left( \op{\varphi} \right)$ for
the Josephson energy terms or $A = \cos \left( \op{\varphi} / 2 \right) i \hat{\gamma}_2 \hat{\gamma}_3$ for
the Majorana energy terms. These matrix elements are strongly suppressed with
increasing $k$ since
\begin{eqnarray}
  \left. \begin{array}{r}
    \left| \langle k | \cos \left( \op{\varphi} / 2 \right) | 0 \rangle
    \right|\\
    \left| \langle k | \cos \left( \op{\varphi} \right) | 0 \rangle \right|
  \end{array} \right\} & = & \frac{c_k}{k!} \left| \langle k | \op{\varphi}^k
  | 0 \rangle \right| \propto \frac{1}{k!}, 
\end{eqnarray}
$\text{}$with $c_k$ on the order of 1 or zero if $k$ is odd. One may therefore
estimate
\begin{eqnarray*}
  g_k & \sim & g_1 \frac{1}{k^2 k!} .
\end{eqnarray*}
Using next that each of the plasmon states is maximally $2 k$-fold degenerate,
we obtain
\begin{eqnarray*}
  \sum_k g_k & \lesssim & g_0  \sum_{k \geqslant 1} k \times \frac{1}{k^2 k!}
  \leqslant e g_0,
\end{eqnarray*}
which again results in a correction factor of order 1 compared to the
time-scale estimate derived for the lowest excited state ($e$ in the above expression is the Euler number).

\section{Screening of charging energy by Majorana tunnel coupling}\label{app:eps-extract}

In this Appendix, we discuss the screening of the charging energy of a wire
segment when coupled to another nanowire segment through a Majorana coupling
of the form (\ref{eq:ht}). Let us concretely consider the situation when the
left valve in the segmented wire structure in {\color{black} Fig.
\ref{fig:model}} is open $( E_{J, L} \gg E_C)$, while the right valve is
closed $( E_{J, R} = 0)$. The question is how the energy gap between the two local
parity states on the right island ($p_{34}=0,1$) changes when the Majorana coupling $E_M$ is
increased by opening the central valve.

We employ for this purpose the following toy Hamiltonian (\ref{eq:hmaj1}),
which only includes the parity degrees of freedom:
\begin{eqnarray}
  H_{\tmop{eff}} & = & i E_M \op{\gamma}_2 \op{\gamma}_3 + i \varepsilon_{P,
  R}  \op{\gamma}_3 \op{\gamma}_4. 
  \label{eq:heffmaj}
\end{eqnarray}
We use here the notation introduced in {\color{black} Sec.
\ref{sec:model}}. The above Hamiltonian crucially relies on
the assumption $E_{J, L}, E_M \gg E_{C, R}$, in which both segments can be
treated as being grounded. Otherwise the Cooper pair condensate can not be
ignored and the full Hamiltonian including all charge states has to be used.
Equation (\ref{eq:heffmaj}) does thus not incorporate a term of the form $\sim
i \varepsilon_{P, L} \op{\gamma}_1 \op{\gamma}_2$ because the corresponding
charging-induced energy splitting $\varepsilon_{P, L} \propto e^{- \sqrt{8
E_{J, L} / E_{C, 1}}}$ is much smaller than $E_M$ and $\varepsilon_{P, R}$
provided $E_{J, L} \gg E_M$. This leaves us first with the Coulomb-induced
coupling $\varepsilon_{P, R}$ between MBS $\gamma_3$ and $\gamma_4$, which is
in principle a function of all parameters of the model, in particular $E_M$
and $E_{J, L}$. Second, we have to account for the tunneling-induced coupling
of MBS $\gamma_2$ and $\gamma_3$, where we neglected fluctuations in the phase
difference of both islands contained in the fractional Josephson energy
(\ref{eq:hfull}), i.e., the effect of the Cooper pair condensate. This works here in order to study the two lowest states, which exhibit no plasma
oscillations.

\begin{center}
  \Figure{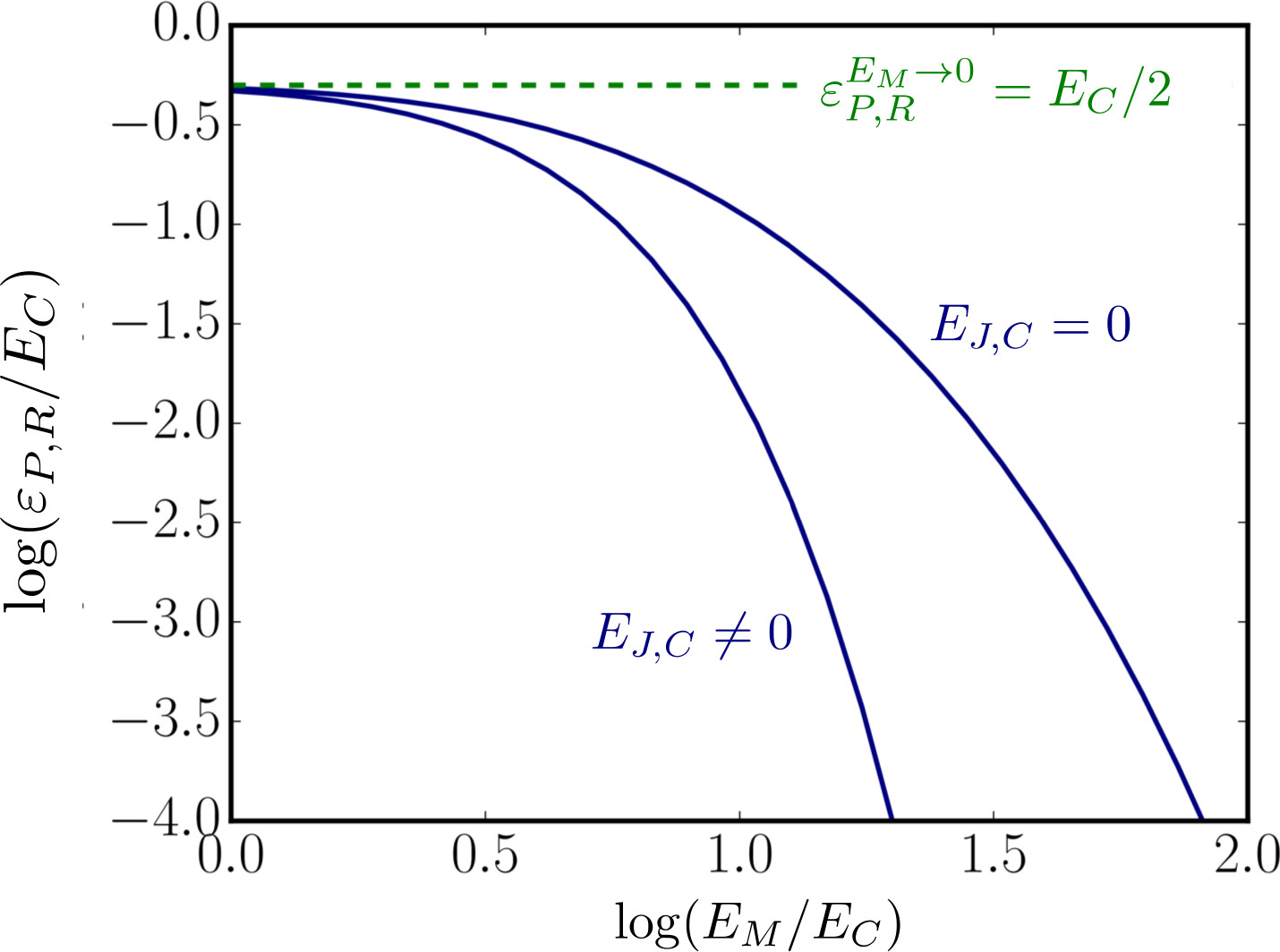}{Numerically
  computed energy gap $\varepsilon_{P, R}$, Eq. (\ref{eq:epsapp}) (solid
  blue lines), between odd and even parity states on the right island as a function
  of the central Majorana coupling $E_M$. We also indicate the parity energy splitting
  $\varepsilon_{P, R}  ( E_M \rightarrow 0) = E_{C, R} / 2$, corresponding to the
  two lowest charge states in the limit of vanishing Majorana coupling (green dashed).
  For the central Josephson coupling, we use $E_{J, C} = 5 E_M^2 / \Delta$ if
  nonzero as indicated and the other parameters are given by $E_{C, R} = E_{C,
  L} = E_C$, $n_{g, L} = n_{g, R} = 0, E_{J, L} = 100 E_C, E_{J, R} = 0$,
  $\Delta = 100 E_C$.\label{fig:eps}}
\end{center}

The Hamiltonian (\ref{eq:heffmaj}) admits two MBS at zero energy. One of them
is $\gamma_1$ (since it is decoupled from the other MBS) and the other one is
\begin{eqnarray}
  \op{\gamma}_0 & = & a \op{\gamma}_2 + b \op{\gamma}_4 \text{ \ = \ }
  \frac{\varepsilon_{P, R} \op{\gamma}_2 + E_M
  \op{\gamma}_4}{\sqrt{\varepsilon^2_{P, R} + E_M^2}} .  \label{eq:gamma0app}
\end{eqnarray}

The ground state is thus degenerate with respect to the occupation of the
fermionic mode $\op{f}_{1 0} = \left( \op{\gamma}_1 + i \op{\gamma}_0 \right)
/ 2$ with number operator $\op{n}_{1 0} = \left( 1 + i \op{\gamma}_1
\op{\gamma}_0 \right) / 2$. The expectation values $\left\langle \op{n}_{1 0}
\right\rangle = 0, 1$ characterize the total parity of the two islands. The
ground state with even total parity, $\left\langle \op{n}_{1 0} \right\rangle
= 0$, satisfies
\begin{eqnarray}
  - 1 = \left\langle i \op{\gamma}_1 \op{\gamma}_0 \right\rangle & = &
  \frac{\varepsilon_{P, R} \left\langle i \op{\gamma}_1 \op{\gamma}_2
  \right\rangle + E_M \left\langle i \op{\gamma}_1 \op{\gamma}_4
  \right\rangle}{\sqrt{\varepsilon^2_{P, R} + E_M^2}} .  \label{eq:parityeq}
\end{eqnarray}

To extract $\varepsilon_{P, R}$ as a function of the parameters, we compute
expectation values on the right hand side numerically. The expectation
value $\left\langle i \op{\gamma}_1 \op{\gamma}_2 \right\rangle$ is connected to that of the operator introduced by Eq.
(\ref{eq:sigmaz}):
\begin{eqnarray}
  \left\langle i \op{\gamma}_1 \op{\gamma}_2 \right\rangle & = & \left\langle
  \op{P}_{o o} \right\rangle - \left\langle \op{P}_{e e} \right\rangle.
\end{eqnarray}
To obtain  $\left\langle i \op{\gamma}_1 \op{\gamma}_4 \right\rangle$, we can exploit the fact that the Hilbert space of the effective Hamiltonian \eq{eq:heffmaj} is two-dimensional when restricted to the two lowest states with even total parity. In this subspace, we can define Pauli matrices
$\hat{\sigma}_z = i \op{\gamma}_1 \op{\gamma}_2 =i \op{\gamma}_3 \op{\gamma}_4  $ and $\hat{\sigma}_x = i \op{\gamma}_2 \op{\gamma}_3= i \op{\gamma}_1 \op{\gamma}_4 $ (but the equalities are valid only in this subspace, not in general). For a pure state, the corresponding Bloch vector must have length 1 and this implies $ \langle \hat{\sigma}_x \rangle ^2 +\langle \hat{\sigma}_z \rangle ^2=1  $ and thus
\begin{eqnarray}
  \left\langle i \op{\gamma}_1 \op{\gamma}_4 \right\rangle & = & - \sqrt{1 -
  \left\langle i \op{\gamma}_1 \op{\gamma}_2 \right\rangle^2} . 
\end{eqnarray}
(The sign follows from the fact that the energy of the ground state must be minimal.)
Equation (\ref{eq:parityeq}) can now be solved for $\varepsilon_{P,
R}$,
\begin{eqnarray}
  \varepsilon_{P, R} & = & \frac{E_M}{\sqrt{1 / \left\langle i \op{\gamma}_1
  \op{\gamma}_2 \right\rangle^2 - 1}},  \label{eq:epsapp}
\end{eqnarray}
implying for the coefficients in Eq. (\ref{eq:gamma0app}):
\begin{eqnarray}
  a \text{ \ = \ } \left\langle i \op{\gamma}_1 \op{\gamma}_4 \right\rangle, &
  & b \text{ \ = \ } \sqrt{1 - \left\langle i \op{\gamma}_1 \op{\gamma}_2
  \right\rangle^2} . 
\end{eqnarray}

The result for the coefficient $a$ has been shown and discussed in
{\color{black} Sec. \ref{sec:majeps}}. We briefly discuss here the dependence
of the parity splitting $\varepsilon_{P, R}$ on $E_M$, which is shown in
{\color{black} Fig. \ref{fig:eps}}. Indeed, $\varepsilon_{P, R}$ is not
constant as a function of $E_M$ but becomes exponentially small for $E_M \gg
E_C$. This shows that the charging energy on the right island can be
effectively suppressed by opening the central junction. Importantly, this does
{\tmem{not}} require the system to be tuned into the single island regime
$E_M \gg E_{J, L}$ (cf. {\color{black} App. \ref{app:asymmetries}}): In
the regime $E_M \ll E_{J, L}$, the left island functions similar to a bulk
superconductor since its Josephson plasma ground state has a distribution in charge
space that is much broader than that of the right island.

In {\color{black} Fig. \ref{fig:eps}}, we also show results down to $E_M \sim
E_{C, R}$ even though Eq. (\ref{eq:heffmaj}) is strictly speaking not valid in
this regime. What we intend to illustrate is that our procedure nevertheless
approaches the correct value for $\varepsilon_{P, R}$ in the limit $E_M /
E_{C, R} \rightarrow 0$, which is simply half the energy difference between
the two lowest charge states $n_R = 0, 1$ (green dashed line).

We thus have argued that gate-controlled transfer of MBS between two nanowire
segments is possible by tuning just a {\tmem{single}} valve -- the one
connecting the two nanowire segments.

\bibliographystyle{apsrev}
\end{document}